\documentclass[a4paper,11pt]{article}
\pdfoutput=1 
\usepackage{jheppub} 
\usepackage[T1]{fontenc} 
\usepackage{amsmath,amssymb,amsfonts,cases}
\usepackage{graphicx,caption,subcaption}
\usepackage{dcolumn}
\usepackage{bm}
\usepackage{bbm}
\usepackage{slashed}
\usepackage{wrapfig}

\newcommand{\wbar}{\overline{w}}
\newcommand{\Gbar}{\overline{G}}
\newcommand{\tr}{\text{tr}}
\newcommand{\beq}{\begin{equation}}
\newcommand{\eeq}{\end{equation}}
\newcommand{\bea}{\begin{eqnarray}}
\newcommand{\eea}{\end{eqnarray}}
\newcommand{\bi}{\begin{itemize}}
\newcommand{\ei}{\end{itemize}}
\newcommand{\ben}{\begin{enumerate}}
\newcommand{\een}{\end{enumerate}}

\def\Tr{{\rm Tr}}

\newcommand{\N}{{\mathcal N}}

\newcommand{\see}{S_{\textrm{EE}}}

\newcommand{\gn}{G_{\textrm{N}}}

\newcommand{\g}{\gamma}

\newcommand{\la}{\lambda}

\newcommand{\nn}{\nonumber}

\newcommand{\vr}{\varrho}
\newcommand{\bbh}{b_{\textrm{BH}}}
\newcommand{\bns}{b_{\textrm{NS}}}
\newcommand{\bfree}{b_{\textrm{free}}}

\def\O{\Omega}

\title{\LARGE Wilson Surface Central Charge from Holographic Entanglement Entropy}
\author[1]{John Estes,}
\author[1]{Darya Krym,}
\author[2]{Andy O'Bannon,}
\author[2]{Brandon Robinson,}
\author[2]{and Ronnie Rodgers}

\affiliation[1]{New York City College of Technology, City University of New York,
300 Jay Street, Brooklyn, NY, 11201, USA}

\affiliation[2]{STAG Research Centre, Physics and Astronomy, University of Southampton, Highfield, Southampton SO17 1BJ, UK}
\emailAdd{jestes@citytech.cuny.edu}
\emailAdd{daryakrym@gmail.com}
\emailAdd{a.obannon@soton.ac.uk}
\emailAdd{B.J.Robinson@soton.ac.uk}
\emailAdd{R.J.Rodgers@soton.ac.uk}

\abstract{We use entanglement entropy to define a central charge associated to a two-dimensional defect or boundary in a conformal field theory (CFT). We present holographic calculations of this central charge for several maximally supersymmetric CFTs dual to eleven-dimensional supergravity in Anti-de Sitter space, namely the M5-brane theory with a Wilson surface defect and three-dimensional CFTs related to the M2-brane theory with a boundary. Our results for the central charge depend on a partition of $N$ M2-branes ending on $M$ M5-branes. For the Wilson surface, the partition specifies a representation of the gauge algebra, and we write our result for the central charge in a compact form in terms of the algebra's Weyl vector and the representation's highest weight vector. We explore how the central charge scales with $N$ and $M$ for some examples of partitions. In general the central charge does not scale as $M^3$ or $N^{3/2}$, the number of degrees of freedom of the M5- or M2-brane theory at large $M$ or $N$, respectively.}


\begin{document}
\maketitle
\flushbottom

\section{Introduction}
\label{sec:intro}

M-theory is currently the leading candidate for an ultra-violet (UV) complete theory of quantum gravity. M-theory is defined as the (presumably unique) UV completion of 11d supergravity (SUGRA), which is a remarkably simple theory~\cite{Becker:2007zj,West:2012vka}: the only bosonic fields are the metric and a three-form, $C_3$. The stable BPS solitons of 11d SUGRA include M2- and M5-branes charged under $C_3$ electrically and magnetically, respectively. A complete formulation of M-theory will necessarily entail understanding these M-branes and other non-perturbative objects of 11d SUGRA more fully.

In a fashion similar to strings ending on D-branes or D-branes ending on other D-branes~\cite{Strominger:1995ac}, M2-branes can end on M2- or M5-branes, and M5-branes can end on other M5-branes. When the distance between two parallel D-branes shrinks to zero, the strings stretched between them become massless point particles and give rise to non-Abelian gauge multiplets. The low-energy worldvolume theory of multiple coincident D-branes is thus a maximally supersymmetric (SUSY) non-Abelian gauge theory~\cite{Witten:1995im}. However, when parallel M2- or M5-branes become coincident, the M2- or M5-branes stretched between them that become massless are extended objects, not point particles---making the low-energy worldvolume theory more challenging to identify.

For M2-branes the low-energy theory turns out to be a conventional quantum field theory (QFT), and in fact a conformal field theory (CFT): the low-energy theory of $N$ coincident M2-branes at a $\mathbb{C}^4/\mathbb{Z}_k$ singularity is a 3d $\N=6$ SUSY $U(N)_k \times U(N)_{-k}$ Chern-Simons-matter theory, called the Aharony-Bergman-Jafferis-Maldacena (ABJM) theory~\cite{Bagger:2006sk,Gustavsson:2007vu,Bagger:2007jr,Bagger:2007vi,Aharony:2008ug}. When the Chern-Simons level $k =1$ or $2$ the SUSY is enhanced to $\N=8$. When $N \gg k^5$ ABJM is holographically dual to 11d SUGRA on $AdS_4 \times S^7$~\cite{Maldacena:1997re}. The CFT and holographic descriptions have provided a wealth of information about M2-branes. For example, at large $N$ the number of worldvolume degrees of freedom scales as $N^{3/2}$~\cite{Klebanov:1996un,Drukker:2010nc}.

For $M$ coincident M5-branes the low-energy worldvolume theory is less well-understood. The 11d SUGRA soliton solution for M5-branes indicates that the worldvolume theory has 6d $\N=(2,0)$ SUSY and that the worldvolume fields should consist of five scalars, a chiral two-form $A_2$, and their fermionic superpartners filling out a tensor multiplet~\cite{Kaplan:1995cp}.  The fields in the tensor multiplet arise as Goldstone modes: the scalars from breaking translations in the five normal directions, $A_2$ from breaking the symmetry of shifting $C_3$ by the exterior derivative of a two-form, and the fermions from breaking half the SUSY. The $\N=(2,0)$ SUSY has $SO(5)_R$ R-symmetry, which acts to rotate the five transverse directions, hence the five scalars transform as a vector of $SO(5)_R$. The existence of the chiral $A_2$ in the tensor multiplet means that the three-form field strength, $F_3= dA_2$, is self-dual on the M5-brane worldvolume, i.e. $F_3 = \star_{6d} F_3$. Furthermore the tensor multiplet fields are believed to be valued in a worldvolume $\mathfrak{u}(M)$ gauge algebra~\cite{Strominger:1995ac}. We will henceforth ignore the overall $\mathfrak{u}(1)\subset \mathfrak{u}(M)$ representing the center-of-mass motion of the coincident M5-branes.

The existence of $A_2$ is also intuitive because an M2-brane ending on an M5-brane produces in the M5-brane worldvolume a 2d 1/2-BPS soliton, i.e. a string, which naturally couples to $A_2$~\cite{Howe:1997ue}. Moreover, $F_3 = \star_{6d} F_3$ implies that this string has equal electric and magnetic charges, and hence is called a ``self-dual string.'' When the distance between two parallel M5-branes goes to zero and hence the M2-branes stretched between them become massless, the result should be a 6d theory of massless self-dual strings~\cite{Strominger:1995ac}.

Since strings are extended objects, such a perspective suggests that the M5-brane theory might be non-local. However, compelling evidence has accumulated that in fact the M5-brane theory is not only local, but is a conventional CFT. In particular, when $M\to\infty$ the 11d SUGRA soliton solution for M5-branes has a near-horizon $AdS_7 \times S^4$ geometry in the decoupling limit, suggesting that 11d SUGRA on $AdS_7 \times S^4$  should be holographically dual to the M5-brane theory~\cite{Maldacena:1997re}. The $AdS_7$ factor indicates that all 6d field theory observables computed holographically are consistent with those of a local CFT. Additionally, non-trivial solutions to 6d conformal bootstrap equations have been found that are consistent with $\N=(2,0)$ SUSY and locality~\cite{Beem:2015aoa}.

From the perspective of conventional, perturbative QFT, the existence of any unitary, interacting QFT in $d>4$ is surprising, since power counting rules out any local Lagrangian. Two additional challenges arise in writing a local Lagrangian for the M5-brane theory. First, imposing self-duality at the level of the Lagrangian in any dimension is difficult. Second, generalizing $\mathfrak{u}(M)$ gauge transformations to a higher-form gauge field, such as the two-form $A_2$, is challenging. Remarkably, for the Abelian case, $M=1$, a classical action for the M5-brane worldvolume fields that has $\mathfrak{u}(1)$ gauge invariance, $\N=(2,0)$ superconformal symmetry, and locality, as well as 6d self-duality of $F_3$ enforced by auxiliary scalar field (acting as a Lagrange multiplier), is known~\cite{Pasti:1997gx,Bandos:1997ui}. However, whether a classical action can be written for $M>1$ remains unknown.

Despite the challenges, the M5-brane theory is extremely important to study, not only because M5-branes are key ingredients of M-theory, but also because the M5-brane worldvolume theory holds a uniquely privileged place among QFTs. In 6d, $\N=(2,0)$ is the maximal amount of SUSY possible, and 6d is the maximal dimension in which superconformal symmetry is possible~\cite{Nahm:1977tg}. A 6d $\N=(2,0)$ SUSY CFT cannot be reached as the infra-red (IR) fixed point of a local renormalization group (RG) flow from a free UV fixed point. The M5-brane theory has no known dimensionless parameters besides $M$ that could be tuned to allow a perturbative expansion. In short, the dimensionality, symmetries, and $M$ determine the M5-brane theory completely. The M5-brane theory is thus an isolated, intrinsically strongly-interacting fixed point. Via compactification the M5-brane theory can describe many lower-dimensional SUSY QFTs and dualities among them, and could potentially be a ``master theory'' containing information about all lower-dimensional QFTs.

In this paper, we will use holography to study the M5-brane theory with a 2d conformal defect as well as 3d CFTs with boundaries (BCFTs) related to the ABJM BCFT, using the following 1/4-BPS intersection of M-branes:
\begin{center}
\begin{tabular}{ |c | c || c | c | c | c | c | c | c | c | c | c | c|}
  \hline			
       &  & $x_0$ & $x_1$ & $x_2$ & $x_3$ & $x_4$ & $x_5$ & $x_6$ & $x_7$ & $x_8$ & $x_9$ & $x_{10}$ \\
\hline
$N$ & M2 & X & X & X &   &   &   &   &   &   &   &  \\
$M$ & M5 & X & X &   & X & X & X & X &   &   &   &  \\ \hline
$M'$ & M5$'$ & X & X &   &   &   &   &   & X & X & X & X \\ \hline
\end{tabular}
\end{center}
Schematically, this table represents a stack of $N$ coincident M2-branes, a stack of $M$ coincident M5-branes, and another stack of $M'$ coincident M5-branes that we label M5$'$ to distinguish them from the first stack. Most importantly, the M2-branes end on the M5- and M5$'$-branes at $x_2=0$.

Consider first setting $M' =0$, that is, the intersection of $N$ M2-branes ending on $M$ M5-branes, with no M5$'$-branes.  Recall that the endpoint of a semi-infinite string ending on a D-brane gives rise to an infinitely massive (s)quark in the D-brane's worldvolume.  Integrating out the heavy (s)quark yields a Wilson line, i.e. the holonomy of the D-brane worldvolume gauge field along the (s)quark worldline~\cite{Maldacena:1998im,Rey:1998ik}. Similarly, the end of a semi-infinite M2-brane ending on an M5-brane produces a self-dual string with infinite tension, which can be integrated out to yield a ``Wilson surface'' operator~\cite{Ganor:1996nf}. In the Abelian case, $M=1$, a precise form is known for the 1/2-BPS Wilson surface operator, $\exp\left( i\int A_2 + \ldots \right)$, with the integral over the surface spanned by the self-dual string and the ellipsis denoting terms required by SUSY, involving the M5-branes' worldvolume scalars, see for example ref.~\cite{Bullimore:2014upa}. In the non-Abelian case, $M>1$, a precise form is unknown but is believed to be schematically $\tr_{\cal{R}} \exp \left( i\int A_2 + \ldots\right)$, with the trace in representation $\cal{R}$ of $\mathfrak{su}(M)$. The representation $\cal{R}$ describes how the $N$ M2-branes are partitioned among the $M$ M5-branes, as we discuss in detail below.

We will consider only flat Wilson surfaces, i.e. Wilson surfaces extended along the $\mathbb{R}^{1,1}$ spanned by $x_0$ and $x_1$. From the M5-brane theory's perspective the Wilson surface is a 1/2-BPS 2d superconformal soliton. The M5-brane theory's bosonic symmetry is $SO(6,2) \times SO(5)_R$, where $SO(6,2)$ is the 6d conformal group. The table above shows that the Wilson surface preserves an $SO(2,2) \times SO(4)_R \times SO(4)_R$ subgroup, where the global 2d conformal group $SO(2,2) \subset SO(6,2)$ leaves invariant the Wilson surface, and the first $SO(4)_R$ rotates $(x_3,x_4,x_5,x_6)$ while the second rotates $(x_7,x_8,x_9,x_{10})$. The R subscripts indicate that these act as R-symmetries on the supercharges preserved by the Wilson surface, forming a 2d ``large'' $\N=(4,4)$ SUSY (``small'' $\N=(4,4)$ has a single $SO(4)_R$).

We will compute a central charge associated with the Wilson surface using holography. The holographic description's geometry is an $AdS_3$ and two $S^3$'s fibered over a Riemann surface~\cite{D'Hoker:2008wc,DHoker:2008rje,Estes:2012vm,Bachas:2013vza} and hence has the expected isometry $SO(2,2)\times SO(4)_R \times SO(4)_R$. The holographic description represents a large $M$ limit with arbitrary $N$. The geometry is asymptotically locally $AdS_7 \times S^4$ for all $N$, and becomes precisely $AdS_7 \times S^4$ when $N=0$.

As is well-known, 2d large $\N=(4,4)$ super-groups actually come in a one-parameter family called $D(2,1;\gamma)\times D(2,1;\gamma)$, where $\gamma$ is the free parameter~\cite{Sevrin:1988ew,Frappat:1996pb}. The most general solutions of 11d SUGRA that have super-isometry $D(2,1;\gamma)\times D(2,1;\gamma)$ and locally asymptote to $AdS_7 \times S^4$ are known~\cite{Bachas:2013vza}. The bosonic subgroup of $D(2,1;\gamma)\times D(2,1;\gamma)$ is $SO(2,2)\times SO(4) \times SO(4)$, and hence these solutions all involve an $AdS_3$ and two $S^3$'s fibered over a Riemann surface~\cite{DHoker:2008wvd}. The solutions describing Wilson surfaces in the M5-brane theory at large $M$ and arbitrary $N$ have $\gamma=-1/2$.

Additionally, the most general solutions of 11d SUGRA with super-isometry $D(2,1;\gamma)\times D(2,1;\gamma)$ are known that locally asymptote to ``half'' of $AdS_4 \times S^7$, in a sense we explain below~\cite{Bachas:2013vza}. These solutions are holographically dual to 3d maximally SUSY BCFTs. The exact BCFTs are not yet known, though some properties are clear from the 11d SUGRA solutions. In particular, in these solutions generically both $M$ and $M'$ are non-zero. The solutions thus describe M2-branes ending on M5- and M5$'$-branes, and hence the BCFTs must be cousins of the maximally SUSY ABJM BCFT.\footnote{Solutions of 11d SUGRA that are candidates for the holographic dual of the maximally SUSY ABJM BCFT appear in ref.~\cite{Bachas:2013vza}, but have potentially dangerous singularities.} Presumably these BCFTs are obtained from the ABJM BCFT by couplings to 2d SUSY multiplets at the boundary, and/or by sources or expectation values of scalar operators away from the boundary, similar to the superconformal interfaces between ABJM theories in refs.~\cite{DHoker:2009lky,Bobev:2013yra}. We will henceforth refer to these theories as ``cousins of the ABJM BCFT.''

For $k=1$ or $2$, ABJM's bosonic symmetry is enhanced to $SO(3,2)\times SO(8)_R$, where $SO(3,2)$ is the 3d conformal group, and in the intersection above the $SO(8)_R$ acts on $(x_3,\ldots,x_{10})$. Maximally superconformal boundary conditions at $x_2=0$ preserve the $SO(2,2) \subset SO(3,2)$ that leaves the boundary invariant, and $SO(4)_R \times SO(4)_R \subset SO(8)_R$ R-symmetry. The maximally SUSY ABJM BCFT's bosonic symmetry is thus $SO(2,2) \times SO(4)_R \times SO(4)_R$, and the super-group of the theory is $D(2,1;\gamma)\times D(2,1;\gamma)$ with $\gamma=1$. The cousins of the maximally SUSY ABJM BCFT that we will study have arbitrary $\gamma<0$, and their holographic duals again involve an $AdS_3$ and two $S^3$'s fibered over a Riemann surface~\cite{Bachas:2013vza}. These solutions correspond to a limit with large $N$ and a large number of M5- and M5$'$-branes. However the values of $M$ and $M'$ are in fact undetermined, intuitively because the M2-branes do not ``know'' how many M5- and M5$'$-branes have zero M2-branes ending on them, and so cannot know the total numbers of M5- and M5$'$-branes. Additionally, sending both $M$ and $M'$ to zero produces a singular solution, presumably because removing a BCFT's boundary is a singular operation.

Using these 11d SUGRA solutions, we will compute a central charge associated with the Wilson surface or 3d BCFT boundary. Crucially, the Wilson surface or 2d boundary is not a 2d CFT, but a 2d defect in, or boundary of, an ambient CFT that has $d>2$, which implies in general that the full Virasoro symmetry is not present. To be more precise, the Wilson surface or boundary breaks the ambient $SO(6,2)$ or $SO(3,2)$ conformal symmetry down to $SO(2,2)$, i.e. the global part of the Virasoro symmetry, in which the usual central charge does not appear. We must therefore define a central charge some other way.

We will use entanglement entropy (EE). Following refs.~\cite{Jensen:2013lxa,Estes:2014hka,Gentle:2015jma}, we will compute holographically the EE of a spherical region centered on the 2d defect or of a semi-circle centered on the 2d boundary. This EE has UV divergences, as expected, so we introduce a UV cutoff and subtract the EE of the ambient CFT. What remains is a term logarithmic in the cutoff, a constant term, and terms that vanish as the cutoff is removed. In analogy with the EE of a single interval in a 2d CFT~\cite{Holzhey:1994we,Calabrese:2004eu}, we identify the coefficient of that logarithmic term as $1/3$ times the central charge. In short, we compute the \textit{change} in the coefficient of the EE's logarithmic term due to the 2d defect or boundary, and use the result to define a central charge. We will denote the resulting Wilson surface or 2d boundary central charge as $b_{6d}$ or $b_{3d}$, respectively. Again in analogy with 2d CFT, we will interpret these as counting massless degrees of freedom supported on the Wilson surface or 2d boundary.\footnote{In a 3d BCFT the boundary central charge defined from EE is proportional to a central charge that appears in the trace anomaly~\cite{Fursaev:2013mxa,Fursaev:2016inw} and obeys a c-theorem, strictly decreasing in a boundary RG flow~\cite{Jensen:2015swa}. Our $b_{3d}$ thus counts massless degrees of freedom at the 2d boundary. However, a similar interpretation of $b_{6d}$ may not always be justified. In particular, examples of defects in higher-d CFTs are known in which central charges defined from EE do \textit{not} decrease along defect RG flows~\cite{Kumar:2016jxy,Kumar:2017vjv,Kobayashi:2018lil}. These examples include certain RG flows on a Wilson surface in a totally symmetric representation ${\cal R}$~\cite{Rodgers:2018mvq}.} Whether and how these central charges defined from EE are related to other potential definitions, for example via the thermodynamic entropy, stress tensor correlators, and so on, we leave as an important open question.

Our main result, for $b_{6d}$, takes a remarkably simple form,
\beq
\label{eq:b6dcompact}
b_{6d} = \frac{3}{5} \left [ 16 \left(\lambda,\varrho\right)-\left(\lambda,\lambda\right)\right],
\eeq
where $\lambda$ is the highest weight vector of the Wilson surface's representation $\cal{R}$, $\varrho$ is the $\mathfrak{su}(M)$ Weyl vector, and $(\cdot\,,\,\cdot)$ is defined with the Killing form on the weight space. The inner product $(\cdot,\,\cdot)$ is invariant under the action of the Weyl group, hence so is $b_{6d}$. In particular, $b_{6d}$ is invariant under complex conjugation of a representation, $\cal{R} \to \overline{\cal{R}}$, which acts as a Weyl reflection. Eq.~\eqref{eq:b6dcompact} is also reminiscent of the results for the M5-brane theory's own central charges, $a$ and $c$, which can similarly be written in terms of purely group theoretic data~\cite{Beem:2014kka,Cordova:2015vwa}. However, eq.~\eqref{eq:b6dcompact} obscures $b_{6d}$'s dependence on $N$ and $M$. In sec.~\ref{sec:cc} we present $b_{6d}$ for some specific $\cal{R}$, such as the rank $N$ totally symmetric and anti-symmetric representations, to explore the dependence on $N$ and $M$. Our results for $b_{3d}$ cannot be written so neatly as eq.~\eqref{eq:b6dcompact}, largely because we do not know the total number of M5- and M5$'$-branes and thus do not know $\mathfrak{su}(M)$ or $\mathfrak{su}(M')$.

However, a key observation about our results for both $b_{6d}$ and $b_{3d}$ is that neither naturally scales as $N^{3/2}$, characteristic of M2-branes at large $N$~\cite{Klebanov:1996un,Drukker:2010nc}, or as $M^3$, characteristic of M5-branes at large $M$ ($a \propto c \propto M^3$ at large $M$)~\cite{Freed:1998tg,Harvey:1998bx,Henningson:1998gx}. In other words, in general for the M-brane intersections we consider, the number of massless degrees of freedom on the Wilson surface or 2d boundary does not scale, in any obvious way, with the total number of degrees of freedom of the M2- or M5-brane theory.

This paper is organized as follows. In sec.~\ref{sec:sugra} we review the 11d SUGRA solutions describing M5-branes with Wilson surfaces or cousins of the ABJM BCFT with $\gamma<0$. In sec.~\ref{sec:hc} we derive an integral for the holographic EE, and then evaluate the integral to extract $b_{6d}$ and $b_{3d}$. In sec.~\ref{sec:cc} we summarize our results, including for specific $\cal{R}$, and compare to existing results. (Readers interested only in our results can skip directly to sec.~\ref{sec:cc}.) We conclude in sec~\ref{sec:discussion} with discussion and suggestions for future research.

The companion paper ref.~\cite{Rodgers:2018mvq} reproduces our results for $b_{6d}$ for the fundamental, totally anti-symmetric, and totally symmetric representations, using probe branes in $AdS_7 \times S^4$, namely a probe M2-brane when $N\ll M$ or M5-brane when $N \ll M^2$. Ref.~\cite{Rodgers:2018mvq} also uses the probe M5-branes to explore RG flows on Wilson surfaces.

\textbf{Note added:} After this paper appeared on the arxiv, ref.~\cite{Jensen:2018rxu} clarified how our central charge $b$ obtained from the EE of a (hemi-)sphere is related to central charges in the defect's contribution to the Weyl anomaly. This clarifies the relationship between unitarity and positivity of $b$ discussed in sec.~\ref{sec:wilson-surface}, among other things.

For a CFT in $\mathbb{R}^{1,d-1}$ with $d\geq 3$ and a 2d conformal defect, the trace of the stress tensor splits into two terms, $\left< T_{~\mu}^{\mu} \right> = \left< T_{~\mu}^{\mu}\right>_{\textrm{bulk}} + \delta^{d-2}(x) \left<T_{~\mu}^{\mu}\right>_{\textrm{defect}}$, where $\left<T_{~\mu}^{\mu}\right>_{\textrm{bulk}}$ is the CFT's trace anomaly, which is $0$ for odd $d$ but may be non-zero for even $d$, $\delta^{d-2}(x)$ is a delta function that localizes to the 2d defect, and~\cite{Graham:1999pm,Henningson:1999xi, Schwimmer:2008yh}
\beq
\label{eq:defecttrace}
  \left<T_{~\mu}^{\mu}\right>_{\textrm{defect}} = - \frac{1}{24\pi} \left(
          c \, \hat{R} + d_1 \, \Pi_{ab}^\mu \Pi_\mu^{ab} - d_2 \, \hat{g}^{ac} \hat{g}^{bd} W_{abcd}
      \right),
\eeq
where $\hat{g}_{ab}$ is the defect's induced metric ($a,b=0,1$), $\hat{R}$ is the Ricci scalar built from $\hat{g}_{ab}$, $\Pi^{\mu}_{ab}$ is the defect's traceless extrinsic curvature, $W_{abcd}$ is the Weyl tensor pulled back to the defect, and the dimensionless numbers $c$, $d_1$, and $d_2$ are the defect central charges. In a 3d BCFT the boundary's contribution to the trace anomaly has the form of eq.~\eqref{eq:defecttrace}, but $W_{abcd}=0$ identically, so $d_2$ does not exist. All of $c$, $d_1$, and $d_2$ can depend on boundary conditions imposed on CFT fields at the defect or boundary and on degrees of freedom supported only at the defect or boundary. Indeed, $c$ obeys a $c$-theorem~\cite{Jensen:2015swa}, and may thus serve as a measure of the number of massless degrees of freedom at the defect or boundary. However, unlike a 2d CFT's central charge, unitarity does not require $c \geq 0$, and in fact even simple, unitary theories can have $c<0$, such as a 3d free, massless scalar BCFT with Dirichlet boundary conditions~\cite{Nozaki:2012qd,Jensen:2015swa, Solodukhin:2015eca}. Whether a lower bound exists on $c$ remains unknown, though for 3d BCFTs $c\geq -\frac{2}{3} \, d_1$ is conjectured~\cite{Herzog:2017kkj}, where unitarity requires $d_1\geq 0$~\cite{Herzog:2017xha,Herzog:2017kkj}. Ref.~\cite{Jensen:2018rxu} showed that if the average null energy condition is valid in a CFT with a 2d conformal defect, then $d_2 \geq 0$ also.

Refs.~\cite{Kobayashi:2018lil,Jensen:2018rxu} showed that
\beq
\label{eq:ee_trace_anomaly}
  b = c - \frac{d-3}{d-1} d_2.
\eeq
For $d=3$, as is the case for the cousins of ABJM, this reproduces the result \(b = c\)~\cite{Fursaev:2013mxa,Fursaev:2016inw}. For $d > 3$, eq.~\eqref{eq:ee_trace_anomaly} implies that $b$ may be negative even when $c$ and $d_2$ are both positive. Indeed, for Wilson surfaces in the M5-brane theory,  ref.~\cite{Jensen:2018rxu} used eq.~\eqref{eq:ee_trace_anomaly}, our result for $b_{6d}$ in eq.~\eqref{eq:b6dcompact}, and the result of ref.~\cite{Gentle:2015jma} for $d_2$ to calculate
\beq
\label{eq:cd2values}
c = 24\left(\lambda,\varrho\right) + 3\left(\lambda,\lambda\right), \qquad d_2 = 24\left(\lambda,\varrho\right) + 6\left(\lambda,\lambda\right),
\eeq
both of which are positive for any ${\cal R}$. However, the linear combination $b$ in eq.~\eqref{eq:ee_trace_anomaly} can be negative for some ${\cal R}$, as we discuss in sec.~\ref{sec:wilson-surface}. These results show that $b<0$ does not signal violation of unitarity.

\section{Review: the SUGRA Solutions}
\label{sec:sugra}

The solutions of 11d SUGRA in ref.~\cite{Bachas:2013vza} that holographically describe Wilson surfaces or cousins of the ABJM BCFT (and which built upon the solutions in refs.~\cite{DHoker:2008rje,D'Hoker:2008wc,Estes:2012vm}) are 1/2-BPS, meaning they support 16 real supercharges, and have super-isometry $D(2,1;\gamma)\times D(2,1;\gamma)$ with $\gamma\in \mathbb{R}$. The super-group $D(2,1;\gamma)\times D(2,1;\gamma)$ has bosonic subgroup $SO(2,1)\times SO(3) \times SO(3)$, where the super-charges anti-commute into a linear combination of the generators of these three bosonic factors. The parameter $\gamma$ determines the relative weights of the coefficients in that linear combination. The super-group $D(2,1;\gamma)\times D(2,1;\gamma)$ is invariant under the simultaneous operations $\gamma \to 1/\gamma$ and swapping the two $SO(3)$'s. Without loss of generality we can thus restrict to $\gamma \in [-1,1]$, and in fact all the solutions we consider below will have $\gamma \in(-1,0)$. These symmetries of the 11d SUGRA solutions match those of 2d large $\N=(4,4)$ SUSY, which admits exactly the same one-parameter family of Lie super-groups~\cite{Sevrin:1988ew,Frappat:1996pb}.\footnote{Ref.~\cite{Sevrin:1988ew} classifies and constructs the Virasoro extension of the super-group.} We leave the full details of these 11d SUGRA solutions to refs.~\cite{Bachas:2013vza,DHoker:2008rje,D'Hoker:2008wc,Estes:2012vm}, and here review only features that we will need in subsequent sections. In particular, we will consider the solutions of refs.~\cite{DHoker:2008rje,Bachas:2013vza}, using the conventions of ref.~\cite{Bachas:2013vza}.\footnote{To clarify, ref.~\cite{D'Hoker:2008wc} constructed solutions for three special values of $\gamma$, ref.~\cite{Estes:2012vm} constructed the solutions for general $\gamma$, and refs.~\cite{DHoker:2008rje,Bachas:2013vza} identified the specific solutions we consider in this paper.}

The $SO(2,2) \times SO(4) \times SO(4)$ isometry implies that the metric involves $AdS_3$ and two $S^3$'s fibered over a Riemann surface, which we take to be the upper half plane,
\beq
\label{eq:metric2}
ds^2 =  f_1^2 \,ds^2_{\text{AdS}_3} + f_2^2 \,ds^2_{S^3} + f_3^2 \,ds^2_{S^3} + 2 \, \O^2 \,|dw|^2,
\eeq
where $w$ and $\wbar$ are the coordinates of the upper half plane (so $\textrm{Im}(w) \geq 0$). The functions $f_1$, $f_2$, $f_3$, and $\O$ depend only on $w$ and $\wbar$. We denote by $ds^2_{S^3}$ the metric for a unit-radius round $S^3$ and $ds^2_{AdS_3}$ the metric for a unit-radius $AdS_3$, that is,
\beq
\label{eq:ads3metric}
ds^2_{AdS_3} = \frac{1}{u^2} \left(du^2 -dt^2 + dx_{\parallel}^2\right),
\eeq
where $u \in [0,\infty)$, with $AdS_3$ boundary at $u = 0$, $t\in(-\infty,\infty)$ is the time coordinate, and $x_{\parallel}\in(-\infty,\infty)$ is the spatial coordinate parallel to the Wilson surface or 2d boundary.

The solutions in ref.~\cite{Bachas:2013vza} that we will use are completely determined by a triple of data $(h,G,\gamma)$, where $h$ is a harmonic function over the upper half plane, and $G$ is a complex-valued function that obeys
\beq
\partial_w G = \frac{1}{2} (G + \Gbar) \, \partial_w \ln h,
\eeq
although $G$ is not completely determined by $h$. If we define the complex-valued functions
\begin{align}
\label{eq:wpm}
&W_+ \equiv |G+i|^2 + \gamma (G \Gbar - 1)\,,&
&W_- \equiv |G-i|^2 + \gamma^{-1} (G \Gbar -1 ),&
\end{align}
where $\gamma$ is the constant parametrizing the supergroup, then in these solutions $f_1$, $f_2$, $f_3$, and $\O$ are given by
\begin{subequations}
\label{eq:metfuncs}
\beq
f_1^6 = \frac{h^2 W_+ W_-}{c_1^6 (G \Gbar - 1)^2}, \qquad f_2^6 = \frac{h^2 (G \Gbar - 1) W_-}{c_2^3 c_3^3 W_+^2}, \qquad f_3^6 = \frac{h^2 (G \Gbar - 1) W_+}{c_2^3 c_3^3 W_-^2},
\eeq
\beq
\O^6 = \frac{1}{8} \frac{|\partial_w h|^6}{c_2^3 c_3^3 h^4} (G \Gbar - 1) W_+ W_-.
\eeq
\end{subequations}
The parameters $c_1$, $c_2$, and $c_3$ are constants that obey $c_1 + c_2 + c_3=0$, so that only two are independent. In fact, a simultaneous re-scaling of $c_1$, $c_2$, and $c_3$ can be absorbed by re-scaling $h$ without changing the solution, so only a single constant is independent. That single constant must map to $\gamma$: the precise relation is $\gamma = c_2/c_3$. As mentioned above, the solutions that we will consider have $\gamma \in (-1,0)$.

When $\g<0$, global regularity of these solutions requires that $h$ and $G$ obey $h>0$ and $|G|<1$ everywhere on the interior of the upper half $w$ plane (all $\textrm{Im}\left(w\right)>0$) and that $h=0$ and $G = \pm i$ on the boundary of the upper half $w$ plane ($\textrm{Im}\left(w\right)=0$). All the solutions that we consider below will obey these conditions.

The four-form $F_4 = dC_3$ of these solutions appears in ref.~\cite{Bachas:2013vza}. We will not present the solution for $F_4$ explicitly, but in sec.~\ref{sec:fluxes} we will discuss the M2- and M5-brane charges determined by the solution for $F_4$.

The invariance of $D(2,1;\gamma)\times D(2,1;\gamma)$ under $\gamma \to 1/\gamma$ and swapping of $SO(3)$ sub-groups appears in these 11d SUGRA solutions as invariance under $\gamma \to 1/\gamma$ and exchange of $W_+ \leftrightarrow W_-$. As clear from eq.~\eqref{eq:metfuncs}, that leaves $f_1$ and $\Omega$ invariant but trades $f_2 \leftrightarrow f_3$, thus effectively interchanging the geometry's two $S^3$ factors.

\subsection{Asymptotically Locally $AdS_7 \times S^4$ Solutions}
\label{subsec:ads7metric}

The 11d SUGRA solutions holographically describing Wilson surfaces in the M5-brane theory at large $M$ are of the form in eq.~\eqref{eq:metric2} with
\beq
\label{eq:ads7metric}
h = - i \left(w-\wbar\right), \qquad G = - i\left( 1 + \sum_{j=1}^{2n+2} (-1)^j \frac{w - \xi_j}{|w - \xi_j|} \right),
\eeq
where the integer $n \geq 0$ and the $\xi_j$ are $2n+2$ real-valued constants determining $G$'s branch points on the boundary of the upper half plane. More specifically, the $\xi_j$ are points on the real line $\textrm{Im}(w)=0$ where $G$ changes sign from $\pm i$ to $\mp i$.

In general, the upper half plane is invariant under the $SL(2,\mathbb{R})$ group of transformations, which is three-dimensional. Implicitly we fixed two $SL(2,\mathbb{R})$ transformations with our choice of $h$. The third transformation is translations, which we can use to fix one of the $\xi_j$. The remaining $2n+1$ values of the $\xi_j$ determine the M2- and M5-brane charges of the solution, as we discuss in sec.~\ref{sec:fluxes}. The solutions describing M2-branes ending on M5-branes have $\gamma=-1/2$, whereas $\gamma=-2$ describes M2-branes ending on M5$'$-branes. However the latter map to the former via $\gamma \to 1/\gamma$ and swapping the two $S^3$'s, as described above. In what follows, we will consider arbitrary $\gamma \in (-1,0)$ in many intermediate steps but will always set $\gamma = -1/2$ in our final results.

Another form of the $G$ in eq.~\eqref{eq:ads7metric} that will be useful for describing the geometry's asymptotics comes from using polar coordinates in the upper half plane, $w \equiv r e^{i \theta}$ with $r \in [0,\infty)$ and $\theta \in [0,\pi]$. If we expand $G$ in Legendre polynomials, $P_{k}\left(\cos \theta\right)$,
\beq
\label{eq:ads7legendre}
G = -i\left( 1+ \sum_{k=1}^\infty m_k \frac{\left( e^{i \theta} P_k\left(\cos\theta\right) - P_{k-1}\left(\cos\theta\right) \right)}{r^k} \right), \qquad m_k \equiv \sum_{j=1}^{2 n+2} (-1)^j \left(\xi_j\right)^k,
\eeq
then the $\xi_j$ determine the expansion coefficients $m_k$, where $n$ is finite but $k \in [0,\infty)$.

Though not immediately obvious, the solutions in eq.~\eqref{eq:ads7metric} are asymptotically locally $AdS_7 \times S^4$. Indeed, asymptotically we can put the metric of these solutions in the Fefferman-Graham (FG) form of $AdS_7 \times S^4$,
\beq
\label{eq:ads7asympmet}
ds^2 = \frac{4 L_{S^4}^2}{z^2} \left( dz^2 -dt^2 + dx_{\parallel}^2 + dr_\perp^2 + \frac{\left(1+\gamma\right)^2}{\gamma^2}r_\perp^2 ds^2_{S^3} \right) + L_{S^4}^2 \left(d\phi^2 + \sin^2\phi\, ds^2_{S^3}  \right) + \ldots,
\eeq
where $z\in[0,\infty)$ is the FG holographic coordinate, with boundary at $z=0$, $r_{\perp}\in[0,\infty)$ is the distance to the defect, $\phi\in[0,\pi]$ is the zenith angle of the asymptotic $S^4$, and the ellipsis represents terms sub-leading in $1/r$. The change of coordinates to FG form admits the following asymptotic expansion,
\begin{subequations}
\label{eq:ads7asympcoords}
\begin{align}
z =& \frac{u}{\sqrt{r}} \left(\frac{1+\gamma}{\sqrt{-\gamma}}\sqrt{2\,m_1} + {\cal O}  \left( \frac{1}{r^2} \right) \right), \\
r_\perp =& \, u \left(1 + {\cal O}  \left( \frac{1}{r} \right) \right), \\
\phi =& \, \theta \left(1 + {\cal O}  \left( \frac{1}{r} \right) \right),
\end{align}
\end{subequations}
which determines the radius of curvature $L_{S^4}$ of the asymptotic $S^4$,
\beq
\label{eq:ads7rad}
L_{S^4}^6 = \frac{1}{c_1^6}\frac{(1 + \gamma)^6}{\gamma^2}m_1^2.
\eeq
The factor $\left(1+\gamma\right)^2/\gamma^2$ in eq.~\eqref{eq:ads7asympmet} indicates that for generic $\gamma$ the dual M5-brane theory lives on a space with a conical singularity at $r_{\perp}=0$. These 11d SUGRA solutions thus suggest that deformations of the M5-brane theory with $D(2,1;\gamma)\times D(2,1;\gamma)$ super-group are only possible, for generic $\gamma$, on spaces with such a conical singularity. Only for $\gamma=-1/2$ does the singularity disappear such that the theory lives on 6d Minkowski space.

Eqs.~\eqref{eq:ads7asympmet} and~\eqref{eq:ads7asympcoords} show clearly that in these solutions we can approach the asymptotic $AdS_7 \times S^4$ boundary in two ways. First, if we fix $u$ and send $r \to \infty$ then $z \to 0$ with fixed $r_{\perp}=u$, which means that we arrive at the boundary a distance $u$ from the defect. Second, if we fix $r$ and send $u \to 0$, then $z \to 0$ but now $r_{\perp} \to 0$, and so we arrive at the boundary precisely on the defect. Recall from eq.~\eqref{eq:ads3metric} that $u \to 0$ is the boundary of the $AdS_3$ factor.

The $AdS_7 \times S^4$ solution is simply the $\gamma=-1/2$ and $n=0$ case, which has two branch points. Using the $SL(2,\mathbb{R})$ translational symmetry we place these two branch points on the real line $\textrm{Im}(w)=0$ symmetrically, at $\textrm{Re}(w)=\pm \xi$, such that the solution has only one free parameter, $\xi$. In this case, in eq.~\eqref{eq:ads7asympcoords} the higher-order powers of $1/r$ vanish, and hence the ellipsis in eq.~\eqref{eq:ads7asympmet} vanishes. We will see below that $\xi$ determines $m_1 = 2 \xi$ and hence $L_{S^4}$, the single free parameter of the $AdS_7 \times S^4$ solution. The cases with $\gamma=-1/2$ and $n>0$ then describe Wilson surfaces in the M5-brane theory.

\subsection{Asymptotically Locally $AdS_4 \times S^7$ Solutions}
\label{subsec:ads4metric}

The 11d SUGRA solutions holographically describing cousins of the ABJM BCFT with $\gamma<0$ are of the form in eq.~\eqref{eq:metric2} with
\beq
\label{eq:ads4metric}
h = - i \left(w-\wbar\right), \qquad G = - i \sum_{j=1}^{2n+1} (-1)^j \frac{w - \xi_j}{|w - \xi_j|},
\eeq
where the integer $n\geq 0$ and the $\xi_j$ are $2n+1$ real-valued constants determining $G$'s branch points on the boundary of the upper half plane, i.e. on the real line $\textrm{Im}(w)=0$. As in sec.~\ref{subsec:ads7metric}, we have implicitly fixed two $SL(2,\mathbb{R})$ transformations of the upper half plane with our choice of $h$, and we can use translation symmetry to fix one of the $\xi_j$. The remaining $2n$ values of the $\xi_j$ then determine the solutions' M2- and M5-brane charges, as we discuss in sec.~\ref{sec:fluxes}. We will consider solutions with $\g\in(-1,0)$.

As in sec.~\ref{subsec:ads7metric}, we can obtain another form of the $G$ in eq.~\eqref{eq:ads4metric} that will be useful for describing the asymptotics by introducing $w \equiv r e^{i \theta}$ with $r \in [0,\infty)$ and $\theta \in [0,\pi]$. Expanding $G$ in Legendre polynomials then gives
\beq
\label{eq:ads4legendre}
G = i \left(e^{i \theta} -  \sum_{k=1}^\infty m_k \frac{\left( e^{i \theta} P_k\left(\cos\theta\right) - P_{k-1}\left(\cos \theta\right) \right)}{r^k} \right), \qquad m_k \equiv \sum_{j=1}^{2 n+ 1} (-1)^j \left(\xi_j\right)^k,
\eeq
where the $\xi_j$ determine the expansion coefficients $m_k$.\footnote{We use the same symbol $m_k$ for the expansion coefficients in both eq.~\eqref{eq:ads7legendre} and~\eqref{eq:ads4legendre}, though the difference should always be clear from the context.}

The solutions in eq.~\eqref{eq:ads7metric} are in fact asymptotically locally ``half'' of $AdS_4 \times S^7$, in the following sense. Asymptotically we can put the metric of these solutions in the FG form of $AdS_4 \times S^7$,
\beq
\label{eq:ads4asympmet}
ds^2 =  \frac{L_{S^7}^2}{4} \left( \frac{dz^2}{z^2} + \frac{-dt^2 + dx_{\parallel}^2 + dx_\perp^2}{z^2} \right) + L_{S^7}^2 \left(d\phi^2 + \cos^2\phi \, ds^2_{S^3} + \sin^2\phi\, ds^2_{S^3} \right)+\ldots,
\eeq
where the ellipsis represents terms sub-leading in $1/r$, and where the asymptotic $S^7$ radius, $L_{S^7}$, is given by
\beq
\label{eq:ads4rad}
L_{S^7}^6 = \frac{16}{c_1^6}\frac{(1 + \gamma)^6}{(-\gamma)^3} (-m_1^2 - m_2).
\eeq
The change of coordinates to FG form has the following asymptotic expansion,
\begin{subequations}
\label{eq:ads4asympcoords}
\begin{align}
z =& \,\frac{u}{r} \left(\frac{1 + \gamma}{\sqrt{\gamma}} \frac{\sqrt{m_1^2+m_2}}{2} + {\cal O}  \left( \frac{1}{r^2} \right) \right), \\
x_{\perp} =& \, u \left(1 + {\cal O}  \left( \frac{1}{r^2} \right) \right), \\
\phi =& \, \frac{\theta}{2} \left(1 + {\cal O}  \left( \frac{1}{r^2} \right) \right),
\end{align}
\end{subequations}
where $\theta \in [0,\pi]$ implies $\phi\in [0,\pi/2]$. Eq.~\eqref{eq:ads4asympcoords} shows that $u \in [0,\infty)$ implies the FG holographic coordinate $z\in[0,\infty)$, with boundary at $z=0$, but crucially $x_{\perp}\in [0,\infty)$. The background geometry for the holographically dual field theory thus has coordinates $t \in (-\infty,\infty)$, $x_{\parallel} \in (-\infty,\infty)$, and $x_{\perp}\in[0,\infty)$, i.e. \textit{half} of 3d Minkowski space with a boundary at $x_{\perp}=0$. In this sense, these solutions are locally asymptotic to ``half'' of $AdS_4 \times S^7$, as advertised.

As a result, no continuous limit exists in which these solutions reduce to the exact $AdS_4 \times S^7$ solution. In fact, the exact $AdS_4 \times S^7$ solution differs from these solutions in several ways. For example, in these solutions the $h$ in eq.~\eqref{eq:ads4metric} has a single pole at $r = \infty$, and $\gamma \in (-1,0)$, whereas in the exact $AdS_4 \times S^7$ solution $h$ has two poles and $\gamma=1$. Since we cannot continuously send $\gamma \to 1$, the BCFTs holographically dual to our solutions therefore appear to be cousins of, but not continuously connected to, the ABJM BCFT. Identifying the dual BCFTs is an important question we leave for future research.

Eqs.~\eqref{eq:ads4asympmet} and~\eqref{eq:ads4asympcoords} also show that, as in sec.~\ref{subsec:ads7metric}, we can approach the asymptotic (half) $AdS_4 \times S^7$ boundary in two ways. First, if we fix $u$ and send $r \to \infty$, then $z \to 0$ with fixed $x_{\perp}=u$, so that we arrive at the $AdS_4$ boundary a distance $u$ from the BCFT's boundary. Second, if we fix $r$ and send $u \to 0$, then $z \to 0$ but now $x_{\perp} \to 0$, and so we arrive at the $AdS_4$ boundary precisely on the BCFT's 2d boundary.

\subsection{Partitions and M-brane Charges}
\label{sec:fluxes}

The M-brane intersection of sec.~\ref{sec:intro} describes $N$ coincident M2-branes ending on $M$ coincident M5-branes and $M'$ coincident M5$'$-branes. Such an intersection is characterized by a partition of $N$ describing which M5- or M5$'$-brane each M2-brane ends on. In this section we will briefly sketch how we extract this partition from the 11d SUGRA solutions reviewed above, leaving the full details to ref.~\cite{Bachas:2013vza}.

The asymptotically locally $AdS_7 \times S^4$ solutions of sec.~\ref{subsec:ads7metric} describe M2-branes ending only on M5-branes, not on M5'-branes, and are fully characterized by a partition of $N$ M2-branes among the $M$ M5-branes. The asymptotically locally $AdS_4 \times S^7$ solutions of sec.~\ref{subsec:ads4metric} describe M2-branes ending on both M5 and M5'-branes, but are still fully characterized by a partition of $N$ M2-branes among $M$ M5-branes. The special features of both solutions will be discussed in more detail below. Our discussion of the partitions will applies to both solutions, with one crucial exception: the partitions of the $AdS_7 \times S^4$ solutions include cases in which some M5-branes have no M2-branes ending on them, whereas the $AdS_4 \times S^7$ solutions do not admit such partitions.

\begin{figure}[t!]
\begin{center}
\includegraphics[width=.8\textwidth]{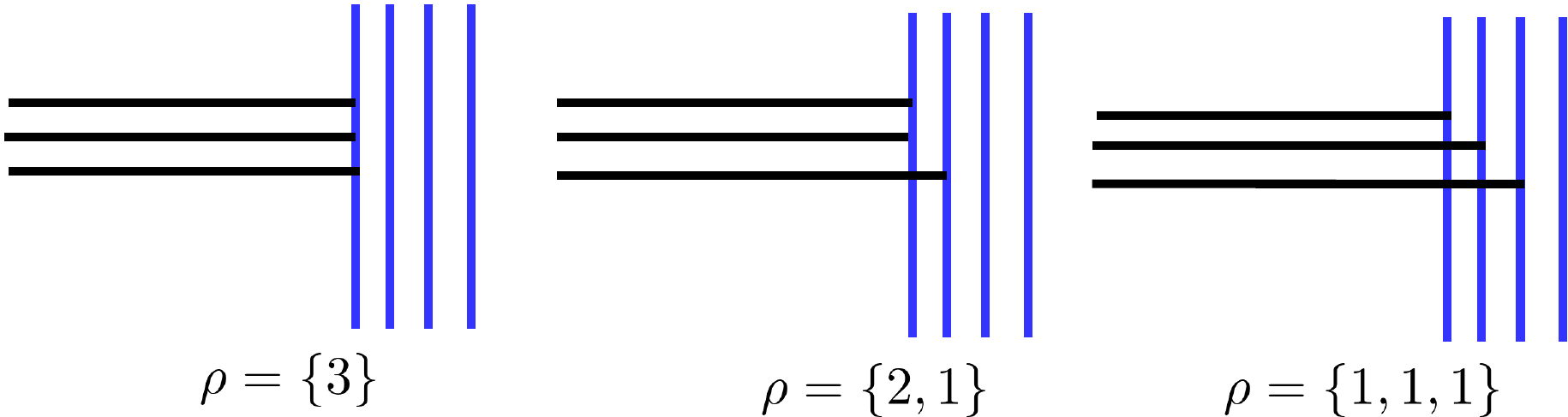}
\caption{\label{fig:part1} The three possible partitions $\rho$ with $N=3$ M2-branes and $M=4$ M5-branes. The horizontal axis represents the $x_2$ direction while the vertical axis represents one of the directions $(x_3,x_4,x_5,x_6)$. The horizontal black lines represent M2-branes while the vertical blue lines represent M5-branes. The M5-branes are ordered such that the number of M2-branes on each M5-brane decreases from left to right. The three possible partitions are then $\rho=\{3\}$ (left), $\rho =\{2,1\}$ (middle), and $\rho=\{1,1,1\}$ (right).}
\end{center}
\end{figure}

\begin{figure}[t!]
\begin{center}
\includegraphics[width=.8\textwidth]{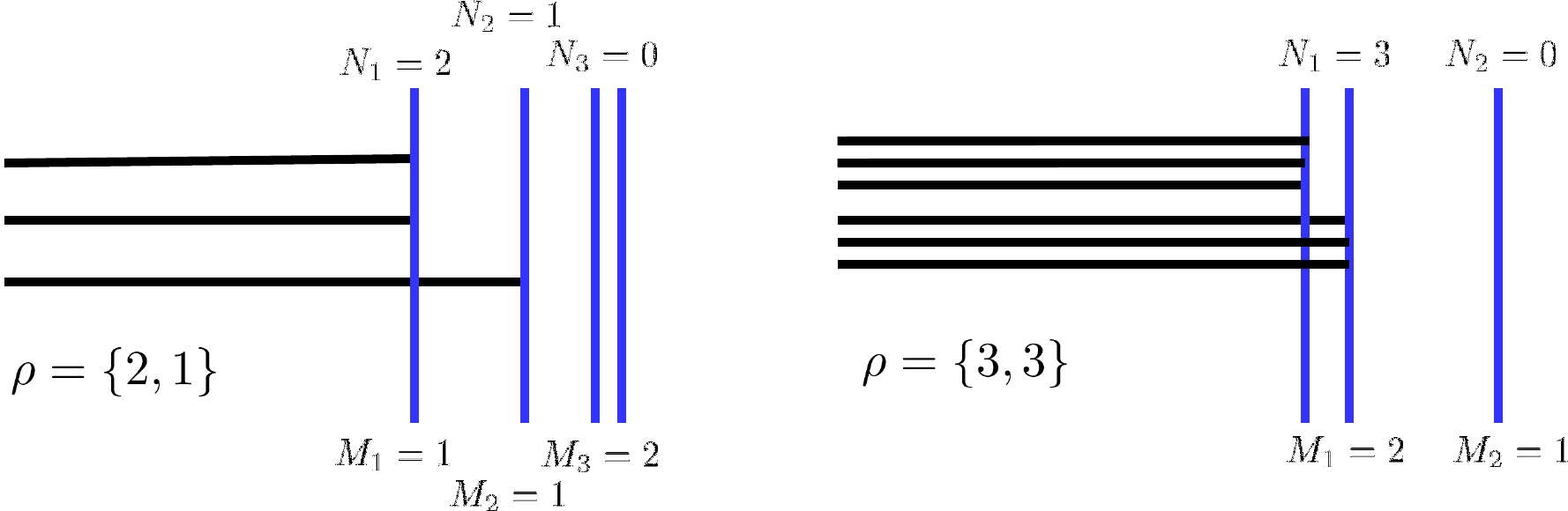}
\caption{\label{fig:part2} Two partitions $\rho$ parametrized as in eq.~\eqref{eq:part}: $\rho=\{2,1\}$ with $M=4$ (left) and $\rho=\{3,3\}$ with $M=3$ (right). The horizontal axis represents the $x_2$ direction while the vertical axis represents one of the directions $(x_3,x_4,x_5,x_6)$. The horizontal black lines represent M2-branes while the vertical blue lines represent M5-branes. The $N_a$ are the distinct integers in $\rho$, each with degeneracy $M_a$. Put differently, $M_a$ is the number of M5-branes with $N_a$ M2-branes ending on them.}
\end{center}
\end{figure}

We will denote the partition as $\rho$, which we will order with entries decreasing from left to right. To illustrate which M2-brane ends on which M5-brane, we will temporarily imagine separating the M2-branes in one of the directions $(x_3,x_4,x_5,x_6)$ and separating the M5-branes in the $x_2$ direction, as shown in figs.~\ref{fig:part1} and~\ref{fig:part2}. Once $\rho$ is specified we will re-collapse the M2- and M5-branes back to the original intersection of coincident stacks.

We will use two parametrizations of $\rho$. In the first parametrization we label $\ell_q$ as the number of M2-branes ending on the $q^{\textrm{th}}$ M5-brane and write the partition as $\rho=\{\ell_1,\ell_2,\ldots\}$ with integers $\ell_1 \geq \ell_2 \geq \ell_3 \ldots$. In this parametrization the total number $M$ of M5-branes is simply the upper limit of $1 \leq q \leq M$, while the total number $N$ of M2-branes is
\beq
\label{eq:nellsum}
N = \sum_{q=1}^{M} \ell_q.
\eeq
We allow for some $\ell_q=0$, representing M5-branes with zero M2-branes ending on them. However, when writing $\rho=\{\ell_1,\ell_2,...\}$ we will omit all zero entries. As examples, fig.~\ref{fig:part1} shows all $\rho$ for the case with $N=3$ and $M=4$.

In the second parametrization we write the partition as
\beq
\label{eq:part}
\rho = \{ \underbrace{N_1, N_1, ... N_1}_{M_1}, \underbrace{N_2, N_2, ... N_2}_{M_2}, ..., \underbrace{N_n, N_n, ... N_n}_{M_n},\underbrace{N_{n+1}, N_{n+1}, ... N_{n+1}}_{M_{n+1}}\},
\eeq
representing $M_1$ M5-branes each with $N_1$ M2-branes ending on them, $M_2$ M5-branes each with $N_2$ M2-branes ending on them, and so on with $N_1 > N_2 > N_3 \ldots$. In other words, the $N_a$ are $n$ \textit{distinct non-zero} integers, each with degeneracy $M_a$. The final entries in $\rho$ are $N_{n+1}=0$, with degeneracy $M_{n+1}$, representing the number of M5-branes with no M2-branes ending on them. We thus specify $\rho$ by specifying the set of integers $\{N_a\}$ and the set of their degeneracies $\{M_a\}$. The partition $\rho$ has a total of $2n+1$ parameters, the $n$ distinct integers in $\{N_a\}$ plus the $n+1$ distinct integers in $\{M_a\}$. In this parametrization the total numbers $N$ of M2-branes and $M$ of M5-branes are
\beq
\label{eq:totnm}
N = \sum_{a=1}^{n+1} M_a N_a, \qquad M = \sum_{a=1}^{n+1} M_a.
\eeq
As in the first parametrization, when writing $\rho$ we will omit all zero entries, that is, we omit the entries $N_{n+1}$. As examples, fig.~\ref{fig:part2} shows the sets $\{N_a\}$ and $\{M_a\}$ for $\rho=\{2,1\}$ with $M=4$ (fig.~\ref{fig:part2} left) and $\rho=\{3,3\}$ with $M=3$ (fig.~\ref{fig:part2} right).

\begin{figure}[t!]
\begin{center}
\includegraphics[width=.8\textwidth]{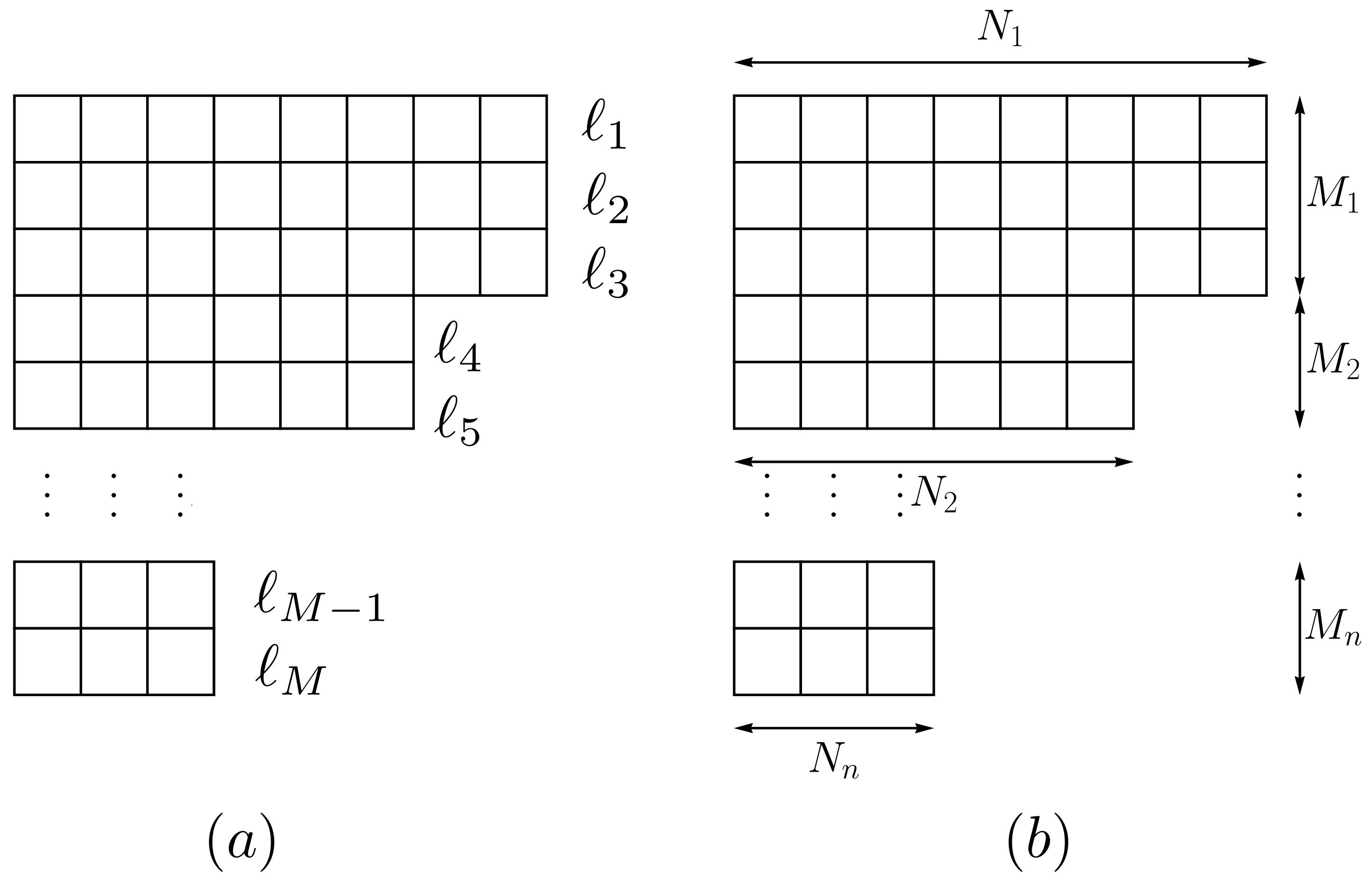}
\caption{\label{fig:young} Extracting the same Young tableau from our two different parametrizations of an ordered partition $\rho$. (a) The parametrization $\rho = \{\ell_1,\ell_2,\ell_3,\ldots\}$ with $\ell_1 \geq \ell_2 \geq \ell_3 \ldots$, where each $\ell_q$ gives the number of boxes in the $q^{\textrm{th}}$ row of the Young tableau, with $1 \leq q \leq M$. In the figure we assumed all the $\ell_q$ are non-zero. (b) The parametrization of eq.~\eqref{eq:part}, where $M_1$ is the number of rows with $N_1$ boxes, $M_2$ is the number of rows with $N_2 < N_1$ boxes, and so on. In physical terms, in the Young tableau each box represents an M2-brane and each row represents an M5-brane.}
\end{center}
\end{figure}

The ordered partition $\rho$ determines a Young tableau, with the physical interpretation that each box in the Young tableau represents an M2-brane and each row represents an M5-brane. In the first parametrization, $\rho = \{\ell_1,\ell_2,\ell_3,\ldots\}$ with $\ell_1 \geq \ell_2 \geq \ell_3 \ldots$, each $\ell_q$ gives the number of boxes in the $q^{\textrm{th}}$ row of the Young tableau, as shown in fig.~\ref{fig:young} (a).  In the second parametrization, eq.~\eqref{eq:part}, $M_1$ is the number of rows with $N_1$ boxes, $M_2$ is the number of rows with $N_2 < N_1$ boxes, and so on, as shown in fig.~\ref{fig:young} (b).

When the M2- and M5-branes re-collapse back to the original intersection of coincident stacks, the gauge algebra on the M5-branes' worldvolume is $\mathfrak{su}(M)$ and the M2-branes' boundary represents the Wilson surface. The partition $\rho$, or the corresponding Young tableau, then determines the Wilson surface's representation $\cal{R}$ of $\mathfrak{su}(M)$. Complex conjugation of the representation, $\cal{R} \to \overline{\cal{R}}$, acts in the first parametrization as $\ell_q \to M - \ell_{M-q}$ and in the second parametrization as
\beq
\label{eq:conjpart}
M_a \rightarrow M_{n+2-a}, \qquad N_a \rightarrow N_1 - N_{2+n-a}.
\eeq

As mentioned above, the asymptotically locally $AdS_4 \times S^7$ solutions are also fully characterized by a partition $\rho$ of $N$. Crucially, however, the partition $\rho$ will have only non-zero entries, as we explain below. In particular, in the parametrization by $\{N_a\}$ and $\{M_a\}$ we will not have a value for $M_{n+1}$. Eq.~\eqref{eq:totnm} will thus give us $N$ but not $M$. Of course, generically the $M_a$ with $a<n+1$ will be non-zero, so we know $M$ must be non-zero, but we will not be able to fix its value. In the parametrization $\rho=\{\ell_1,\ell_2,\ell_3,\ldots\}$ all the $\ell_q$ will be non-zero, and the total number $N$ will still be a sum of the $\ell_q$ as in eq.~\eqref{eq:nellsum}, though the upper limit of the sum will be $\sum_{a=1}^n M_a<M$. In a similar fashion, the asymptotically locally $AdS_4 \times S^7$ solutions will have $M'\neq 0$ whose value we will not be able to fix. Without $M$ or $M'$ we will not be able to identify gauge algebras $\mathfrak{u}(M)$ or $\mathfrak{u}(M')$. The partition $\rho$ will still specify a Young tableau, as described above, but without an algebra we will not be able to identify a representation ${\cal R}$ with the Young tableau.

\subsubsection{Asymptotically Locally $AdS_7 \times S^4$ Solutions}
\label{sec:ads7part}

In the asymoptotically locally $AdS_7 \times S^4$ solutions reviewed in sec.~\ref{subsec:ads7metric}, no explicit M2- or M5-brane sources appear in the 11d SUGRA equations. However, the solutions have non-zero flux of the 11d SUGRA 4- or 7-form wrapping closed, non-contractible 4- or 7-cycles, respectively. Presumably the M2- and M5-branes have been replaced by these fluxes, or ``dissolved into flux,'' similar to how D3-branes are replaced by five-form flux in the $AdS_5 \times S^5$ solution of type IIB SUGRA (i.e. how open string degrees of freedom are replaced by closed string degrees of freedom). As explained in ref.~\cite{Bachas:2013vza}, from these fluxes we can recover an ordered partition $\rho$ and hence a Wilson surface representation $\cal{R}$, as follows.

\begin{figure}[t!]
\begin{center}
\includegraphics[width=.6\textwidth]{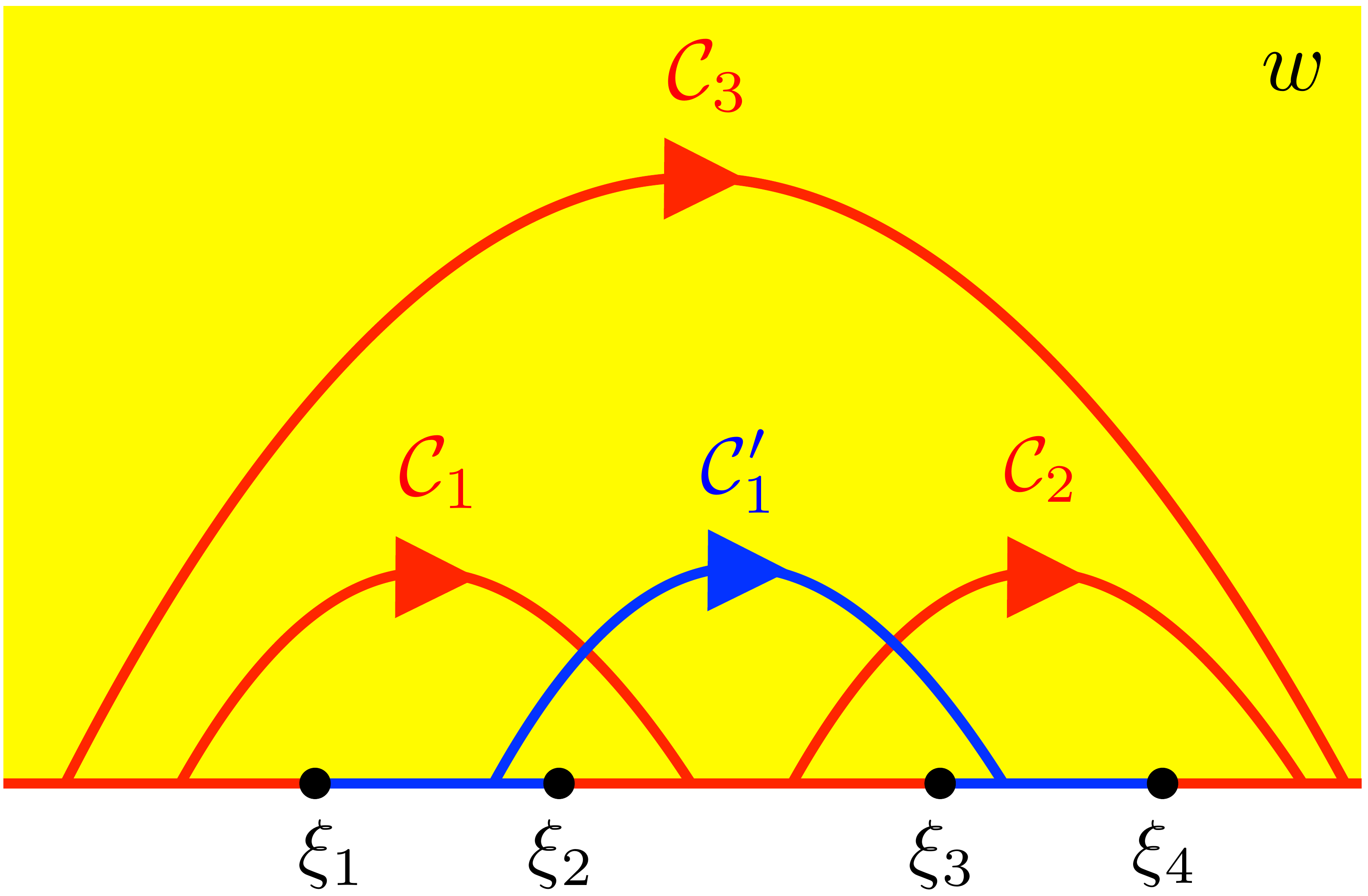}
\caption{\label{fig:ads7plane} Sketch of the upper-half complex $w$ plane for the asymptotically locally $AdS_7 \times S^4$ solutions with metric in eq.~\eqref{eq:metric2} with the $h$ and $G$ of eq.~\eqref{eq:ads7metric}. We show an example in which $G$ has four branch points on the $\textrm{Re}\left(w\right)$ axis, $\xi_1$, $\xi_2$, $\xi_3$, $\xi_4$. The geometry has two $S^3$'s. On the $\textrm{Re}\left(w\right)$ axis, in each red segment one of these $S^3$'s collapses to a point while the other does not, while in each blue segment the behavior of the two $S^3$'s is switched. The geometry thus has three non-contractible 4-cycles, ${\cal C}_1$, ${\cal C}_1'$, and ${\cal C}_2$. The 4-cycle ${\cal C}_3={\cal C}_1+{\cal C}_2$ can be continuously deformed to $r \to \infty$, and indeed, in that region becomes the $S^4$ of the asymptotically locally $AdS_7 \times S^4$ region. Taking the product of a non-contractible 4-cycle in which one $S^3$ collapses with the $S^3$ that does not collapse defines a non-contractible 7-cycle. Each \textit{pair} of branch points thus defines a non-contractible 4- and 7-cycle.}
\end{center}
\end{figure}

How do we identify closed, non-contractible 4- and 7-cycles in the geometry eq.~\eqref{eq:metric2} with the $h$ and $G$ of eq.~\eqref{eq:ads7metric}? As an example, fig.~\ref{fig:ads7plane} shows the upper-half complex $w$ plane for $n=1$, that is, with $2n+2=4$ branch points on the $\textrm{Re}\left(w\right)$ axis, $\xi_1$, $\xi_2$, $\xi_3$, $\xi_4$. Fig.~\ref{fig:ads7plane} also shows that geometry's three independent non-contractible 4-cycles, the curves ${\cal C}_1$, ${\cal C}_1'$, and ${\cal C}_2$, and a single, non-independent 4-cycle, ${\cal C}_3$, obeying ${\cal C}_3={\cal C}_1+{\cal C}_2$. However, unlike the other 4-cycles, ${\cal C}_3$ can be continuously deformed to $r \to \infty$.

To see why the curves ${\cal C}_1$, ${\cal C}_1'$, and ${\cal C}_2$ determine closed, non-contractible 4-cycles, recall from sec.~\ref{sec:sugra} that the geometry has two $S^3$'s, and that regularity requires $h=0$ and $G=\pm i$ on the $\textrm{Re}\left(w\right)$ axis. At each point on the $\textrm{Re}\left(w\right)$ axis one of the two $S^3$'s collapses to a point, while the other does not. Specifically, from eqs.~\eqref{eq:metric2},~\eqref{eq:wpm}, and~\eqref{eq:metfuncs} we see that $h=0$ and $G=+i$ implies $f_2=0$ while $f_3\neq 0$, and so one $S^3$ collapses to a point while the other does not. Conversely, $h=0$ and $G=-i$ implies $f_2 \neq 0$ and $f_3=0$ interchanging the behavior of the two $S^3$'s. When we cross a branch point $\xi_j$, $G$ flips sign from $G = \pm i$ to $G = \mp i$, which implies that different $S^3$'s collapse to a point on the two sides of $\xi_j$. As a result, any curve in the upper-half complex $w$ plane that begins on the $\textrm{Re}\left(w\right)$ axis, jumps over \textit{two} branch points, and then ends on the $\textrm{Re}\left(w\right)$ axis, when combined with the $S^3$ that collapses at the curve's endpoints, forms a closed, non-contractible 4-cycle. In contrast, a curve jumping over an odd number of branch points does not define a closed 4-cycle. Each closed, non-contractible 4-cycle also defines a closed, non-contractible 7-cycle: given a 4-cycle in which one $S^3$ collapses, simply take the product with the $S^3$ that does not collapse. In short, every \textit{consecutive pair} of branch points defines a closed, non-contractible 4-cycle and 7-cycle.\footnote{A word of caution: in ref.~\cite{Bachas:2013vza} the notation ${\cal C}_a$ refers only to a curve in the upper-half $w$ plane, while in this paper it refers to a 4-cycle, i.e. the curve times the $S^3$ that collapses at the curve's endpoints. Correspondingly, in this paper a 7-cycle will be denoted ${\cal C}_a \times S^3$.}

As mentioned above, in fig.~\ref{fig:ads7plane} the 4-cycle ${\cal{C}}_3$ can be decomposed as ${\cal C}_3={\cal C}_1+{\cal C}_2$, and can be continuously deformed to $r \to \infty$. As mentioned below eq.~\eqref{eq:ads7asympcoords}, fixing $u$ and sending $r \to \infty$ takes us to the asymptotic $AdS_7\times S^4$ boundary a distance $u$ from the defect, and so we identify the 4-cycle at $r \to \infty$ as the $S^4$ of the asymptotic local $AdS_7 \times S^4$.

The total number of non-contractible 4-cycles in these geometries is $2n+1$, i.e. the number of branch points minus one. The number of non-contractible cycles thus matches the number of free parameters in the solution: as mentioned in sec.~\ref{subsec:ads7metric}, we can fix one of the $2n+2$ branch points using translations, leaving $2n+1$ free parameters.

Integrating $F_4$ over a non-contractible 4-cycle yields a charge that we interpret as a number of dissolved M5- or M5$'$-branes. We will not write the explicit form for $F_4$, which appears in ref.~\cite{Bachas:2013vza}, but we will need the explicit forms for these charges in terms of the locations of branch points $\xi_j$. In the upper-half complex $w$ plane and starting from the left-most cycle, let ${\cal C}_a$, with $a = 1,2,\ldots,n+1$, denote the independent non-contractible 4-cycles all involving the same collapsing $S^3$. In the $n=1$ example above these are the 4-cycles ${\cal C}_1$ and ${\cal C}_2$. Anticipating their role in a partition $\rho$, we refer to the integrals of the 4-form flux over ${\cal C}_a$ as $M_a$, i.e.
\beq
\label{eq:madef}
M_a \equiv \frac{1}{2 (4\pi^2 \gn)^{1/3}} \int_{{\cal C}_a} F_4,
\eeq
with $\gn$ the 11d Newton's constant.  For $\gamma\in(-1,0)$, $M_a$ turns out to be proportional to (minus) the distance between two consecutive branch points~\cite{Bachas:2013vza}:\footnote{As mentioned below eq.~\eqref{eq:metfuncs}, these solutions are invariant under simultaneous re-scaling of $c_1$ and $h = - i (w - \overline{w})$. Re-scaling $h$ clearly means re-scaling $w$ and hence the branch points $\xi_j$. As a result, the charges $M_a$, $M_b'$, and $N_a$ in eqs.~\eqref{eq:m5flux},~\eqref{eq:m5primeflux}, and~\eqref{eq:m2flux}, respectively, are invariant under the re-scaling.}
\beq
\label{eq:m5flux}
M_a = -\frac{1}{c_1^3} \frac{\left(1+\gamma\right)^3}{\gamma}  \frac{\left(2\pi\right)^{4/3}}{\gn^{1/3}} \left(\xi_{2a}-\xi_{2a-1}\right).
\eeq
We have chosen an orientation so that for positive $c_1$ and $\gamma\in(-1,0)$, the $M_a$ are positive.  In the upper-half complex $w$ plane and starting from the left-most cycle, let ${\cal C}_b'$, with $b=1,2,\ldots,n$, denote the independent non-contractible 4-cycles that all involve the same collapsing $S^3$ orthogonal to the $S^3$ that collapses in the ${\cal C}_a$ cycles. In the $n=1$ example above, this is the 4-cycle ${\cal C}_1'$. Since the $S^3$ that collapses in the ${\cal C}_b'$ is orthogonal to that of the ${\cal C}_a$, we refer to the integral of the 4-form flux over ${\cal C}_b'$ as M5$'$-brane charge, which we denote $M_b'$. We define $M_b'$ with the same conventions as eq.~\eqref{eq:madef}. For $\gamma\in(-1,0)$, $M_b'$ also turns out to be proportional to the distance between two consecutive branch points~\cite{Bachas:2013vza},
\beq
\label{eq:m5primeflux}
M_b' = \frac{1}{c_1^3} \frac{\left(1+\gamma\right)^3}{\gamma^2} \frac{\left(2\pi\right)^{4/3}}{\gn^{1/3}} \left(\xi_{2b+1}-\xi_{2b}\right).
\eeq
Again, we have chosen the orientation so that for positive $c_1$ and $\gamma\in(-1,0)$, the $M_b'$ are positive.
Such M5$'$-brane charge presumably arises from brane polarization, a.k.a. the Myers effect~\cite{Myers:1999ps}: the M2-branes source $C_3$ in a background with non-zero $F_4$ from the M5-branes, so that the Wess-Zumino (WZ) term $\propto \int C_3 \wedge F_4 \wedge F_4$ in the 11d SUGRA action produces non-trivial $F_4$ flux orthogonal to that of the M5-branes, i.e. M5$'$-brane flux.\footnote{The presence of M5$'$-branes is also related to the fact that totally anti-symmetric representations can be described by a probe M5$'$-brane in $AdS_7 \times S^4$, wrapping $AdS_3 \in AdS_7$ and $S^3 \in S^7$~\cite{Lunin:2007ab}.}

The 4-cycle at $r \to \infty$ identified as the $S^4$ in the asymptotically locally $AdS_7 \times S^4$ region can be decomposed as $\sum_{a=1}^{n+1}{\cal C}_a$. Correspondingly, the 4-form flux through that cycle determines the total number of M5-branes, $M=\sum_{a=1}^{n+1} M_a$. Using eq.~\eqref{eq:m5flux}, we can relate $M$ to the coefficient $m_1$ defined in the Legendre polynomial expansion of eq.~\eqref{eq:ads7legendre},
\beq
\label{eq:ads7m1exp}
m_1 \equiv \sum_{a=1}^{n+1} (\xi_{2a} - \xi_{2a-1}) = -c_1^3 \, \frac{\gamma}{\left(1+\gamma\right)^3} \frac{\gn^{1/3}}{\left(2\pi\right)^{4/3}} \, M.
\eeq
Plugging eq.~\eqref{eq:ads7m1exp} into eq.~\eqref{eq:ads7rad} we thus recover the usual relation between $M$ and the radius $L_{S^4}$ of the asymptotic $S^4$,
\beq\label{eq:ls4-to-m}
L_{S^4}^3 = - \frac{1}{c_1^3 }\frac{(1 + \gamma)^3}{\gamma} \, m_1 = \frac{\gn^{1/3}}{\left(2\pi\right)^{4/3}} \, M.
\eeq

In contrast to the 4-form flux, the integrated 7-form flux, which we interpret as the number of dissolved M2-branes, is ambiguous: due to the WZ term $\propto \int C_3 \wedge F_4 \wedge F_4$ in the 11d SUGRA action, when $F_4\neq 0$ we cannot define 7-form flux that is simultaneously local, conserved, quantized, and invariant under large gauge transformations of $C_3$~\cite{Marolf:2000cb}. At best we can define 7-form flux that has three of these properties but not the fourth. Following ref.~\cite{Bachas:2013vza} we will use the 7-form flux that gives the Page charge, which is local, conserved, and quantized, but not gauge invariant. Explicitly, we will integrate $\star_{11d} F_4 + \frac{1}{2} C_3 \wedge F_4$ over a closed, non-contractible 7-cycle, where the $C_3 \wedge F_4$ term comes from the WZ term, and clearly produces a M2-brane charge that is not invariant under large gauge transformations of $C_3$.

With that choice, to extract a fixed value of M2-brane charge we must fix a gauge of $C_3$, which we do as follows. In the solutions of ref.~\cite{Bachas:2013vza}, $C_3$ can be written as a sum of three terms, one with legs along $AdS_3$, and two with legs along the two $S^3$'s, respectively. Let $C_3'$ denote the term with legs along the $S^3$ that collapses in a 4-cycle ${\cal C}_b'$ but not in a 4-cycle ${\cal C}_a$. To gauge-fix, we demand that $C_3'=0$ when the $S^3$ collapses at one endpoint of one of the ${\cal C}_b'$.\footnote{Regularity demands that $C_3$ vanish whenever it wraps a vanishing cycle, so our gauge choice simply enforces regularity. However, due to the presence of M5-branes we can make $C_3$ well defined only on a patch. The entire geometry can be covered with $n+1$ patches corresponding to $n+1$ gauge choices.} Since the geometry has $n$ 4-cycles ${\cal C}_b'$, each with two endpoints, we have $n+1$ choices of such endpoints, namely the left end-point of each 4-cycle plus the right endpoint of the right-most 4-cycle. For any such choice, in the solutions of ref.~\cite{Bachas:2013vza} the M2-brane charges are ordered from smallest to largest as we move from left to right in the upper-half complex $w$ plane, but generically include negative values. Of the $n+1$ gauge choices for $C'_3$, only one allows for the M2-brane charges to be all positive: choosing $C_3'=0$ at the right endpoint of ${\cal C}_n'$ produces all positive M2-brane charges, with the right-most M2-brane charge vanishing. To see why, deform all the ${\cal C}_a$ to lie along the real axis and note that $\star_{11d}F_4$ vanishes along the real axis since it wraps a vanishing 7-cycle. The only contribution then comes from the WZ contribution $C_3 \wedge F_4$. Since we have set $C_3' = 0$ along the cycle ${\cal C}_{n+1}$, the WZ contribution also vanishes along this cycle.

With $C_3$ gauge-fixed, and anticipating their role in a partition $\rho$, we denote the integrals of 7-form flux through ${\cal C}_a$ as $M_a N_a$, which we interpret as the total number of M2-branes ending on the M5-branes associated with ${\cal C}_a$,\footnote{Our $M_a N_a$ in eq.~\eqref{eq:manadef} was defined as $N_a$ in ref.~\cite{Bachas:2013vza}.}
\beq
\label{eq:manadef}
M_a N_a = \frac{1}{4 (4\pi^2 \gn)^{2/3}} \int_{{\cal C}_a  \times S^3} C'_3 \wedge F_4.
\eeq
  For the solutions in ref.~\cite{Bachas:2013vza}, and using eq.~\eqref{eq:m5primeflux}, $N_a$ turns out to be the sum of M5$'$-brane fluxes on all 4-cycles from ${\cal C}'_a$ to ${\cal C}'_n$,
\beq
\label{eq:m2flux}
N_a = \sum_{b=a}^n M_b' =  \frac{1}{c_1^3} \frac{\left(1+\gamma\right)^3}{\gamma^2} \frac{\left(2\pi\right)^{4/3}}{\gn^{1/3}} \sum_{b=a}^n\left(\xi_{2b+1}-\xi_{2b}\right),
\eeq
and where the vanishing of the right-most 7-form flux implies $N_{n+1}=0$.

Validity of the SUGRA approximation requires curvatures to be small compared to the Planck scale. This requires the branch points $(\xi_{2b}, \xi_{2b+1})$ to be far apart for all $b$, and therefore $N_{a} \gg N_{a+1}$ for all $a$, and also $N_n \gg 1$ (recall the $N_a$ are ordered). However, with $AdS_7 \times S^4$ asymptotics, when the $N_a$ are of order unity the system should be well-described by probe branes. Taking the probe limit in our results for $b_{6d}$ indeed reproduces calculations using probe branes, so our results seem to apply over a wider regime than might be expected. Similar statements about the range of validity of our results apply to what follows (i.e. the $N_b'$ in eq.~\eqref{eq:m2pflux}).

The total number of dissolved M2-branes is the sum of 7-form fluxes of all the ${\cal C}_a$, $N = \sum_{a=1}^n M_a N_a$. Eq.~\eqref{eq:m2flux} shows that the 7-form fluxes through the non-contractible 7-cycles is fixed by 4-form fluxes, and hence are not independent parameters. The total number of free parameters thus remains $2n+1$, though we have a choice to package some of them as either the $M_b'$ or the $N_a$. In what follows we will choose the latter.

In short, eqs.~\eqref{eq:m5flux} and~\eqref{eq:m2flux} allow us to extract the $2n+1$ free parameters of a partition $\rho$, namely the sets of $n$ distinct non-zero integers $\{N_a\}$ and the $n+1$ distinct degeneracies $\{M_a\}$, from the $2n+1$ free parameters, i.e. the branch points $\xi_j$, of the asymptotically locally $AdS_7 \times S^4$ solutions.  We therefore identify the same partition $\rho$ determining the brane intersection, the 11d SUGRA solution, and the Wilson surface's representation ${\cal R}$.

For a Wilson line in Yang-Mills theory all observables are invariant under the combined operations of complex conjugation of the representation, ${\cal R} \to \overline {\cal R}$, and orientation reversal of the Wilson line. We expect the same to be true for a Wilson surface in the M5-brane theory. Indeed, we can demonstrate that these combined operations can be realized in these 11d SUGRA solutions by the combined operations of a gauge transformation of $C_3$, which produces all negative M2-brane charges, followed by an orientation reversal of the 7-cycles, which flips the M2-brane charges back to being positive.

To implement these combined operations in the geometry, we start with a geometry parametrized by a partition $\rho$ corresponding to a representation ${\cal R}$. We perform the unique gauge transformation of $C_3'$ that makes all the M2-brane charges negative, namely requiring $C_3'=0$ at the left endpoint of ${\cal C}_1$. Under this gauge transformation, the $N_a$ are all shifted as $N_a \rightarrow N_a - N_1$.  The shifted $N_a$ are thus all negative, and of course the shifted $N_1$ vanishes. To obtain positive M2-brane charges we reverse the orientation of the 7-cycles, thereby flipping the signs of the shifted $N_a$. We can do so for example by reversing the orientation of the two $S^3$'s and the direction of integration in the upper-half $w$ plane (reversing the arrows in fig.~\ref{fig:ads7plane}). Finally, it is convenient, but not necessary, to make the coordinate transformation $\textrm{Re}\left(w\right) \rightarrow - \textrm{Re}\left(w\right)$. This yields a geometry parametrized by a set of branch points $\xi_j^c \equiv \xi_{2n+3-j}$ in the same gauge we defined above eq.~\eqref{eq:m2flux}. Computing the M5- and M2-brane charges via eqs.~\eqref{eq:m5flux} and~\eqref{eq:m2flux}, respectively, then reveals that the charges have been shifted as $M_a \to M_{n+2 - a}$ and $N_a \to N_1 - N_{2+n-a}$. From eq.~\eqref{eq:conjpart} we recognize this shift as complex conjugation of the representation, ${\cal R} \to \overline{{\cal R}}$.

Crucially, the geometry is invariant under these combined operations. As we review in sec.~\ref{sec:hc}, the CFT's EE is proportional to the area of a minimal surface in the dual geometry~\cite{Ryu:2006bv,Ryu:2006ef,Nishioka:2009un} (eq.~\eqref{eq:rt} below). That surface only ``knows'' about the geometry, and hence is invariant under any operations that leave the geometry invariant. Indeed, our results for EE, and in particular $b_{6d}$, are invariant under ${\cal R} \to \overline{{\cal R}}$, as mentioned below eq.~\eqref{eq:b6dcompact}.

The operations leading to complex conjugation in these 11d SUGRA solutions have a simple interpretation in the corresponding brane intersection, as Hanany-Witten moves~\cite{Hanany:1996ie}. As a simple example, consider a brane intersection of the type in figs.~\ref{fig:part1} and~\ref{fig:part2}, with $N$ M2-branes ending on $N$ distinct M5-branes (out of the $M$ total number of M5-branes), producing a totally anti-symmetric representation of rank $N$. Imagine we move an M5$'$-brane from infinity on the left to a finite value of $x_2$, to the left of the M5-branes. If we send this M5$'$-brane to infinity on the right, then when the M5$'$-brane passes through the stack of M5-branes the $N$ M2-branes will be destroyed and $M-N$ anti-M2-branes will be created between the M5$'$-brane and each M5-brane that did not have an M2-brane ending on it. An orientation reversal $x_2 \to -x_2$ then maps the anti-M2-branes to M2-branes, while the M5- and M5$'$-branes are not mapped to anti-branes. Clearly the total number of M2-branes is ambiguous, as in the 11d SUGRA solutions above. Moreover, such a Hanany-Witten move clearly corresponds to complex conjugation of the representation, which for a totally anti-symmetric representation of rank $N$ indeed acts as $N \to M-N$.

\subsubsection{Asymptotically Locally $AdS_4 \times S^7$ Solutions}
\label{sec:ads4part}

Similar to the asymptotically locally $AdS_7 \times S^4$ solutions, in the asymptotically locally $AdS_4 \times S^7$ solutions reviewed in sec.~\ref{subsec:ads4metric} no explicit M2- or M5-brane sources appear. However the solution does have non-zero 4- and 7-form fluxes supported on non-contractible 4- and 7-cycles, respectively, representing dissolved M2-, M5-, and M5$'$-branes. We can extract an ordered partition $\rho$ from these fluxes in a fashion very similar to the asymptotically locally $AdS_7 \times S^4$ solutions in sec.~\ref{sec:ads7part}.

\begin{figure}[t!]
\begin{center}
\includegraphics[width=.6\textwidth]{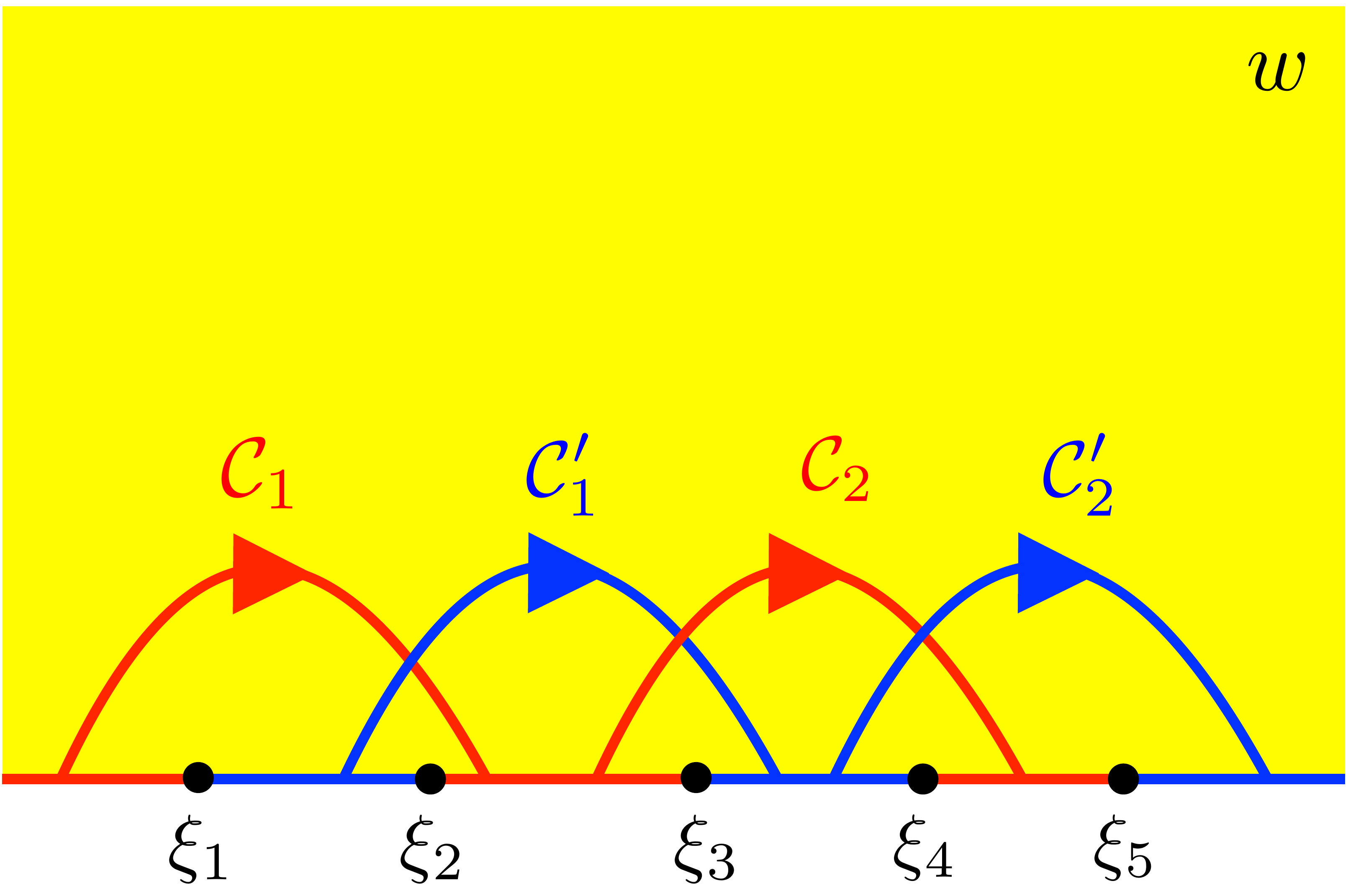}
\caption{\label{fig:ads4plane} Sketch of the upper-half complex $w$ plane for the asymptotically locally $AdS_4 \times S^7$ solutions with metric in eq.~\eqref{eq:metric2} with the $h$ and $G$ of eq.~\eqref{eq:ads4metric}. We show an example in which $G$ has five branch points on the $\textrm{Re}\left(w\right)$ axis, $\xi_1,\ldots,\xi_5$. Like the asymptotically locally $AdS_7 \times S^4$ solutions (i.e. the example in fig.~\ref{fig:ads7plane}), each \textit{pair} of branch points defines a non-contractible 4- and 7-cycle. In this example we labeled the four independent non-contractible 4-cycles ${\cal C}_1$, ${\cal C}_1'$, ${\cal C}_2$, and ${\cal C}'_2$. However, unlike the asymptotically locally $AdS_7 \times S^4$ solutions, in these solutions at $r \to \infty$ the geometry is asymptotically locally $AdS_4 \times S^7$, which has the non-contractible 7-cycle $S^7$ and no non-contractible 4-cycle except $AdS_4$.}
\end{center}
\end{figure}

The procedure to identify closed, non-contractible 4- and 7-cycles in the asymptotically locally $AdS_4 \times S^7$ solutions is nearly identical to the asymptotically locally $AdS_7 \times S^4$ solutions in sec.~\ref{sec:ads7part}, with one crucial difference: where the latter solutions had an even number of branch points, the asymptotically locally $AdS_4 \times S^7$ solutions have an odd number. Explicitly, the $G$ in eq.~\eqref{eq:ads7metric} involves a sum over $2n+2$ of the branch points $\xi_j$, while the $G$ in eq.~\eqref{eq:ads4metric} is a sum over $2n+1$ of the $\xi_j$. As an example, fig.~\ref{fig:ads4plane} shows the upper-half complex $w$ plane for $n=2$, meaning $2n+1=5$ branch points on the $\textrm{Re}\left(w\right)$ axis, $\xi_1,\ldots,\xi_5$. Fig.~\ref{fig:ads4plane} also shows the geometry's four independent closed, non-contractible 4-cycles, namely ${\cal C}_1$, ${\cal C}_1'$, ${\cal C}_2$, and ${\cal C}'_2$. By exactly the same arguments as in sec.~\ref{sec:ads7part}, each consecutive pair of branch points defines a closed, non-contractible 4- and 7-cycle.

Another key difference with the asymptotically locally $AdS_7 \times S^4$ solutions is that none of the non-contractible 4-cycles can be continuously deformed to $r \to \infty$, as obvious in the example of fig.~\ref{fig:ads4plane}. Indeed, as $r \to \infty$ these solutions are asymptotically locally $AdS_4 \times S^7$, which has no non-contractible 4-cycles (besides $AdS_4$) and one non-contractible 7-cycle, $S^7$. The latter cannot be decomposed into the other 7-cycles, ${\cal C}_a \times S^3$ and ${\cal C}_b'\times S^3$, which have the topology of $S^4 \times S^3$ and hence have non-contractible 4- and 3-cycles, unlike $S^7$.

We define the closed, non-contractible 4-cycles as in sec.~\ref{sec:ads7part} by the collapse of one or the other $S^3$, denoted ${\cal C}_a$ and ${\cal C}_b'$ with $a,b=1,2,\ldots,n$. The total number of such ${\cal C}_a$ and ${\cal C}_b'$ in these geometries is $2n$. The number of non-contractible 4-cycles thus matches the number of free parameters in the solution: as mentioned in sec.~\ref{subsec:ads4metric}, we can fix one of the $2n+1$ branch points using translations, leaving $2n$ free parameters.

Integrating $F_4$ over a non-contractible 4-cycle gives us 4-form charge, which we again interpret as a number of dissolved M5- or M5$'$-branes. As in sec.~\ref{sec:ads7part}, $M_a$ is defined as the integral of the 4-form flux over ${\cal C}_a$ in eqs.~\eqref{eq:madef}, and similarly for $M_{b}'$ and ${\cal C}_b'$. In these solutions the expressions for $M_a$ and $M_b'$ in terms of the $\xi_j$ are in fact identical to those in eqs.~\eqref{eq:m5flux} and~\eqref{eq:m5primeflux}, respectively.

As in sec.~\ref{sec:ads7part}, and again following ref.~\cite{Bachas:2013vza}, we will use the 7-form that gives the Page charge, $\star_{11d} F_4 + \frac{1}{2} C_3 \wedge F_4$. We thus need to fix a gauge for $C_3$. Following ref.~\cite{Bachas:2013vza} we make the unique gauge choice in which the gauge potential $C_3$ vanishes in the asymptotic region and produces regular 7-form flux through the $S^7$ at $r \to \infty$, whose integral represents the total number of dissolved M2-branes, $N$. Specifically, with $C_3'$ defined in sec.~\ref{sec:ads7part}, we demand that $C_3'=0$ at the left end-point of ${\cal C}_1'$, and that the part of $C_3$ with legs along the $S^3$ that collapses in a 4-cycle ${\cal C}_a$ vanishes at the right endpoint of ${\cal C}_n$. With these gauge choices the WZ contribution vanishes and
\beq
N = \frac{1}{2 (4\pi^2 \gn)^{1/3}} \int_{{\cal S}^7} \star_{11d} F_4.
\eeq
Additionally, with these gauge choices the expression for $N_a$, the 7-form flux through a 7-cycle ${\cal C}_a \times S^3$, is identical to eq.~\eqref{eq:m2flux}. We interpret $N_a$ as the number of M2-branes that end on the M5-brane associated with ${\cal C}_a$. A crucial difference from sec.~\ref{sec:ads7part}, however, is that now all of these 7-form charges are non-zero, and in particular there are only $n$ independent charge $N_a$, as opposed to the $n+1$ independent charges of sec.~\ref{sec:ads7part}. Although the $S^7$ is not a sum of the 7-cycles ${\cal C}_a \times S^3$, the total number $N$ of M2-branes nevertheless turns out to be the sum of 7-form charges, $N = \sum_{a=1}^n M_a N_a$~\cite{Bachas:2013vza}.

Using eqs.~\eqref{eq:m5flux},~\eqref{eq:m5primeflux}, and~\eqref{eq:m2flux} for $M_a$, $M_b'$, and $N_a$, respectively, we can relate $N$ to the radius $L_{S^7}$ of the $r \to \infty$ asymptotic $S^7$ in eq.~\eqref{eq:ads4rad}. In particular, we need $-m_1^2-m_2$, with the coefficients $m_1$ and $m_2$ defined via the Legendre polynomial expansion of $G$ in eq.~\eqref{eq:ads4legendre},
\beq
\label{eq:ads4m1m2exp}
-m_1^2 - m_2 = 2 \sum_{a=1}^n \sum_{b=a}^n (\xi_{2a} - \xi_{2a-1}) (\xi_{2b+1} - \xi_{2b}) = 2 c_1^6\,\frac{\left(-\gamma\right)^3}{\left(1+\gamma\right)^6} \,\frac{\gn^{2/3}}{\left(2\pi\right)^{8/3}}\, \sum_{a=1}^n M_a N_a.
\eeq
Plugging eq.~\eqref{eq:ads4m1m2exp} into eq.~\eqref{eq:ads4rad} and using $N = \sum_{a=1}^n M_a N_a$, we recover the usual relation between $N$ and the radius $L_{S^7}$ of the asymptotic $S^7$,
\beq
\label{eq:s7radius}
L_{S^7}^6 = \frac{16}{c_1^6}\frac{(1 + \gamma)^6}{(-\gamma)^3} (-m_1^2 - m_2)
= 32 \,\frac{\gn^{2/3}}{\left(2\pi\right)^{8/3}}\, N.
\eeq

In short, eqs.~\eqref{eq:m5flux} and~\eqref{eq:m2flux} allow us to extract the $2n$ free parameters of a partition $\rho$, the sets of $n$ distinct non-zero integers $\{N_a\}$ and $n$ degeneracies $\{M_a\}$, from the $2n$ free parameters of the asymptotically locally $AdS_4 \times S^7$ solutions, the branch points $\xi_j$.

In contrast to sec.~\ref{sec:ads7part}, only $n$ degeneracies $\{M_a\}$ appear because the geometry has only $n$ 4-cycles ${\cal C}_a$. In particular, as mentioned below eq.~\eqref{eq:conjpart}, we cannot determine a value for $M_{n+1}$ from these solutions, and hence cannot determine $M = \sum_{a=1}^{n+1} M_a$. However, $\sum_{a=1}^n M_a < M$ will be non-zero, implying $M \neq 0$. Analogous statements apply for the $M_b'$ and $M'$. In 11d SUGRA terms, these solutions have no asymptotically $AdS_7 \times S^4$ regions that would allow us to fix $M_{n+1}$ or $M'_{n+1}$. As a result, we will not be able to identify $\mathfrak{su}(M)$ or $\mathfrak{su}(M')$, so while we will have a partition $\rho$ and corresponding Young tableau, we will not be able to identify a representation ${\cal R}$.

\begin{figure}[t!]
\begin{center}
\includegraphics[width=.85\textwidth]{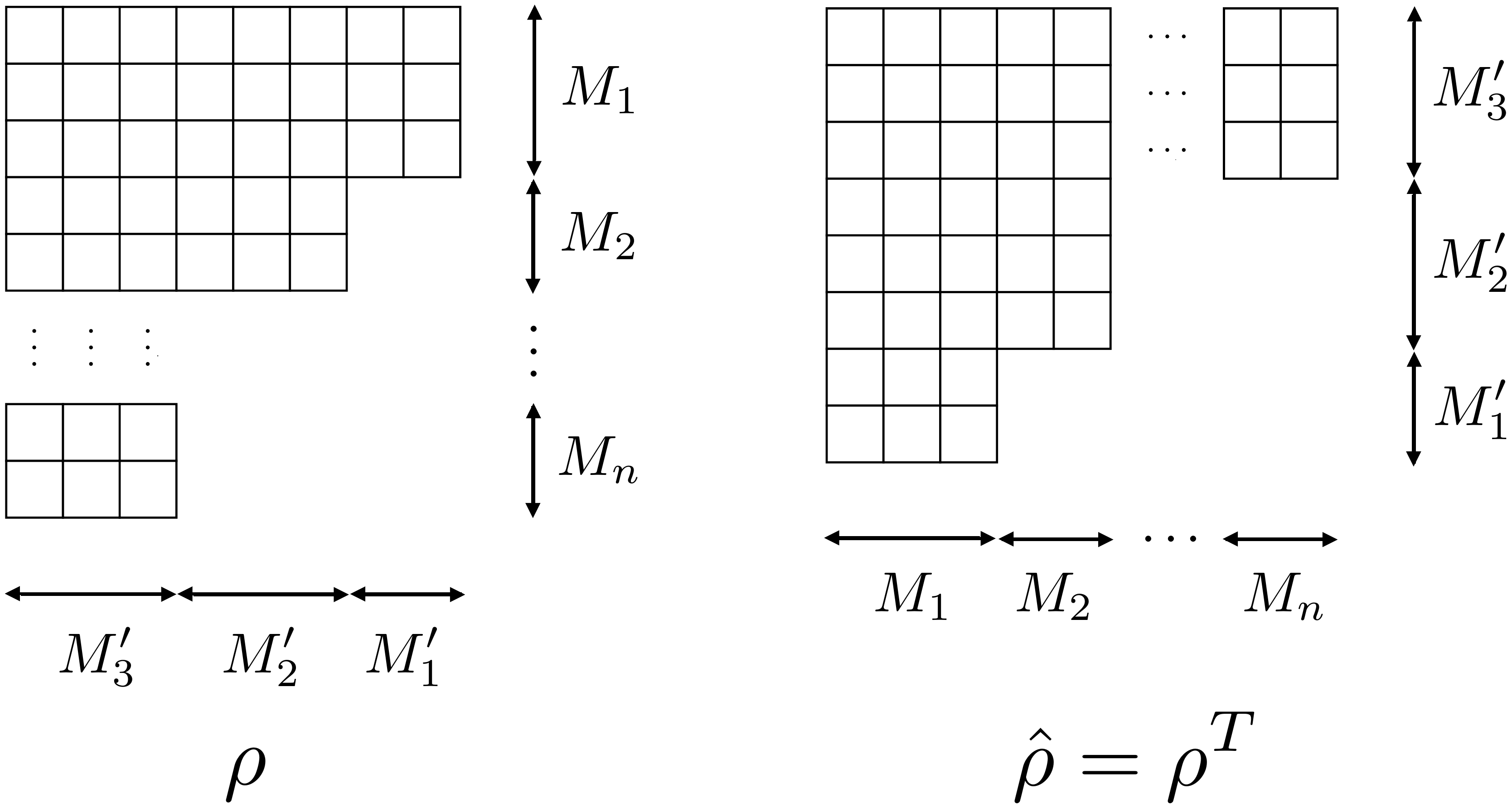}
\caption{\label{fig:young2} The asymptotically $AdS_4 \times S^7$ solutions are fully characterized by a partition $\rho$ that can be specified by the sets of M2- and M5-brane charges $\{N_a\}$ and $\{M_a\}$, respectively, with $a = 1,2,\ldots,n$, as shown on the left in fig.~\ref{fig:young} . However, $\rho$ can also be specified by the sets of M5- and M5$'$-brane charges  $\{M_a\}$ and $\{M_b'\}$, with $b=1,2,\ldots,n$, as shown on the left above. In that parametrization, the $M_b'$ are degeneracies of columns. In the transposed partition $\rho^T$, shown on the right above, the $M_b'$ are degeneracies of rows, as they are in $\hat{\rho}$ in eq.~\eqref{eq:hatrhodef}, hence we identify $\hat{\rho}=\rho^T$.}
\end{center}
\end{figure}

Also in contrast to sec.~\ref{sec:ads7part}, the M5-branes and M5$'$-branes appear here on equal footing: no 4-cycle in the asymptotic $r \to \infty$ region selects one over the other. As a result, instead of labelling the solutions by the partition $\rho$ defined by the charges $N_a$ and $M_a$ coming from the ${\cal C}_a$ cycles, we could have labeled the solutions by another partition $\hat \rho$ defined by the charges $M'_b$ and $N'_b$ coming from the ${\cal C}'_b$ cycles,
\beq
\label{eq:hatrhodef}
\hat \rho = \{ \underbrace{N'_n, N'_n, ... N'_n}_{M'_n}, \underbrace{N'_{n-1}, N'_{n-1}, ... N'_{n-1}}_{M'_{n-1}}, ... \,\underbrace{N'_1, N'_1, ... N'_1}_{M'_1} \} .
\eeq
The $M'_b$ are defined as in eq.~\eqref{eq:m5primeflux} and the $N'_b$ are defined by an analogue of eq.~\eqref{eq:m2flux},
\beq
\label{eq:m2pflux}
N'_b = \sum_{a=1}^b M_a =  -\frac{1}{c_1^3} \frac{\left(1+\gamma\right)^3}{\gamma}  \frac{\left(2\pi\right)^{4/3}}{\gn^{1/3}} \sum_{a=1}^b \left(\xi_{2a}-\xi_{2a-1}\right).
\eeq
Requiring the branch points to be well separated so that the SUGRA approximation is reliable, we find the condition \(N_{b+1}' \gg N_b'\) for all \(b\).
The $2n$ free parameters of these solutions are fully determined by either $\rho$ or $\hat \rho$, hence the two must be related. To see how, we use the fact that we can fix any $2n$ free parameters we like, and hence can determine $\rho$ using $\{M_a\}$ and $\{M_b'\}$. In that parametrization, the $M_b'$ are the degeneracies of \textit{columns}, as shown on the left in fig.~\ref{fig:young2}. In the transposed partition, $\rho^T$, the $M_b'$ become the degeneracies of rows, as in $\hat{\rho}$, so we immediately identify $\rho^T = \hat{\rho}$.

Indeed, we can show that the solution determined by $\rho$ and $\gamma$ is identical, up to a choice of orientation, to the solution determined by $\hat{\rho}=\rho^T$ and $\gamma^{-1}$, as follows. As mentioned above, taking $\gamma \to 1/\gamma$ leaves $f_1$ and $\Omega$ invariant but maps $f_2 \leftrightarrow f_3$, effectively interchanging the geometry's two $S^3$'s and hence swapping the M5- and M5$'$-branes. (The transformation of the 4-form fluxes under $\gamma \to \gamma^{-1}$ is consistent with this statement.)  Performing two operations, namely $\gamma \to \gamma^{-1}$ followed by trading the two $S^3$'s, and hence the M5- and M5$'$-branes, is thus a symmetry of the solution. However, the orientation of the 7-cycles, and hence the sign of the M2-brane fluxes, is reversed in the process.  The sign can be reversed by an overall orientation reversal, as in the $AdS_7 \times S^4$ case.  The effect on $\rho$ is to map $\rho \to \rho^T$, which because the M5- and M5$'$ were swapped, we identify as $\hat{\rho}$. In short, the solution determined by $\rho$ and $\gamma$ is identical to the solution determined by $\hat{\rho}=\rho^T$ and $\gamma^{-1}$, up to an overall orientation reversal, as advertised. We will see in section \ref{sec:abjm-bcft-cc} that our result for EE, and in particular for $b_{3d}$, is indeed invariant, up to an overall sign, under the simultaneous operations $\gamma \rightarrow \gamma^{-1}$ and $\rho \rightarrow \rho^T = \hat{\rho}$.

This symmetry should appear in the holographically dual BCFT. Of course, as mentioned in sec.~\ref{sec:intro}, the exact BCFTs dual to these 11d SUGRA solutions remain unknown. However, the symmetry relating solutions determined by $\rho$ or $\hat{\rho}$ suggests that the BCFTs arise from M2-branes ending on both M5- and M5$'$-branes, as in the brane intersection of sec.~\ref{sec:intro}. If we separate the M5-branes from one another (the Coulomb branch), then a partition $\rho$ will determine which M5-brane each M2-brane ends on. Alternatively, if we separate the M5$'$-branes from one another (the Higgs branch), then the partition $\hat \rho = \rho^T$ will determine which M5$'$-brane each M2-brane ends on. Simultaneously separating both M5- and M5$'$-branes (moving onto the Coulomb and Higgs branches simultaneously) should not be possible. The BCFT should have a duality that simultaneously swaps the separated M5-branes with separated M5$'$-branes, sends $\gamma \to \gamma^{-1}$, and sends $\rho \to \rho^T = \hat{\rho}$.

\section{Holographic Entanglement Entropy}
\label{sec:hc}

In this section we calculate $b_{6d}$ and $b_{3d}$ holographically from EE, following Ryu and Takayanagi's (RT's) prescription~\cite{Ryu:2006bv,Ryu:2006ef,Nishioka:2009un}. A very similar calculation for the asymptotically locally $AdS_7 \times S^4$ solutions appears in ref.~\cite{Gentle:2015jma}. In this section we will follow ref.~\cite{Gentle:2015jma} very closely.

To compute EE in a QFT vacuum we fix time $t$, separate space into two regions by an ``entangling surface,'' and trace out states outside the entangling surface, thus obtaining a reduced density matrix for the region inside. The EE is this reduced density matrix's von Neumann entropy. Generically EE diverges due to strong UV correlations near the entangling surface, so to extract physical information we must introduce a UV regulator.

Holographically, in an asymptotically AdS geometry, RT's prescription for the EE for a sub-region of the CFT is
\beq
\label{eq:rt}
\see = \frac{A}{4 \gn},
\eeq
where $A$ is the area of the minimal surface in the holographically dual geometry that approaches the entangling surface at the AdS boundary. Computing $\see$ is thus a two-step process. First, determine the minimal area surface by writing the area functional and solving the associated Euler-Lagrange equations. Second, plug that solution back into the area functional and integrate to obtain $A$. The UV divergences of $\see$ appear as divergences in $A$ near the AdS boundary. In AdS spacetime, the standard regulator is thus a cutoff on the FG holographic coordinate: rather than integrating to the AdS boundary $z=0$ we integrate only to $z= \varepsilon>0$.

In $AdS_3$ eq.~\eqref{eq:rt} reproduces known results for 2d CFTs~\cite{Ryu:2006bv,Ryu:2006ef,Nishioka:2009un,Holzhey:1994we,Calabrese:2004eu}. For example, when the entangling surface consists of two points a distance $\ell$ apart, the minimal surface in $AdS_3$ is a semi-circle at fixed $t$ with diameter $\ell$ centered on the $AdS_3$ boundary. In that case eq.~\eqref{eq:rt} gives
\beq
\label{eq:ee2d}
\see^{2d} = \frac{c}{3} \ln \left(\frac{\ell}{\varepsilon}\right) + \mathcal{O}\left(\varepsilon^0\right),
\eeq
with CFT central charge $c$. Henceforth, we will  use a superscript to distinguish $\see$ in different dimensions, such as the superscript $2d$ on $\see^{2d}$. Crucially, re-scaling the cutoff $\varepsilon$ changes the $\mathcal{O}\left(\varepsilon^0\right)$ terms, while the coefficient of $\ln \left(\frac{\ell}{\varepsilon}\right)$, namely $c/3$, is cutoff-independent and hence physical. In $AdS_4$ eq.~\eqref{eq:rt} produces the form expected for a 3d CFT~\cite{Ryu:2006bv,Ryu:2006ef,Nishioka:2009un}. For example when the entangling surface is a circle of radius $\ell$, the minimal surface in $AdS_4$ is a hemisphere at fixed $t$ with radius $\ell$ centered on the $AdS_4$ boundary. In that case eq.~\eqref{eq:rt} gives
\beq
\label{eq:ee3d}
\see^{3d} = c_1 \, \frac{\ell}{\varepsilon} + c_0 + \mathcal{O}\left(\varepsilon\right),
\eeq
where $c_1$ and $c_0$ are constants. Re-scalings of the cutoff change $c_1$ but not $c_0$, so only the latter is physical. Indeed $c_0$ is proportional to minus the logarithm of the Euclidean CFT partition function on $S^3$~\cite{Casini:2011kv}. The $AdS_4 \times S^7$ solution of 11d SUGRA gives $c_0 \propto - N^{3/2}$~\cite{Klebanov:1996un,Drukker:2010nc}. In $AdS_7$ eq.~\eqref{eq:rt} produces the form expected for a 6d CFT~\cite{Ryu:2006bv,Ryu:2006ef,Nishioka:2009un}. For example, when the entangling surface is an $S^4$ of radius $\ell$, the minimal surface in $AdS_7$ is a five-dimensional hemisphere at fixed $t$ with radius $\ell$ centered on the $AdS_7$ boundary. In that case eq.~\eqref{eq:rt} gives
\beq
\label{eq:ee6d}
\see^{6d} = c_4 \, \frac{\ell^4}{\varepsilon^4} + c_2 \, \frac{\ell^2}{\varepsilon^2} + c_L \ln \left(\frac{\ell}{\varepsilon}\right) + \mathcal{O}\left(\varepsilon^0\right),
\eeq
where $c_4$, $c_2$ and $c_L$ are constants. Only $c_L$ is invariant under re-scalings of the cutoff, hence only $c_L$ is physical. Indeed, $c_L\propto - a$, where $a$ is a central charge of the 6d CFT~\cite{Casini:2011kv}. The $AdS_7 \times S^4$ solution of 11d SUGRA gives $c_L \propto - a \propto  -M^3$~~\cite{Freed:1998tg,Harvey:1998bx,Henningson:1998gx}.

\begin{figure}[t!]
\begin{center}
\includegraphics[width=.4\textwidth]{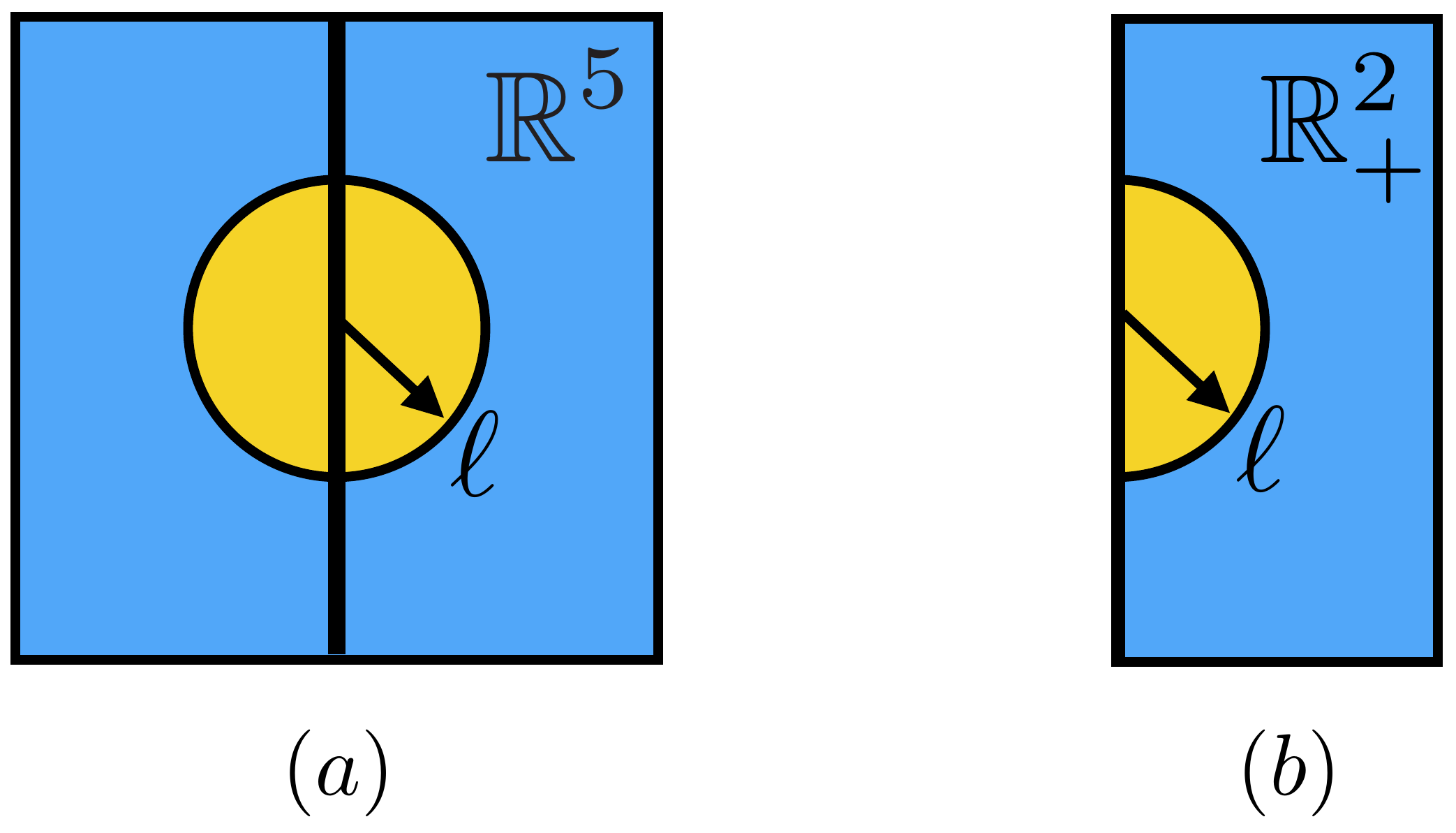}
\caption{\label{fig:eesurf} Schematics of our entangling surfaces. (a) For the M5-brane theory with a Wilson surface, the blue square represents $\mathbb{R}^5$ at fixed $t$, the vertical black line represents the Wilson surface, and the black circle represents our entangling surface, an $S^4$ of radius $\ell$ centered on the Wilson surface. (b) In cousins of the ABJM BCFT, the blue rectangle represents half of $\mathbb{R}^2$ at fixed $t$, denoted $\mathbb{R}^2_+$, with the boundary at left, and our entangling surface is the semi-circle of radius $\ell$ centered on the boundary.}
\end{center}
\end{figure}

Following refs.~\cite{Jensen:2013lxa,Estes:2014hka,Gentle:2015jma}, in our (B)CFTs we choose (hemi-)spherical entangling surfaces centered on the 2d defect or boundary, as follows. For the 11d SUGRA solutions reviewed in sec.~\ref{subsec:ads7metric}, dual to the M5-brane theory with a Wilson surface,  our entangling surface will be an $S^4$ of radius $\ell$ centered on the Wilson surface, as shown in fig.~\ref{fig:eesurf} (a). For the 11d SUGRA solutions reviewed in sec.~\ref{subsec:ads4metric}, dual to cousins of the ABJM BCFT, our entangling surface will be a semi-circle centered on the CFT's boundary, as shown in fig.~\ref{fig:eesurf} (b).

As mentioned above, the first step in the holographic calculation of $\see$ is to find the minimal surface in the holographically dual geometry at fixed $t$ that approaches our entangling surface at the asymptotic boundary. Luckily, this first step has already been done for us in refs.~\cite{Jensen:2013lxa,Estes:2014hka}. Actually refs.~\cite{Jensen:2013lxa,Estes:2014hka}'s results are much more general: for \textit{any} holographic dual of a CFT with conformal defect or boundary, and for a (hemi-)spherical entangling surface centered on the defect or boundary, refs.~\cite{Jensen:2013lxa,Estes:2014hka} found the solution for the \textit{global} minimum of the area functional.

We can immediately adapt the solution of refs.~\cite{Jensen:2013lxa,Estes:2014hka} to our case: the minimal surface at fixed $t$ wraps both $S^3$'s and the upper-half complex $w$ plane, and in the $AdS_3$ subspace is given by $x_{\parallel}^2 + u^2 = \ell^2$. A quick check of this solution is that for $r \to \infty$ and fixed $u$, which takes us to the asymptotic boundary at $x_{\perp}=u$, the surface becomes $x_{\parallel}^2 + x_{\perp}^2 = \ell^2$, which is indeed the equation for an $S^4$ in $\mathbb{R}^5$ or semi-circle in half of $\mathbb{R}^2$. For fixed $r$ and $u \to 0$, which takes us to the $AdS_3$ boundary, the surface reduces to $x_{\parallel} = \pm \ell$, representing the endpoints of an interval of length $\ell$ on the 2d defect or boundary.

As mentioned above, the second step in the holographic calculation of $\see$ is to plug the solution for the minimal surface into the area functional and then integrate to obtain the minimal area and hence $\see$. In our cases, plugging the solution for the minimal surface into the area functional produces
\beq
\label{eq:integral}
\see = \frac{2 \,\textrm{Vol}\left(S^3\right)^2}{4 \gn} \int dw \, d\overline{w} \left( \frac{\Omega^2}{f_1^2} \, f_1^3 f_2^3 f_3^3 \right) \int \frac{du\,\ell}{u \sqrt{\ell^2 - u^2}},
\eeq
where $\textrm{Vol}\left(S^3\right)=2\pi^2$ is the volume of a unit-radius $S^3$, and the integrals are over the upper-half complex $w$ plane and $u \in [0,\ell]$, resdpectively. The latter integration only covers one branch of the minimal surface $x_{\parallel}=\pm \sqrt{\ell^2 - u^2}$, hence the overall factor of $2$.

If we switch to the FG coordinates of eq.~\eqref{eq:ads7asympmet} or~\eqref{eq:ads4asympmet} and introduce an FG cutoff $z=\varepsilon$, then the $\see$ in eq.~\eqref{eq:integral} exhibits $\varepsilon\to 0$ divergences, as expected. In particular, $\see$ for the asymptotically locally $AdS_7 \times S^4$ solutions looks like the EE of the 6d CFT, of the form of $\see^{6d}$ in eq.~\eqref{eq:ee6d}, plus a contribution from the 2d defect that has the form of $\see^{2d}$ in eq.~\eqref{eq:ee2d}. This structure is common in holographic calculations of EE for CFTs with defects or boundaries~\cite{Jensen:2013lxa,Estes:2014hka,Gentle:2015jma}. As mentioned in sec.~\ref{sec:intro}, following refs.~\cite{Estes:2014hka,Gentle:2015jma} we will subtract the 6d CFT's contribution (using the same cutoff) and then extract the coefficient of any remaining logarithmic term. In other words, we will extract the \textit{change} in the coefficient of the logarithmic term due to the Wilson surface. In SUGRA terms, we will subtract the area of the minimal surface in $AdS_7 \times S^4$ bounded by a sphere of radius $\ell$ on the boundary of $AdS_7$, described above. Stated precisely, we will compute
\beq
\label{eq:b6ddef}
b_{6d}= 3 \,\ell \, \frac{d}{d\ell}\left(\see-\see^{6d}\right),
\eeq
where because $\see-\see^{6d}$ is of the form of $\see^{2d}$ in eq.~\eqref{eq:ee2d}, $\ell \, \frac{d}{d\ell}$ extracts the coefficient of the logarithmic term, and the factor of $3$ simply accounts for the normalization of the central charge in eq.~\eqref{eq:ee2d}.

For the asymptotically locally $AdS_4 \times S^7$ solutions our result for $\see$ looks like \textit{half} the EE of a 3d CFT, meaning $1/2$ times the $\see^{3d}$ in eq.~\eqref{eq:ee3d}, plus a contribution from the CFT's 2d boundary of the form of $\see^{2d}$ in eq.~\eqref{eq:ee2d}. Intuitively, in CFT terms the $1/2$ factor appears because introducing the boundary ``cuts off'' half of the 3d CFT, or in SUGRA terms because these solutions are only asymptotic locally to ``half of'' $AdS_4 \times S^7$, as discussed in sec.~\ref{subsec:ads4metric}. Again following refs.~\cite{Estes:2014hka,Gentle:2015jma}, we will subtract the 3d contribution (using the same cutoff), and then extract the coefficient of the logarithmic term. In other words, we will extract the \textit{change} in the coefficient of the logarithmic term due to the CFT's 2d boundary---which of course comes entirely from the 2d boundary, since $\see^{3d}$ has no logarithmic term. In SUGRA terms, we will subtract $1/2$ the area of the minimal surface in $AdS_4 \times S^7$ that approaches a circle of radius $\ell$ at the boundary, described above. Stated precisely, we will compute
\beq
\label{eq:b3ddef}
b_{3d} = 3 \,\ell \, \frac{d}{d\ell}\left(\see-\frac{1}{2}\see^{3d}\right).
\eeq

The FG cutoff $z = \varepsilon$ preserves the symmetry of a Minkowski slice at fixed $z$, dual to the CFT's Poincar\'e symmetry. However in our CFTs the 2d defect or boundary breaks Poincar\'e symmetry to the subgroup that leaves the defect or boundary invariant. In what follows we will thus not use the FG cutoff $z = \varepsilon$, rather we will use the cutoff prescription of refs.~\cite{Estes:2014hka,Gentle:2015jma} for the $u$ and $w$ coordinates, which preserves the reduced symmetry.

The prescription of refs.~\cite{Estes:2014hka,Gentle:2015jma} actually involves two cutoffs. First is an FG cutoff in $AdS_3$, that is, in eq.~\eqref{eq:integral} we perform the $u$ integration from a cutoff $u=\varepsilon_u>0$ to $u=\ell$,
\beq
\label{eq:uint}
\int_{\varepsilon_u}^{\ell} du \frac{\ell}{u \sqrt{\ell^2 - u^2}} = \ln\left(\frac{2\ell}{\varepsilon_u}\right)+{\cal{O}}\left(\varepsilon_u^2\right),
\eeq
so that the integral for $\see$ in eq.~\eqref{eq:integral} becomes
\beq
\label{eq:integral2}
\see = \frac{2 \, \textrm{Vol}\left(S^3\right)^2}{4 \gn} \int dw \, d\overline{w} \left( \frac{\Omega^2}{f_1^2} \, f_1^3 f_2^3 f_3^3 \right) \left[\ln\left(\frac{2\ell}{\varepsilon_u}\right)+{\cal{O}}\left(\varepsilon_u^2\right)\right].
\eeq
We define ${\cal{I}}$ as the remaining integral,
\beq\label{eq:Iastintegral}
{\cal{I}}\equiv \frac{2 \, \textrm{Vol}\left(S^3\right)^2}{4 \gn} \int dw \, d\overline{w} \left( \frac{\Omega^2}{f_1^2} \,  f_1^3 f_2^3 f_3^3 \right).
\eeq
To write ${\cal I}$ explicitly we plug in $\textrm{Vol}\left(S^3\right)=2\pi^2$,
\begin{align}
\label{eq:metproducts}
&\frac{\Omega^2}{f_1^2} = \frac{1}{2}
\frac{c_1^2}{c_2 c_3} \frac{|\partial_w h|^2}{h^2} (G \overline{G} - 1) \,, &
&f_1 f_2 f_3 = \pm \frac{h}{c_1 c_2 c_3}\,,&
\end{align}
and in the second equation of eq.~\eqref{eq:metproducts} we choose the sign to guarantee a positive integrand, given that $|G|<1$ as mentioned below eq.~\eqref{eq:wpm}. Eq.~\eqref{eq:Iastintegral} then becomes
\beq
\label{}
{\cal I} = \frac{2 (2 \pi^2)^2}{4 \gn} \frac{1}{c_1 c_2^4 c_3^4}\frac{1}{2}\int dw \, d\overline{w} \, |\partial_w h|^2 \, h \left(1-G \overline{G}\right).
\eeq
Using $h = - i \left(w-\overline{w}\right)$ from eqs.~\eqref{eq:ads7metric} and~\eqref{eq:ads4metric}, we have $|\partial_w h|^2 = 1$. Introducing polar coordinates $w \equiv r e^{i \theta}$, so that $dw\,d\overline{w} = 2\,d\theta \,dr \, r$ and $h = -i (w - \overline{w})=2r \sin \theta$, we find
\begin{align}
\label{eq:integral3}
{\cal I} =& \frac{2 (2 \pi^2)^2}{4 \gn} \frac{2}{c_1 c_2^4 c_3^4}  \int_0^{\pi} d \theta \sin\theta
\int_0^{r_c} dr \, r^2 (1 - G \overline{G}),
\end{align}
where we have made the endpoints of integration explicit, including a large-$r$ cutoff, $r_c$. The prescription of refs.~\cite{Estes:2014hka,Gentle:2015jma} is to choose $r_c$ in a way that preserves the subgroup of the Poincar\'e group that leaves the 2d defect or boundary invariant. Crucially, a constant $r_c$ does not preserve those symmetries, rather $r_c$ must be a more complicated function whose form depends on the details of the 11d SUGRA solution. In the next two subsections we will compute $r_c$ and then extract  ${\cal I}$ in eq.~\eqref{eq:integral3}.

As mentioned above, in principle we would like to extract $b_{6d}$ or $b_{3d}$ from a term in $\see$ that is $\propto \ln(\ell/\varepsilon)$, with FG cutoff $\varepsilon$. However, how do we do so using the cutoffs $\varepsilon_u$ and $r_c$? The result for the integral ${\cal I}$ will be a sum of terms, including terms with positive powers of $r_c$, a term independent of $r_c$, and terms with negative powers of $r_c$. In eq.~\eqref{eq:integral2} these all multiply $\ln(\ell/\varepsilon_u)$. The terms with positive powers of $r_c$, which are clearly cutoff-dependent and hence unphysical, will turn out to be identical to those of the undeformed $AdS_7 \times S^4$ or (half of) $AdS_4 \times S^7$ solutions, and so will cancel in the background subtraction $\see - \see^{6d}$ or $\see - \frac{1}{2} \see^{3d}$. The terms with negative powers of $r_c$ clearly vanish as $r_c \to \infty$ and so can be safely ignored. We will thus be left with the term independent of $r_c$, or more precisely what remains of that term after the background subtraction, which still multiplies $\ln(\ell/\varepsilon_u)$. Applying $3 \ell \frac{d}{d\ell}$, as in eqs.~\eqref{eq:b6ddef} and~\eqref{eq:b3ddef}, then extracts this coefficient of $\ln(\ell/\varepsilon_u)$, which is thus our $b_{6d}$ or $b_{3d}$. In short, we will apply eqs.~\eqref{eq:b6ddef} and~\eqref{eq:b3ddef} as advertised, though the form of divergences will look very different in terms of $\varepsilon_u$ and $r_c$ as compared to the usual FG cutoff $\varepsilon$. For a more detailed comparison of these cutoffs, see ref.~\cite{Gentle:2015jma}.

\subsection{Asymptotically Locally $AdS_7 \times S^4$ Solutions}

For the asymptotically locally $AdS_7 \times S^4$ solutions reviewed in sec.~\ref{subsec:ads7metric} we follow ref.~\cite{Gentle:2015jma} very closely, but with three major differences. First, where ref.~\cite{Gentle:2015jma} set $\gamma= -1/2$ we will leave $\gamma$ arbitrary as long as practicable. Second, where ref.~\cite{Gentle:2015jma} set $m_1=2$, we will leave $m_1$ arbitrary. Third and most importantly, we will translate our result into the data $\{N_a\}$ and $\{M_a\}$ of a partition/Young tableau, as described in sec.~\ref{sec:fluxes}.

As discussed above, we implement the double cutoff of refs.~\cite{Estes:2014hka,Gentle:2015jma}. First, in the asymptotic large-$r$ region we put the metric in a form that makes manifest the symmetries of the 2d defect,
\beq
\label{eq:ads7metriccutoff}
ds^2 = 4 L_{S^4}^2 \left( \frac{dv^2}{v^2}+ \frac{ds^2_{\text{AdS}_3} + \frac{(1+\gamma)^2}{\gamma^2} ds^2_{S^3}}{v^2} \right) + L_{S^4}^2 \left(d\tilde \theta^2 + \sin^2(\tilde \theta) ds^2_{S^3}  \right) + \ldots,
\eeq
where asymptotically at large $r$,
\begin{subequations}
\begin{align}
\label{eq:AdS7FGtrans}
v =& \, \frac{1+\gamma}{\sqrt{-\gamma}} \frac{\sqrt{2m_1}}{\sqrt{r}} \left(1 + \frac{(1 + 2 \gamma)(-3 + \cos(2\theta))m_1^2 + 12 \gamma m_2 \cos\theta}{48 \gamma m_1}  \frac{1}{r} + {\cal O}\left(\frac{1}{r^2}\right) \right),\\
\tilde \theta =& \, \theta + \frac{(1 + 2\gamma)m_1^2 \cos\theta + 3 \gamma m_2}{6 \gamma m_1} \frac{\sin\theta}{r}  + {\cal O}\left(\frac{1}{r^2}\right),
\end{align}
\end{subequations}
where $v \in [0,\infty)$, $\tilde{\theta} \in [0,\pi]$, and the ellipsis represent terms orthogonal to $v$ and sub-leading in $1/r$. In these coordinates, we can approach the asymptotic local $AdS_7 \times S^4$ boundary in two ways. First is to fix $u$ and send $v \to 0$, where the latter is equivalent via eq.~\eqref{eq:AdS7FGtrans} to $r \to \infty$, which takes us to the asymptotically local $AdS_7 \times S^4$ boundary at a point away from the 2d defect. For $\gamma = -1/2$ the dual field theory's metric is that of $AdS_3 \times S^3$, which is conformal to 6d Minkowski space, while for $\gamma\neq -1/2$ a conical singularity appears, as in eq.~\eqref{eq:ads7asympmet}. Second is to fix $v$ and send $u \to 0$, which takes us to the $AdS_3$ boundary, i.e. to a point on the 2d defect.

The prescription of refs.~\cite{Estes:2014hka,Gentle:2015jma} is to introduce an FG cutoff for $v$, that is a cutoff $v = \varepsilon_v>0$, which translates to a cutoff $r_c$ that depends on $\varepsilon_v$ and $\theta$. Explicitly, in eq.~\eqref{eq:AdS7FGtrans} we set $v = \varepsilon_v$ and then invert to find $r_c(\varepsilon_v,\theta)$ in a small $\varepsilon_v$ expansion,
\begin{align}
\label{eq:AdS7cutoff}
&r_c(\varepsilon_v,\theta) = \frac{2 (1+ \gamma)^2 m_1}{-\gamma} \frac{1}{\varepsilon_v^2} + \frac{(1+ 2 \gamma)m_1}{-8  \gamma} + \frac{m_2}{2m_1} \cos\theta - \frac{(1+2 \gamma) m_1}{-24 \gamma} \cos(2\theta) \nonumber\\
& + \left\{ \frac{89 m_1^4+212 \gamma(1 + \gamma)m_1^4 + 729 \gamma^2 m_2^2 - 576 \gamma^2 m_1 m_3}{9216 \gamma(1+\gamma)^2 m_1^3}
+ \frac{(1+2\gamma)m_2}{24 (1 + \gamma^2)m_1}\cos\theta \sin^2\theta \right .  \nonumber\\&
+ \left . \frac{3m_1^4 + 8 \gamma m_1^4 + 8 \gamma^2 m_1^4 - 36 \gamma^2 m_2^2 + 80 \gamma^2 m_1 m_3}{768 \gamma(1+\gamma)^2 m_1^3} \cos(2\theta) + \frac{(1+2\gamma)^2 m_1}{3072 \gamma(1+\gamma)^2} \cos(4\theta) \right\} \varepsilon_v^2 \nonumber \\ & + {\cal O}(\varepsilon_v^4),
\end{align}
where we have kept an additional order as compared to eq.~\eqref{eq:AdS7FGtrans}, which will be necessary to extract $b_{6d}$ from $\see$. When $\gamma = - 1/2$, eq.~\eqref{eq:AdS7cutoff} simplifies considerably,
\begin{align}
\label{eq:rcads7m12}
r_c(\varepsilon_v,\theta) =& \frac{m_1}{\varepsilon_v^2} + \frac{m_2}{2m_1} \cos(2\theta) +\left( \frac{4 m_1 m_3 - m_1^4 - 5 m_2^2}{32 m_1^3} + \frac{20 m_1 m_3 - 9 m_2^2 + m_1^4}{96 m_1^3} \cos(2\theta) \right) \varepsilon_v^2 \nonumber \\ & + {\cal O}(\varepsilon_v^4).
\end{align}

Plugging the expression for $G$ in eq.~\eqref{eq:ads7metric} into eq.~\eqref{eq:integral3} gives
\begin{align}
\label{eq:ads7int}
{\cal I} =& \frac{2 (2 \pi^2)^2}{4 \gn} \frac{2}{c_1 c_2^4 c_3^4}  \sum_{j=1}^{2n+2} (-1)^j  \int_0^{\pi} d \theta \sin\theta
\int_0^{r_c(\varepsilon_v,\theta)} dr \, r^2 \frac{2(r \cos(\theta) - \xi_j)}{\sqrt{r^2 + \xi_j^2 - 2 r \xi_j \cos(\theta)}} \nonumber\\
&+ \frac{2 (2 \pi^2)^2}{4 \gn} \frac{2}{c_1 c_2^4 c_3^4} \sum_{j,k=1}^{2n+2} (-1)^{j+k}   \int_0^{\pi} d \theta \sin\theta
\int_0^{r_c(\varepsilon_v,\theta)} dr \, r^2 \frac{r e^{i \theta} - \xi_j}{|r e^{i \theta} - \xi_j|}\frac{r e^{-i \theta} - \xi_k}{|r e^{i \theta} - \xi_k|},
\end{align}
where $r_c(\varepsilon_v,\theta)$ is the cutoff in eq.~\eqref{eq:AdS7cutoff}. The integrals in eq.~\eqref{eq:ads7int} are performed in ref.~\cite{Gentle:2015jma},\footnote{In ref.~\cite{Gentle:2015jma} the first and second lines of eq.~\eqref{eq:ads7int} are denoted $J_1$ and $J_2$, respectively.} and are very similar to those in the asymptotically locally $AdS_4 \times S^7$ case performed in the appendix, so here we only quote the result:
\begin{align}
\label{eq:ads7inttemp}
{\cal I} =& -\frac{8(2 \pi^2)^2 L_{S^4}^9}{4 \gn \gamma (1 + \gamma)}
\Bigg(\frac{8 (1+\gamma)^4}{3\gamma^2}\frac{1}{\varepsilon_v^4}+\frac{2(1+\gamma)^2(3+16\gamma)}{15\gamma^2}\frac{1}{\varepsilon_v^2} -\frac{85-8\gamma(52+115\gamma)}{5040\gamma^2}\nonumber\\
&+\frac{m_2^2}{10m_1^4}-\frac{2m_3}{15m_1^3} -\frac{1}{3m_1^3}\sum_{\underset{j<k}{j,k=1}}(-1)^{j+k}|\xi_j-\xi_k|^3\Bigg),
\end{align}
where we dropped terms that vanish as $\varepsilon_v \to 0$. We will continue to do so in what follows. If we set $\gamma=-1/2$ then eq.~\eqref{eq:ads7inttemp} becomes
\begin{align}
\label{eq:ads7int2}
{\cal I} = \frac{2^6\pi^4 L_{S^4}^9}{3 \,\gn}
\Bigg(\frac{1}{\varepsilon_v^4}-\frac{1}{\varepsilon_v^2} - \frac{3}{40}
+\frac{3 m_2^2 - 4 m_1 m_3}{20 m_1^4}-\frac{1}{2m_1^3}\sum_{\underset{j<k}{j,k=1}}^{2n+2} (-1)^{j+k}|\xi_j-\xi_k|^3\Bigg).
\end{align}

 As expected, eq.~\eqref{eq:ads7int2} contains terms that diverge as $\varepsilon_v \to 0$. To extract $b_{6d}$ using eq.~\ref{eq:b6ddef}, we will need the result for ${\cal I}$ for the exact $AdS_7 \times S^4$ solution, which we denote ${\cal I}_{6d}$. As mentioned at the end of sec.~\ref{subsec:ads7metric}, the $AdS_7 \times S^4$ solution has two branch points at $\textrm{Re}\left(w\right)=\pm\xi$. These can be eliminated by conformally mapping the upper half plane to a semi-infinite strip via $w = \xi \cosh Z$, where $Z=X+iY$ with $X\in [0,\infty]$ and $Y\in[0,\pi]$. In the semi-infinite strip coordinates,
 \beq
 h=-i\xi\left(\cosh Z-\cosh\overline{Z}\right), \qquad G=-i \left( 1+2\,\frac{\sinh(\frac{Z-\overline{Z}}{2})}{\sinh \overline{Z}} \right).
 \eeq
 In semi-infinite strip coordinates the metric takes the form
\begin{align}
\label{AdS7metric}
ds^2=4L_{S^4}^2\left( \cosh^2 \frac{X}{2} \, ds^2_{AdS_3}+\sinh^2 \frac{X}{2} \, ds^2_{S^3} + \frac{dX^2}{4} \right) + L_{S^4}^2\, ds^2_{S^4}.
\end{align}
Mapping eq.~\eqref{AdS7metric} to the asymptotic form in eq.~\eqref{eq:ads7metriccutoff} gives at leading order $v=2 e^{-X/2}$. As a result, the cutoff $\varepsilon_v$ maps to a cutoff $X_c = -2\log(\varepsilon_v/2)$. Plugging eq.~\eqref{AdS7metric} into eq.~\eqref{eq:Iastintegral} and performing the integration with the cutoff $X_c$, we find
\begin{align}
\label{eq:ads7I}
&{\cal I}_{6d} =\frac{2^6\pi^4 L_{S^4}^9} {3 \,\gn} \left(\frac{1}{\varepsilon_v^4}-\frac{1}{\varepsilon_v^2}+\frac{3}{8}\right).
\end{align}

Comparing eqs.~\eqref{eq:ads7int2} and~\eqref{eq:ads7I}, we see that all divergent terms cancel in ${\cal I} - {\cal I}_{6d}$, as advertised. Extracting $b_{6d}$ via eq.~\eqref{eq:b6ddef} then gives
\begin{align}
\label{eq:b6draw}
b_{6d}=&
\frac{2^6\pi^4 L_{S^4}^9}{\gn}
\Bigg\{- \frac{9}{20}
+\frac{3\, m_2^2 - 4 \, m_1 m_3}{20 m_1^4}+\frac{1}{2m_1^3}\sum_{\underset{j<k}{j,k=1}}^{2n+2} (-1)^{j+k}(\xi_j-\xi_k)^3\Bigg\}.
\end{align}

\vspace{-0.25cm}If we use the scaling symmetry to set $m_1=2$, then eq.~\eqref{eq:b6draw} reproduces the result of ref.~\cite{Gentle:2015jma}. However, we can go farther, and write $b_{6d}$ of eq.~\eqref{eq:b6draw} in terms of the partition data $\{N_a\}$ and $\{M_a\}$, as follows. On the right-hand-side of eq.~\eqref{eq:b6draw}, we use eq.~\eqref{eq:ads7m1exp} to replace the factors of $m_1$ in the denominators with factors of $M$. Next we consider the combination $3 m_2^2 - 4 m_1 m_3$ in the second term on the right-hand side of eq.~\eqref{eq:b6draw}. While $m_2$ and $m_3$ are not individually invariant under translations of the $\xi_j$, the combination $3 m_2^2 - 4 m_1 m_3$ is invariant.  That is important, since such shifts are equivalent to a coordinate transformation, under which our final expression must be invariant. Using the definition of the $m_k$ in terms of the $\xi_j$ in eq.~\eqref{eq:ads7legendre}, we can write
\begin{align}
&3 m_2^2 - 4 m_1 m_3 = - \left(\sum_{a=1}^{n+1} \xi_{2a} - \xi_{2a-1} \right)^4
\nonumber\\
&-12 \left(\sum_{a=1}^{n+1} \xi_{2a} - \xi_{2a-1} \right) \left( \sum_{a=1}^n \sum_{b=a}^{n+1} (\xi_{2b} - \xi_{2b-1}) (\xi_{2a} - \xi_{2a-1}) \sum_{c=a}^n (\xi_{2c+1} - \xi_{2c}) \right)
\nonumber\\
&+12 \left(\sum_{a=1}^{n+1} \xi_{2a} - \xi_{2a-1} \right) \left( \sum_{a=1}^n \sum_{b=1}^{a} (\xi_{2b} - \xi_{2b-1}) (\xi_{2a} - \xi_{2a-1}) \sum_{c=a}^n (\xi_{2c+1} - \xi_{2c}) \right)
\nonumber\\
&+ 12  \left( \sum_{a=1}^n (\xi_{2a} - \xi_{2a-1}) \sum_{b=a}^n (\xi_{2b+1} - \xi_{2b}) \right)^2
\nonumber\\
&-12 \left(\sum_{a=1}^{n+1} \xi_{2a} - \xi_{2a-1} \right) \sum_{a=1}^n
(\xi_{2a} - \xi_{2a-1}) \left( \sum_{b=a}^n (\xi_{2b+1} - \xi_{2b}) \right)^2,
\end{align}
which can be proven using recursion. Using eqs.~\eqref{eq:m5flux} and~\eqref{eq:m5primeflux} to replace the $\xi_j$ with $M_a$ and $M_b'$, and then using eq.~\eqref{eq:m2flux} to replace the $M_b'$ with the $N_a$, we find
\begin{align}
\label{eq:ads7msconv1}
3 m_2^2 - 4 m_1 m_3 =&  \, c_1^{12} \frac{\gn^{4/3}}{\left(2\pi\right)^{16/3}} \frac{\gamma^4}{\left(1+\gamma\right)^{12}}\bigg[ - M^4 + 12 \gamma M \left( \sum_{a=1}^n \sum_{b=a}^{n+1} M_b M_a N_a \right)\\&
-12 \gamma M \left( \sum_{a=1}^n \sum_{b=1}^{a} M_b M_a N_a \right)+ 12 \gamma^2  \left( \sum_{a=1}^n M_a N_a \right)^2
-12 \gamma^2  M \sum_{a=1}^n M_a N_a^2  \bigg].\nn
\end{align}
All that remains in eq.~\eqref{eq:b6draw} is the sum over $(\xi_j- \xi_k)^3$. We decompose this sum as
\begin{align}
\sum_{\underset{j<k}{j,k=1}}^{2n+2} (-1)^{j+k} (\xi_j - \xi_k)^3 =& \bigg[
\left( \sum_{a=1}^{n+1} (\xi_{2a}-\xi_{2a-1}) \right)^3  \\
&+ 6 \left( \sum_{b=1}^{n} (\xi_{2b}-\xi_{2b-1}) \sum_{a=b}^{n} (\xi_{2a+1} - \xi_{2a}) \sum_{c=b}^{n+1} (\xi_{2c}-\xi_{2c-1}) \right)  \nonumber\\
&- 6 \left( \sum_{b=1}^{n} (\xi_{2b}-\xi_{2b-1}) \sum_{a=b}^{n} (\xi_{2a+1} - \xi_{2a}) \sum_{c=1}^b (\xi_{2c}-\xi_{2c-1}) \right)
\bigg],\nonumber
\end{align}
which can be proven using recursion. Again using eqs.~\eqref{eq:m5flux} and~\eqref{eq:m5primeflux} to replace the $\xi_j$ with $M_a$ and $M_b'$, and then using eq.~\eqref{eq:m2flux} to replace the $M_b'$ with the $N_a$, we find
\begin{align}
\label{eq:ads7msconv2}
\sum_{\underset{j<k}{j,k=1}}^{2n+2} (-1)^{j+k} (\xi_j - \xi_k)^3  = -c_1^9 \, \frac{\gn}{\left(2\pi\right)^4} \frac{\gamma^3}{\left(1+\gamma\right)^9} &\bigg[ M^3 - 6 \gamma \left( \sum_{a=1}^{n} \sum_{b=a}^{n+1} N_a M_a M_b \right)
\nn\\ &+ 6 \gamma \left( \sum_{a=1}^{n} \sum_{b=1}^a N_a M_a M_b \right) \bigg].
\end{align}
Plugging eqs.~\eqref{eq:ads7msconv1} and~\eqref{eq:ads7msconv2} into eq.~\eqref{eq:b6draw}, using eq.~\eqref{eq:totnm}, and setting $N_{n+1} = 0$ and $\gamma=-1/2$ then gives
\beq
b_{6d} = \frac{3}{5}\left[\frac{N^2}{M} -\sum_{a=1}^{n+1} \left(M_a N_a^2 - 8 \sum_{b=a}^{n+1} M_a N_a M_b + 8 \sum_{b=1}^{a} M_a N_a M_b \right)\right],
\eeq
which can be simplified by rearranging the summations to give our main result,
\beq
\label{eq:b6d-sugra-1}
b_{6d} = \frac{3}{5}\left[8MN + \frac{N^2}{M} +\sum_{a=1}^{n} \left(8 M_a^2 N_a -M_a N_a^2 -16\sum_{b=1}^{a}M_b M_a N_a\right)\right].
\eeq

\subsection{Asymptotically Locally $AdS_4 \times S^7$ Solutions}

As in the previous case, in the asymptotic large-$r$ region we put the metric in a form that makes manifest the symmetries of the 2d boundary,
\begin{align}\label{AdS4:altFG}
ds^2 = \frac{L_{S^7}^2}{4} \left( \frac{dv^2}{v^2} + \frac{ds^2_{\text{AdS}_3}}{v^2} \right) + L_{S^7}^2 \left(d\tilde{\theta}^2 + \sin^2\tilde{\theta} \, ds^2_{S^3} + \cos^2\tilde{\theta} \, ds^2_{S^3}  \right) + \ldots,
\end{align}
where asymptotically at large $r$
\begin{subequations}
\begin{align}
\label{eq:AdS4FGtrans}
v =& \frac{(1+\gamma) \sqrt{-m_1^2-m_2}}{2 \sqrt{-\gamma}} \frac{1}{r} \left(1 + \frac{-m_1^3+m_3}{3(m_1^2 + m_2)} \cos\theta \, \frac{1}{r} + {\cal O}\left(\frac{1}{r^2}\right) \right)\\
\tilde{\theta} =& \frac{\theta}{2} + \frac{-m_1^3+m_3}{6(m_1^2+m_2)} \sin\theta \, \frac{1}{r}  + {\cal O}\left(\frac{1}{r^2}\right),
\end{align}
\end{subequations}
where $v \in [0,\infty)$, $\tilde{\theta} \in [0,\pi/2]$, and the ellipsis represent terms orthogonal to $v$ and sub-leading in $1/r$. In these coordinates, we can approach the asymptotic local (half) $AdS_4 \times S^7$ boundary in two ways. First is to fix $u$ and send $v \to 0$, where the latter is equivalent via eq.~\eqref{eq:AdS4FGtrans} to $r \to \infty$, which takes us to the asymptotically local $AdS_4 \times S^7$ boundary at a point away from the 2d boundary. Second is to fix $v$ and send $u \to 0$, which takes us to the $AdS_3$ boundary, i.e. to a point on the 2d boundary.

We introduce an FG cutoff $\varepsilon_v$, which we plug into eq.~\eqref{eq:AdS4FGtrans} and invert to find
\begin{align}
\label{eq:AdS4cutoff}
r_c(\varepsilon_v,\theta) = \frac{(1+\gamma) \sqrt{-m_1^2-m_2}}{2 \sqrt{-\gamma}} \frac{1}{\varepsilon_v} -  \frac{-m_1^3+m_3}{3(m_1^2 + m_2)} \cos\theta + {\cal O}\left(\varepsilon_v\right).
\end{align}
Plugging the expression for $G$ given in eq.~\eqref{eq:ads4metric} into eq.~\eqref{eq:integral3} we obtain
\begin{align}
\label{integralcomplex}
{\cal I}  =& \frac{ (2 \pi^2)^2}{4 \gn} \frac{4}{c_1 c_2^4 c_3^4}  \int_0^{\pi} d\theta \sin\theta \int_{0}^{r_c(\varepsilon_v,\theta)} dr \, r^2 \left(1 - \sum_{j,k=1}^{2n+1} (-1)^{j+k} \frac{r e^{i\theta} - \xi_j}{|r e^{i\theta} - \xi_j|} \frac{r e^{-i\theta} - \xi_k}{|r e^{i\theta} - \xi_k|} \right).
\end{align}
These integrals are very similar to those in the asymptotically local $AdS_7 \times S^4$ case. We perform the integrals in the appendix, with the result
\beq\label{eq:ads4int2}
{\cal I}  =\frac{(2 \pi^2)^2}{4 \gn} \frac{L_{S^7}^9}{24} \frac{1}{\varepsilon_v} - \frac{(2 \pi^2)^2}{4 \gn} \frac{8}{3}\frac{1}{ c_1 c_2^4 c_3^4}\sum_{\underset{j<k}{j,k=1}}^{2n+1} (-1)^{j+k}(\xi_j-\xi_k)^3.
\eeq

As expected, eq.~\eqref{eq:ads4int2} diverges as $\varepsilon_v \to 0$. To extract $b_{3d}$ using eq.~\ref{eq:b3ddef}, we will need the result for ${\cal I}$ for the exact $AdS_4 \times S^7$ solution, which we denote ${\cal I}_{3d}$. In $AdS_3$ slicing, the $AdS_4 \times S^7$ metric takes the form
\beq
\label{AdS4metric}
ds^2=L_{S^7}^2 \left(dx^2 + \cosh^2 (2x) \, ds^2_{AdS_3}+dy^2+\sin^2 y\, ds^2_{S^3} + \cos^2 y\, ds^2_{S^3}\right),
\eeq
where $x \in (-\infty,\infty)$ and $y \in [0,\pi/2]$. By matching the large-$|x|$ asymptotics of this $AdS_4 \times S^7$ metric to that of eq.~\eqref{AdS4:altFG}, we find $x \in (-x_c,x_c)$ with cutoff $x_ c = -\frac{1}{2}\ln\varepsilon_v$. Plugging eq.~\eqref{AdS4metric} into eq.~\eqref{eq:Iastintegral} and performing the integration with that cutoff, we find
\begin{align}\label{vacuum}
{\cal I}_{3d} = \frac{2 (2 \pi^2)^2}{4 \gn} \frac{L_{S^7}^9}{24} \frac{1}{\varepsilon_v}.
\end{align}

Comparing eqs.~\eqref{eq:ads4int2} and~\eqref{vacuum}, we see that the divergence cancels in ${\cal I} - {\cal I}_{3d}$, as advertised. Extracting $b_{3d}$ via eq.~\eqref{eq:b3ddef} then gives
\begin{align}\label{cads4}
b_{3d} = - \frac{ (2 \pi^2)^2}{\gn} \frac{2}{c_1 c_2^4 c_3^4}\sum_{\underset{j<k}{j,k=1}}^{2n+1} (-1)^{j+k}(\xi_j-\xi_k)^3.
\end{align}
To rewrite this expression in terms of the partition data $\{N_a\}$ and $\{M_a\}$, we decompose the sum in eq.~\eqref{cads4} as
\begin{align}
\sum_{\underset{j<k}{j,k=1}}^{2n+1} (-1)^{j+k}(\xi_j-\xi_k)^3 = 3 \bigg[
- 2 \left( \sum_{b=1}^n \sum_{c=1}^b (\xi_{2c}-\xi_{2c-1}) (\xi_{2b}-\xi_{2b-1}) \sum_{a=b}^n (\xi_{2a+1} - \xi_{2a}) \right)  \nonumber\\
+ \left( \sum_{b=1}^n (\xi_{2b}-\xi_{2b-1})^2 \sum_{a=b}^n (\xi_{2a+1} - \xi_{2a}) \right)
- \sum_{b=1}^n (\xi_{2b}-\xi_{2b-1}) \left(  \sum_{a=b}^n (\xi_{2a+1} - \xi_{2a}) \right)^2
\bigg],
\end{align}
which can be proven using recursion. Using eqs.~\eqref{eq:m5flux} and~\eqref{eq:m5primeflux} to replace the $\xi_j$ with $M_a$ and $M_b'$, and then using eq.~\eqref{eq:m2flux} to replace the $M_b'$ with the $N_a$, we find
\begin{align}
\label{eq:ads4cubicconv}
\sum_{\underset{j<k}{j,k=1}}^{2n+1} (-1)^{j+k}(\xi_j-\xi_k)^3= 3 c_1^9 \frac{\gn}{\left(2\pi\right)^4} \frac{\gamma^4}{\left(1+\gamma\right)^9}\bigg[
-& 2 \left( \sum_{a=1}^n \sum_{b=1}^a M_b  M_a N_a \right) \\
+& \left( \sum_{a=1}^n M_a^2 N_a \right)
+ \gamma \sum_{a=1}^n M_a N_a^2\nn
\bigg].
\end{align}
Plugging eq.~\eqref{eq:ads4cubicconv} into eq.~\eqref{cads4} then gives
\beq
\label{eq:b3dtemp}
b_{3d} = \frac{3}{2} \frac{c^8_1}{c_2^4c_3^4} \frac{\gamma^4}{\left(1+\gamma\right)^9} \left[ -  \sum_{a=1}^n M_a^2 N_a- \gamma \sum_{a=1}^n M_a N_a^2+2 \sum_{a=1}^n \sum_{b=1}^a M_b  M_a N_a\right].
\eeq
Using $c_1+c_2+c_3=0$ and $c_2 = \gamma c_3$, we find
\beq
\label{eq:celim}
\frac{c^8_1}{c_2^4c_3^4} \frac{\gamma^4}{\left(1+\gamma\right)^9} = \frac{\left(c_2+c_3\right)^8 }{\gamma^4c_3^8}\frac{\gamma^4}{\left(1+\gamma\right)^9} = \frac{1}{1+\gamma},
\eeq
and hence we obtain our main result for $b_{3d}$,
\begin{align}
\label{b3dads4}
b_{3d} = \frac{3}{2}\frac{1}{1+\gamma} \left[-\sum_{a=1}^n M_a^2 N_a - \gamma \sum_{a=1}^n M_a N_a^2 +2\sum_{a=1}^n \sum_{b=1}^a M_b  M_a N_a \right].
\end{align}

\section{The Central Charge}
\label{sec:cc}

In this section, we explore our results for $b_{6d}$ and $b_{3d}$ in several ways. In sec.~\ref{sec:wilson-surface} we show that our result for $b_{6d}$ in eq.~\eqref{eq:b6d-sugra-1} can be written in the compact form of eq.~\eqref{eq:b6dcompact}, that is, in terms of $\lambda$, the highest weight vector of the representation ${\cal R}$, and $\varrho$, the Weyl vector of $\mathfrak{su}(M)$. We also determine how $b_{6d}$ scales with $M$ and $N$ for some specific choices of ${\cal R}$. Similarly, in sec.~\ref{sec:abjm-bcft-cc} we determine how $b_{3d}$ scales with $N$ for some specific choices of partition $\rho$. In sec.~\ref{sec:other-comparisons} we briefly survey some previous calculations of self-dual string central charges, and discuss how and why these results differ from ours.

\subsection{Wilson Surface}
\label{sec:wilson-surface}

Our first goal is to show that our result for $b_{6d}$ in eq.~\eqref{eq:b6d-sugra-1},
\beq
\label{eq:b6dsugra}
b_{6d} = \frac{3}{5}\left[8MN + \frac{N^2}{M} +\sum_{a=1}^{n+1} \left(8M_a^2N_a -M_aN_a^2 -16\sum_{b=1}^{a}M_bM_a N_a\right)\right],
\eeq
can we re-written in the form of eq.~\eqref{eq:b6dcompact},
\beq
\label{eq:b6dcompact2}
b_{6d} = \frac{3}{5}\left[16(\la,\,\vr)-(\la,\,\la)\right],
\eeq
where $\la$ is the highest weight vector of the representation $\mathcal{R}$, $\vr$ is the Weyl vector of $\mathfrak{su}(M)$, and $(\cdot ,\,\cdot)$ is the inner product on the weight space. To show the equivalence between the two expressions for $b_{6d}$ we will actually work backwards: starting from eq.~\eqref{eq:b6dcompact2} we will re-write various sums until we reach eq.~\eqref{eq:b6dsugra}.

We start with the parametrization of the partition $\rho=\{\ell_1,\ell_2,\ldots\}$, with integers $\ell_1 \geq \ell_2 \geq \ell_3 \ldots$, where $\ell_q$ with $1 \leq q \leq M$ is the number of M2-branes ending on the $q^{\textrm{th}}$ M5-brane, as illustrated on the left in fig.~\ref{fig:young}. In this parametrization the inner products in eq.~\eqref{eq:b6dcompact2} are simple to write in terms of the Dynkin indices of the representation $\cal{R}$, $\la_q = \ell_q - \ell_{q+1}$,
\beq
\label{eq:b6dsums1}
(\la,\,\vr) = \frac{1}{2}\sum_{q=1}^{M}(M-q)q\la_q,\qquad (\la,\,\la) = \frac{1}{M}\sum_{q=1}^{M}(M-q)\la_q\left[- q\la_q+2\sum_{p=1}^{q}p\la_p \right].
\eeq
As is clear from the definition of $\la_q$, non-zero contributions to the sums in eq.~\eqref{eq:b6dsums1} only come from cases where the number of boxes in the Young tableau changes from one row to the next. The non-zero contributions are thus more conveniently described using the parametrization of the partition $\rho$ in eq.~\eqref{eq:part}, in terms of the set of distinct integers $\{N_a\}$ with degeneracies $\{M_a\}$ for $a = 1, 2,\ldots,n+1$ and $N_{n+1}=0$. In this parametrization, the non-zero contributions to the sums in eq.~\eqref{eq:b6dsums1} come from $\lambda_a = N_a - N_{a+1}$, and the row number can be written as $q = \sum_{b=1}^a M_b$. Plugging these expressions into eq.~\eqref{eq:b6dsums1} gives
\begin{subequations}
\label{eq:rep-data}
\begin{align}
(\la,\,\vr) &= \frac{1}{2}\sum_{a=1}^{n+1}\sum_{b=1}^a\left(M - \sum_{c=1}^aM_c\right) M_b \left(N_a - N_{a+1}\right),\\
\begin{split}
(\la,\,\la) &= \frac{1}{M}\sum_{a=1}^{n+1}\sum_{b=1}^a \left(M - \sum_{c=1}^aM_c\right)\left(N_a- N_{a+1}\right)\left[2\sum_{d=1}^b M_d(N_b-N_{b+1}) -M_b(N_a-N_{a+1})\right]\!\!.
\end{split}
\end{align}
\end{subequations}
Expanding the sums in eq.~\eqref{eq:rep-data} then leads to
\begin{subequations}
\label{eq:rep-data2}
\begin{align}
(\la,\,\vr) &= \frac{1}{2}\sum_{a=1}^{n+1} \left(M M_a N_a + M_a^2 N_a - 2 \sum_{b=1}^aM_b M_a N_a\right),\\
(\la,\,\la) &= \sum_{a=1}^{n+1} M_a N_a^2 - \frac{1}{M}\left(\sum_{a=1}^{n+1} M_a N_a\right)^2,
\end{align}
\end{subequations}
which we simplify using the fact that the total number of M2-branes is $N= \sum_{a=1}^{n+1} M_a N_a$:
\begin{subequations}
\label{eq:rep-data3}
\begin{align}
(\la,\,\vr) &= \frac{1}{2}MN+\frac{1}{2}\sum_{a=1}^{n+1} M_a^2 N_a - \sum_{a=1}^{n+1} \sum_{b=1}^aM_b M_a N_a,\\
(\la,\,\la) &= \sum_{a=1}^{n+1} M_a N_a^2 - \frac{N^2}{M}.
\end{align}
\end{subequations}
Plugging eq.~\eqref{eq:rep-data3} into eq.~\eqref{eq:b6dcompact2} we find
\bea
\label{eq:b6dfinal}
\frac{3}{5}\left[16(\la,\,\vr)-(\la,\,\la)\right] & = & \frac{3}{5}\left[8 MN+8\sum_{a=1}^{n+1} M_a^2 N_a -16\sum_{a=1}^{n+1} \sum_{b=1}^a M_b M_a N_a - \sum_{a=1}^{n+1} M_a N_a^2+\frac{ N^2}{M}\right]\nn \\ & = & \frac{3}{5}\left[8MN + \frac{N^2}{M} +\sum_{a=1}^{n+1} \left(8M_a^2N_a -M_aN_a^2 -16\sum_{b=1}^{a}M_bM_a N_a\right)\right]\nn \\ & = & b_{6d},
\eea
as advertised. We can alternatively express $b_{6d}$ in terms of the quadratic Casimir of the representation, $\mathcal{Q}_\la \equiv 2(\la,\,\vr)+(\la,\,\la)$,
\beq
b_{6d} = \frac{24}{5}\left(\mathcal{Q}_\la - \frac{9}{8}(\la,\,\la)\right).
\eeq

The inner product $(\cdot,\,\cdot)$ is invariant under the action of the Weyl group. These compact forms of $b_{6d}$ thus make manifest that $b_{6d}$ is invariant under the action of the Weyl group on $\lambda$ and $\varrho$, including in particular the Weyl reflection affected by the complex conjugation of the representation, ${\cal R} \to \overline{{\cal R}}$. Such invariance is expected, given that the 11d SUGRA solutions are invariant under ${\cal R} \to \overline{{\cal R}}$, as explained at the end of sec.~\ref{sec:ads7part}. Such invariance is also expected in the field theory: ${\cal R} \to \overline{{\cal R}}$ combined with orientation reversal of the Wilson surface must leave all observables invariant. The EE of our spherical region is invariant under the orientation reversal alone, and hence must also be invariant under ${\cal R} \to \overline{{\cal R}}$ alone, not just under the combined operation. As a result, $b_{6d}$ is invariant under ${\cal R} \to \overline{{\cal R}}$.

These compact forms of $b_{6d}$ do not make immediately obvious how $b_{6d}$ scales with the total numbers $M$ of M5-branes or $N$ of M2-branes. To see such scaling explicitly, we consider two relatively simple examples of ${\cal R}$. First, consider a totally symmetric representation of rank $N$, which corresponds to a Young tableau with a single row of $N$ boxes. In terms of the partition data,  $M_1=1$, $N_1 = N$, and $N_a=0$ for $2\leq a\leq n+1$, which describes all $N$ M2-branes ending on a single M5-brane. In this case we find\footnote{
  \textbf{Note added:} A natural question is whether the value $N = 8 M$ at which $b_{6d}$ in eq.~\eqref{eq:b6dsymm} vanishes has any physical significance. In particular, does $b_{6d}=0$ imply that the symmetric-representation Wilson surface with $N=8M$ is somehow equivalent to a trivial-representation object, like a ``baryon vertex''? We suspect not, because as shown in eq.~\eqref{eq:ee_trace_anomaly}, $b_{6d} = c - \frac{3}{5} d_2$ with $c$ and $d_2$ defined as coefficients in the defect trace anomaly eq.~\eqref{eq:defecttrace}, and taking the values in eq.~\eqref{eq:cd2values}. In particular, both $c$ and $d_2$ are positive for the symmetric representation with $N=8M$, and $b_{6d}$ vanishes because of a cancellation between them, whereas we would expect both of them to vanish independently for a trivial-representation object.}
\bea
\label{eq:b6dsymm}
b_{6d} & = & \frac{3}{5}\left[8 MN + \frac{N^2}{M} - N^2 - 8 N\right] \qquad \textrm{(symmetric)} \\ & = & \frac{3}{5} \frac{N}{M} \left(8M-N\right)\left(M-1\right),\nn
\eea
where the second line helps facilitate probe limits. For example, the limit $N \ll M^2$ gives $b_{6d} \approx \frac{24}{5} N \left(M-N/8\right)$. As shown in ref.~\cite{Rodgers:2018mvq}, this probe limit result agrees with a calculation of EE using probe M5-branes in $AdS_7 \times S^4$ wrapping an $AdS_3 \times S^3$ submanifold inside $AdS_7$ and sitting at a point on the $S^4$. This is true even when $N \sim \mathcal{O}(1)$, in which case we would expect the SUGRA approximation to break down, as discussed below eq.~\eqref{eq:m2flux}.

Our second example is a totally anti-symmetric representation of rank $N$, which corresponds to a Young tableau with a single column of $N$ boxes. In terms of the partition data, $M_1=N$ and $N_1=1$, which describes $N$ M5-branes each having one M2-brane ending on them. In this case we find
\bea
\label{eq:b6dantisymm}
b_{6d} & = & \frac{3}{5}\left[8 MN + \frac{N^2}{M} - 8 N^2 - N\right] \qquad \textrm{(anti-symmetric)} \\ & = & \frac{3}{5}\frac{N}{M}(M-N)(8M-1), \nn
\eea
where again the second line helps facilitate probe limits. For example, the limit $N \ll M^2$ gives $b_{6d} \approx \frac{24}{5} N \left(M-N\right)$. As shown in ref.~\cite{Rodgers:2018mvq}, this limit agrees with a calculation of EE using probe M5-branes in $AdS_7 \times S^4$ wrapping an $AdS_3$ submanifold of $AdS_7$ and $S^3$ submanifold of $S^4$, again for all values of $N$. For a totally anti-symmetric representation ${\cal R}$, complex conjugation ${\cal R} \to \overline{\cal R}$ acts as $N \leftrightarrow M-N$. Our result for $b_{6d}$ in eq.~\eqref{eq:b6dantisymm} is clearly invariant under $N \leftrightarrow M-N$, as expected. Furthermore, a totally anti-symmetric representation of rank $M$, meaning $N=M$, is equivalent to a trivial representation, and indeed plugging $N=M$ into eq.~\eqref{eq:b6dantisymm} gives $b_{6d}=0$.\footnote{
  \textbf{Note added:} In terms of the trace anomaly coefficients $c$ and $d_2$ defined in eq.~\eqref{eq:defecttrace}, $b_{6d} = c - \frac{3}{5} d_2$ vanishes for an antisymmetric representation with $N = M$ because $c$ and $d_2$ each individually vanish in that case, as expected for a trivial representation.}

Assuming that $b_{6d}$ counts degrees of freedom localized to the Wilson surface, a na\"ive expectation is that in the large-$M$ and/or $N$ limit $b_{6d}$ should scale with the number of degrees of freedom of the M5- or M2-brane theory, $M^3$ or $N^{3/2}$, respectively~\cite{Berman:2007bv}. We can force our result for $b_{6d}$ to scale as $M^3$ or $N^{3/2}$ by choosing a representation such that $N$ scales as a certain power of $M$. For example, for a totally symmetric representation of rank $N \propto M^{3/2}$, in eq.~\eqref{eq:b6dsymm} at large $M$ the term $-8N^2 \propto -8M^3$ with all other terms sub-leading. However, in general none of our results for $b_{6d}$ naturally scale as $M^3$ or $N^{3/2}$. Instead, in our results for $b_{6d}$ most terms scale linearly or quadratically with $M$ or $N$.

As displayed above, our result for $b_{6d}$ can become negative. For example, for a totally symmetric representation of rank $N>8M$, eq.~\eqref{eq:b6dsymm} clearly shows that $b_{6d}<0$. In a standard 2d CFT, unitarity and normalizability of the ground state require the central charge to be non-negative. However, whether unitarity imposes a lower bound on $b_{6d}$ is currently unknown. In similar cases, such as 3d BCFTs, unitarity allows negative values of the boundary central charge. For example, in the free massless scalar 3d BCFT with a Dirichlet boundary condition---a perfectly unitary theory---the boundary central charge is negative~\cite{Nozaki:2012qd,Jensen:2015swa,Fursaev:2016inw}. More generally, for unitary 3d BCFTs  ref.~\cite{Herzog:2017kkj} conjectured a lower bound on the boundary central charge that was negative. The fact that Wilson surfaces in the M5-brane theory and their holographic duals have no other known violations of unitarity leads us to suspect strongly that $b_{6d}<0$ does not necessarily signal unitarity violation.

\subsection{Cousins of the ABJM BCFT}
\label{sec:abjm-bcft-cc}

As discussed in sec.~\ref{sec:ads4part}, in the asymptotically locally $AdS_4 \times S^7$ solutions we cannot identify $M$, so while we can identify a partition $\rho$ and corresponding Young tableau, we cannot identify $\mathfrak{su}(M)$ or a representation ${\cal R}$. We can thus write our result for $b_{3d}$ in eq.~\eqref{b3dads4},
\beq
\label{eq:b3d-cousins}
b_{3d} =\frac{3}{2}\frac{1}{1+\gamma}\sum_{a=1}^n \left(-M_a^2N_a -\gamma M_a N_a^2 +2\sum_{b=1}^a M_b M_a N_a\right),
\eeq
only in terms of $N$, and not $M$. As mentioned below eq.~\eqref{eq:ads4metric}, we consider asymptotically locally $AdS_4 \times S^7$ solutions with $\gamma \in (-1,0)$.

As explained at the end of sec.~\ref{sec:ads4part}, the asymptotically locally $AdS_4 \times S^7$ solutions are invariant, up to an orientation reversal, under the combined operations $\gamma \to \gamma^{-1}$ and $\rho \to \hat{\rho} = \rho^T$. To see how these operations act on $b_{3d}$, we re-write eq.~\eqref{eq:b3d-cousins} in terms of the sets $\{N_b'\}$ and $\{M_b'\}$,
\begin{align}
\label{eq:ads4transposeidentity}
b_{3d} =& \frac{3}{2}\frac{1}{1+\gamma}\bigg[
\sum_{b=1}^n M_{n+1-b}' \left( \sum_{a=b}^n M_{n+1-a} \right)^2
- \gamma \sum_{a=1}^n M_a \left(\sum_{b=a}^n M_b'\right)^2
\bigg]
\nonumber\\
=& \frac{3}{2}\frac{1}{1+\gamma} \bigg[
\sum_{b=1}^n M_{b}' N_b'{}^2 - \gamma \sum_{a=1}^n M_a N_a^2
\bigg],
\end{align}
which can be proven using recursion. Eq.~\eqref{eq:ads4transposeidentity} makes clear that the combined operations of $\gamma \to \gamma^{-1}$ and $\rho \to \hat{\rho}=\rho^T$, which acts as $M_a \to M'_{n-a}$ and $N_a \to N'_{n-a}$, sends $b_{3d} \to - b_{3d}$. Equivalently, in the parametrization $\rho=\{\ell_1,\ell_2,\ldots\}$ and $\hat{\rho} = \rho^T = \{\ell_1',\ell_2',\ldots\}$,
\beq
b_{3d}= \frac{3}{2}\frac{1}{1+\gamma} \left[ - \gamma \sum_{q=1}^n \ell_q^2 + \sum_{\hat{q}=1}^n \ell_{\hat{q}}'{}^2 \right].
\eeq
In this parametrization, $\rho \to \hat{\rho}=\rho^T$ acts as $\ell_q \leftrightarrow \ell_{\hat{q}}'$, which when combined with $\gamma \to \gamma^{-1}$ clearly sends $b_{3d} \to - b_{3d}$.

To see how $b_{3d}$ in eq.~\eqref{eq:b3d-cousins} scales with $N$ we consider two examples of partitions analogous to our two examples for $b_{6d}$ in eqs.~\eqref{eq:b6dsymm} and~\eqref{eq:b6dantisymm}. First, we consider the analogue of the rank $N$ totally symmetric representation, which corresponds to a Young tableau with a single row of $N$ boxes. In terms of partition data, this example has $M_1=1$, $N_1 = N$, and $N_a=0$ for $a\geq 2$, which gives
\beq
\label{eq:b3d-cousins-symm}
b_{3d} = \frac{3}{2(1+\gamma)}N\left(1-\gamma N\right). \qquad \textrm{(``symmetric'')}
\eeq
The large-$N$ limit of eq.~\eqref{eq:b3d-cousins-symm} gives $b_{3d}\approx -\frac{3}{2}\frac{\gamma}{1+\gamma} N^2$. As a second example, we consider the analogue of a rank $N$ the totally anti-symmetric representation, which corresponds to a Young tableau with a single column of $N$ boxes.  In terms of partition data, this example has $N_{1}=1$ and $M_1 =N$, which gives
\beq
\label{eq:b3d-cousins-asymm}
b_{3d} = \frac{3}{2(1+\gamma)}N\left(N-\gamma\right). \qquad \textrm{(``anti-symmetric'')}
\eeq
The large-$N$ limit of eq.~\eqref{eq:b3d-cousins-asymm} gives $b_{3d}\approx \frac{3}{2}\frac{1}{1+\gamma}N^2$. Clearly in both cases at large $N$ we find that $b_{3d}$ scales as $N^2$ rather than $N^{3/2}$. Of course, as discussed in sec.~\ref{sec:intro}, in these cases the 3d BCFTs are only cousins of the ABJM BCFT, and in particular have a different supergroup from that of the maximally SUSY ABJM BCFT. We therefore have no reason \textit{a priori} to expect $b_{3d}$ to scale as $N^{3/2}$ at large $N$.

As with $b_{6d}$ in sec.~\ref{sec:wilson-surface}, the $b_{3d}$ in eq.~\eqref{eq:b3d-cousins} can become negative. The examples in eqs.~\eqref{eq:b3d-cousins-symm} and~\eqref{eq:b3d-cousins-asymm} never are, because $\gamma\in(-1,0)$. However, as we saw that $\gamma \rightarrow \gamma^{-1}$ combined with $\rho\rightarrow \hat{\rho}=\rho^T$ sends $b_{3d}\rightarrow -b_{3d}$, so any partition $\rho$ giving $b_{3d}$ has a corresponding $\rho^T$ giving $-b_{3d}$ within the same $D(2,1;\gamma)\times D(2,1;\gamma)$ super-group. For the reasons discussed at the end of sec.~\ref{sec:wilson-surface}, even if some choices of $M_a$ and $N_a$ make eq.~\eqref{eq:b3d-cousins} negative, we strongly suspect that $b_{3d}<0$ does not necessarily signal unitarity violation.

\subsection{Comparisons to Other Calculations}
\label{sec:other-comparisons}

In this sub-section we will review two previous calculations of the self-dual string central charge, one by Berman and Harvey (BH)~\cite{Berman:2004ew} in sec.~\ref{sec:bh} and one by Niarchos and Siampos (NS)~\cite{Niarchos:2012cy} in sec.~\ref{sec:ns}. We will denote these as $\bbh$ and $\bns$, respectively. In sec.~\ref{sec:freeabjm} we will compute the central charge in the maximally SUSY ABJM BCFT in the free limit, which we will denote as $\bfree$.

Put simply, we find that none of the calculations of the central charge of the self-dual string or Wilson surface---$b_{6d}$ and $b_{3d}$, BH, NS, and free ABJM---agree perfectly with any of the others, though some share certain features. Indeed, all of these calculations extract a central charge from different quantities (EE, anomalies, and thermodynamic entropy), and use different limits of $M$ and $N$, and so we have little reason \textit{a priori} to expect agreement. Nevertheless, aside from the $M^3$ and $N^{3/2}$ scalings of the ambient CFTs, these are some of the few readily available results that we have for comparison.

\subsubsection{R-symmetry Anomaly}
\label{sec:bh}

In ref.~\cite{Berman:2004ew} BH considered $N$ M2-branes ending on $M$ M5-branes and performed an anomaly inflow analysis. In particular, BH demanded that the total R-symmetry chiral anomaly on the self-dual string's 2d worldsheet vanishes. Let us briefly review their arguments.

The 2d R-symmetry is $SO(4)_{R_T}\times SO(4)_{R_N}$, where $SO(4)_{R_T}$ acts on the directions tangent to the M5-brane, $(x_3,x_4,x_5,x_6)$, and $SO(4)_{R_N}$ acts on the directions normal to the M5-brane, $(x_7,x_8,x_9,x_{10})$. If we gauge each $SO(4) \simeq SU(2)_+ \times SU(2)_-$, where the $\pm$ subscripts indicate 2d chirality, with corresponding field strengths $F_{\pm}=dA_{\pm}$, then the Euler class of the gauge bundle over the 2d self-dual string worldsheet is $\chi= \frac{1}{4\pi^2}\left(\Tr F_+^2 - \Tr F_-^2\right)$. Written as an exact 4-form, $\chi= d \Phi$ where the 3-form $\Phi$ has the gauge transformation $\delta \Phi = d \phi$ for some 2-form, $\phi$. The 2d chiral R-symmetry has an anomaly that receives contributions from both 2d and 6d degrees of freedom, the latter via anomaly inflow. An anomaly descent calculation shows that the 2d contribution to the anomaly is $2\pi \bbh \int (\phi(A_{N}) - \phi (A_{T}))$, where $A_T$ and $A_N$ are the $SO(4)_{R_T}$ and $SO(4)_{R_N}$ connections, respectively, on the normal bundle to the self-dual string. BH define a central charge as the coefficient $\bbh$. They determine $\bbh$ by demanding that the total 2d chiral R-symmetry anomaly vanishes, i.e. that the 6d anomaly inflow contribution cancels the 2d contribution.

With a single M5-brane, $M=1$, the 2d R-symmetry chiral anomaly is $\pi N\int (\phi(A_{N}) - \phi (A_{T}))$, and so $\bbh = N/2$. The $SO(4)_{R_N}$ anomaly, $\pi N \int \phi(A_{N})$, is cancelled by a local counterterm on the M5-brane's wordvolume, $-\int dA_2 \wedge \phi(A_N)$, where as in sec.~\ref{sec:intro}, $A_2$ is the M5-brane's worldvolume 2-form gauge field. The $SO(4)_{R_T}$ anomaly, $-\pi N \int \phi(A_{T})$, is cancelled by the gauge variation of the self-dual string's minimal coupling to $A_2$.

With multiple M5-branes, $M>1$, BH move onto the Coulomb branch, separating some M5-branes out of the stack of $M$ coincident M5-branes. In that case, the M5-branes' effective action includes the Ganor-Intriligator-Motl term~\cite{Ganor:1998ve,Intriligator:2000eq}, $\alpha \int dA_2 \wedge \phi(A_T)$, where $\alpha$ is known. Starting from a stack of M5-branes with ADE Lie algebra $\mathfrak{g}$ and breaking to a subgroup with ADE Lie algebra $\mathfrak{h}$ times $\mathfrak{u}(1)$, $\alpha = \frac{1}{4}(\textrm{dim}\,\mathfrak{g} - \textrm{dim}\,\mathfrak{h}-1)$. The pullback of the Ganor-Intriligator-Motl term to the 2d self-dual string worldsheet produces a term with an $SO(4)_{R_N}$ anomaly that must cancel the 2d contribution, which allows BH to identify
\beq
\bbh = \frac{1}{2} N \alpha.
\eeq
If we separate a single M5-brane from the stack then $\mathfrak{g} =\mathfrak{su}(M)$ and $\mathfrak{h}=\mathfrak{su}(M-1)$ (as mentioned in sec.~\ref{sec:intro}, we ignore the overall center-of-mass $\mathfrak{u}(1)$), so $\alpha = \frac{1}{2} \left(M-1\right)$ and hence
\beq
\label{eq:bbhone}
\bbh = \frac{1}{4} N \left(M-1\right) \qquad \textrm{for} \quad \mathfrak{su}(M) \to \mathfrak{su}(M-1) \times \mathfrak{u}(1).
\eeq
If we separate all $M$ M5-branes from each other such that $\mathfrak{g} =\mathfrak{su}(M)$ and $\mathfrak{h}=\mathfrak{u}(1)^{M-1}$, then $\alpha = \frac{1}{4} \left(M^2 - M - 1\right)$ and hence
\beq
\label{eq:bbhmax}
\bbh = \frac{1}{8} N \left(M^2-M-1\right) \qquad \textrm{for} \quad \mathfrak{su}(M) \to \mathfrak{u}(1)^{M-1}.
\eeq

If we compare $\bbh$ in eqs.~\eqref{eq:bbhone} and~\eqref{eq:bbhmax} to $b_{6d}$ of the totally symmetric or anti-symmetric representations in eq.~\eqref{eq:b6dsymm} or~\eqref{eq:b6dantisymm}, respectively, then the only obvious similarities are terms scaling as $MN$ and $N$, with different numerical coefficients. However, we can identify at least four reasons why $b_{6d}$ and $\bbh$ need not agree. First, whether $b_{6d}$ and $\bbh$ are the same quantity is unclear. In a 2d CFT the chiral R-symmetry anomaly coefficient is proportional to the central charge $c$~\cite{Benini:2012cz}, and BH assume the same remains true for a 2d defect in a higher-d CFT. However, that has not been demonstrated, and moreover, whether and how either quantity is related to the defect's EE is unclear. Second, we calculated $b_{6d}$ for any representation ${\cal R}$. However, $\bbh$ seems to involve no data about a representation. Which representation(s) are appropriate in comparing $b_{6d}$ and $\bbh$ (if any) is unclear. Third, our $b_{6d}$ was computed at the conformal point where all M5-branes are coincident, whereas $\bbh$ was computed on the Coulomb branch. Fourth, we computed $b_{6d}$ in the SUGRA limit of large $M$, whereas the calculation of $\bbh$ is in principle valid for any $M$. At best we might expect agreement between $b_{6d}$ and $\bbh$'s large-$M$ limit, and indeed, both our results for $b_{6d}$ in eqs.~\eqref{eq:b6dsymm} and~\eqref{eq:b6dantisymm} and the results for $\bbh$ in eqs.~\eqref{eq:bbhone} and~\eqref{eq:bbhmax} scale as $MN$ at large $M$ but with different numerical coefficients.

If we compare $b_{3d}$ of the totally ``symmetric'' or ``anti-symmetric'' partitions in eqs.~\eqref{eq:b3d-cousins-symm} or~\eqref{eq:b3d-cousins-asymm}, then the only obvious similarities are terms scaling as $N$, with different numerical coefficients. However, as before, BH compute a different quantity, which appears to involve no information about a partition. We can also identify two further reasons why $b_{3d}$ and $\bbh$ need not agree. First, as discussed in secs.~\ref{sec:intro} and~\ref{sec:ads4part}, $b_{3d}$ arises from a case with both M5- and M5$'$-branes, while $\bbh$ arises from a case with no M5$'$-branes. Second, as discussed in sec.~\ref{sec:intro} and~\ref{subsec:ads4metric}, $b_{3d}$ arises in a case with super-group $D(2,1;\gamma)\times D(2,1;\gamma)$ with $\gamma\in(-1,0)$, while BH presumably have $\gamma = 1$.

\subsubsection{Blackfolds}
\label{sec:ns}

In refs.~\cite{Niarchos:2012pn,Niarchos:2012cy} NS found a blackfold description~\cite{Emparan:2009at,Emparan:2011hg,Emparan:2011br} of a fully localized intersection of M2- and M5-branes. Specifically, in ref.~\cite{Niarchos:2012pn} in an effective M5-brane theory they found a 1/4-BPS ``spike'' solution representing $N$ M2-branes ending on the $M$ M5-branes. In ref.~\cite{Niarchos:2012cy} they generalized the spike solution to non-zero temperature $T$, in the limit $N \ll M^2$, and computed the thermal entropy density $S$. In the low-$T$ limit, relative to the system's spatial size, they found
\beq
\label{eq:nsent}
S = \frac{8}{135}\frac{\Gamma(\frac{1}{3})\,\Gamma(\frac{1}{6})\,\sqrt{\pi}}{C^8}\,\frac{N^2}{M}\, T+ {\cal O}(T^4),
\eeq
where the constant $C\approx 1.2$ arises as a matching parameter when ``gluing'' the spike to the M5-branes. NS then compare the term $\propto T$ in eq.~\eqref{eq:nsent} with the Cardy entropy density of a 2d CFT, $\frac{\pi}{3} \, c \, T$~\cite{Cardy:1986ie}, and hence identify a central charge,
\beq
\label{eq:bns}
\bns = \frac{8}{45}\frac{\Gamma(\frac{1}{3})\,\Gamma(\frac{1}{6})}{C^8\,\sqrt{\pi}}\,\frac{N^2}{M} \approx 0.35\,\frac{N^2}{M}.
\eeq

Our result for $b_{6d}$ shares at least one superficial similarity with $\bns$, namely $b_{6d}$ in eq.~\eqref{eq:b6dsugra} includes a term $\frac{3}{5} \frac{N^2}{M} \approx 0.6 \frac{N^2}{M}$, similar to $\bns$. However, whether a limit exists---including in particular NS's limit $N \ll M^2$---in which that term dominates over the others in $b_{6d}$ is unclear. Moreover, we can identify at least three reasons why $b_{6d}$ and $\bns$ need not agree. First, whether $b_{6d}$ and $\bns$ are the same quantity is unclear. Specifically, for a 2d defect there appears to be no universal relation between the defect central charge defined from EE and any Cardy-like contribution to $S$~\cite{Jensen:2018rxu}. Second, whether NS's result for $S$ should be interpreted as a Cardy entropy is highly suspect: NS's result is a \textit{low}-$T$ limit, relative to the system's spatial size, whereas Cardy's result is the \textit{high}-$T$ limit. In a 2d CFT modular invariance relates the high- and low-$T$ limits of $S$, so that $c$ determines both limits. Whether that remain true for a 2d defect is unclear---especially since modular invariance is generically absent for a 2d defect. Lastly, $\bns$ seems to involve no data about a representation, which again leaves open which representation(s) to choose in order to compare $b_{6d}$ and $\bns$, if any.

The $b_{3d}$ of the totally ``symmetric'' or ``anti-symmetric'' partitions in eqs.~\eqref{eq:b3d-cousins-symm} or~\eqref{eq:b3d-cousins-asymm} does not appear to have any similarities with $\bns$. As in the comparison to BH in sec.~\ref{sec:bh}, $b_{3d}$ and $\bbh$ need not agree: they are different quantities, $\bns$ involves no information about a partition, NS's solution has no M5$'$-branes, and NS's solution at $T=0$ presumably has $\gamma=1$ rather than $\gamma\in(-1,0)$.

Ref.~\cite{Niarchos:2012cy} pointed out that if we define the combination
\beq
  \Lambda \equiv \frac{M^2}{N},
\eeq
then the central charge~\eqref{eq:bns} can be written
\beq
  b_\mathrm{NS} \approx 0.35 \frac{N^{3/2}}{\sqrt{\Lambda}} \approx 0.35 \frac{M^3}{\Lambda^2},
\eeq
depending on whether $\Lambda$ is substituted for $M$ or $N$, respectively. Hence, for fixed $\Lambda$, $b_\mathrm{NS}$ scales both as $N^{3/2}$, characteristic of M2-branes, and $M^3$, characteristic of M5-branes. However, the same is not true for our results, in general. For example, making the same substitution in result eq.~\eqref{eq:b6dantisymm} for the central charge of a Wilson surface in an antisymmetric representation, we find (using $M \gg 1$)
\beq
  b_{6d} = \frac{3}{5} \left(\sqrt{\Lambda} N^{3/2} - N^2 \right)
  = \frac{3}{5} \left(\frac{M^3}{\Lambda} - \frac{M^4}{\Lambda^2} \right).
\eeq
which includes a term scaling as $N^{3/2}$ or $M^{3}$, but also another term scaling with a power of $N$ or $M$ that is not necessarily subleading in the SUGRA approximation. Indeed these terms are required to ensure that $b_{6d} = 0$ for the antisymmetric representation with $N = M$, corresponding to a trivial representation.

\subsubsection{Free Limit of the ABJM BCFT}
\label{sec:freeabjm}

A Wilson surface has two equivalent descriptions. The first is from the M5-brane theory perspective, as a non-local operator in the 6d $\N=(2,0)$ SUSY CFT. The second is from the M2-brane perspective, as the boundary of the ABJM BCFT with maximally SUSY boundary conditions~\cite{Berman:2009xd}, and possibly coupled to 2d SUSY multiplets~\cite{Niarchos:2015lla}. In this section we will compute the boundary central charge for the maximally SUSY ABJM BCFT in the free limit, which we denote $\bfree$. We will assume no 2d SUSY multiplets are present, that is, we will calculate the contributions to $\bfree$ only from the fields of ABJM.

As mentioned in sec.~\ref{sec:intro}, the ABJM theory~\cite{Aharony:2008ug} is a 3d $\N=6$ SUSY $U(N)_k \times U(N)_{-k}$ Chern-Simons matter CFT, and is the low-energy description of $N$ M2-branes at a $\mathbb{C}^4/\mathbb{Z}_k$ singularity. When $k=1,2$ the SUSY is enhanced to $\N=8$. The field content is two $\N=2$ vector multiplets, two $\N=2$ adjoint hypermultiplets, and two $\N=2$ bi-fundamental hypermultiplets in complex conjugate representations of the gauge group. The on-shell degrees of freedom include, from the adjoint multiplets, the Chern-Simons gauge fields and their fermionic super-partners, and from the bi-fundemental hypermultiplets, eight real scalar fields describing the positions of the M2-branes in the eight transverse directions, and their fermionic super-partners. The theory's 't Hooft coupling is $N/k$, hence the theory becomes weakly-coupled when $k \gg N$ and free when $k \to \infty$ with $N$ fixed.

The maximally SUSY boundary conditions on the ABJM fields that describe M2-branes ending on M5-branes appear for example in ref.~\cite{Berman:2009xd}, which we now briefly review. Four of the scalars have Dirichlet boundary conditions, representing the fact that the M2-branes cannot move away from the M5-branes in the directions  $(x_7,x_8,x_9,x_{10})$. The other four scalars obey a Basu-Harvey-type equation~\cite{Basu:2004ed}, which in the free limit $k \to \infty$ with $N$ fixed reduces to a Neumann boundary condition, representing the fact that the M2-branes can move freely along the M5-branes in the directions $(x_3,x_4,x_5,x_6)$. The bi-fundamental gauge indices on the scalars must also encode the partition $\rho$ describing which M5-brane each M2-brane ends on. The boundary conditions on the Chern-Simons gauge fields and fermionic super-partners follow from SUSY.

We also want the boundary conditions to preserve conformal symmetry. In the free limit, for scalar fields Dirichlet and Neumann boundary conditions each preserve conformal symmetry. For Chern-Simons gauge fields conformal invariance requires a boundary term producing a Wess-Zumino-Witten (WZW) model at the 2d boundary. For Dirac fermions conformal symmetry requires acting with a projector that produces a single chiral fermionic mode at the 2d boundary. The maximally superconformal boundary conditions ensure that the WZW and 2d chiral fermion degrees of freedom together preserve parity. For what follows we will need no further details about the boundary conditions.

For 3d BCFTs the central charge defined from EE appears also as a central charge in the trace anomaly~\cite{Fursaev:2013mxa,Fursaev:2016inw}, in essentially the same way as a 2d CFT. To be precise, in a 3d BCFT the only non-zero contribution to the trace anomaly comes from the 2d boundary~\cite{Schwimmer:2008yh}, and includes a term $\frac{b}{24\pi} \int \hat{R}$ localized at the boundary, where $\hat{R}$ is the Ricci scalar of the boundary's induced metric. The coefficient of that Ricci scalar term obeys a $c$-theorem for boundary RG flows, and hence serves as a measure of the number of degrees of freedom localized at the boundary~\cite{Jensen:2015swa}. In 3d the contribution to the trace anomaly's 2d Ricci scalar term is known for Chern-Simons gauge fields, free Dirac fermions, and free scalars with Dirichlet and Neumann boundary conditions~\cite{Jensen:2015swa,Nozaki:2012qd,Fursaev:2016inw}. We can thus calculate $b$ for the maximally SUSY ABJM BCFT in the free limit simply by summing the known results for the on-shell fields in the SUSY multiplets mentioned above.

Crucially, for a free Dirac fermion in 3d, $b=0$, so in the maximally SUSY ABJM BCFT in the free limit the fermion's contribution to $\bfree$ is zero. For a free real scalar in 3d the values of $b$ for Dirichlet and Neumann boundary conditions are equal and opposite. In the maximally SUSY ABJM BCFT in the free limit, four scalars have Dirichlet boundary conditions and the other four have Neumann, so the scalars' net contribution to $\bfree$ is also zero. As a result, $\bfree$ will be blind to the partition $\rho$ describing which M5-brane each M2-brane ends on. We are left with only the $U(N)_k \times U(N)_{-k}$ Chern-Simons gauge fields, whose contribution to $\bfree$ is two copies of the WZW central charge with the same $N$ but levels $k$ and $-k$. Starting from a Chern-Simons theory with gauge Lie algebra $\mathfrak{g}$ at level $k$, the central charge of the corresponding WZW model is
\beq
c_{\textrm{WZW}} = \frac{k \, \text{dim} \,\mathfrak{g}}{k +g^{\vee}},
\eeq
where $g^{\vee}$ is $\mathfrak{g}$'s dual Coxeter number. For $\mathfrak{g}=\mathfrak{u}(N)$, we have $\text{dim }\mathfrak{g}=N^2$ and $g^{\vee}=N$. The maximally SUSY ABJM BCFT in the free limit therefore has a boundary central charge
\beq
\label{eq:bfree}
\bfree = \lim_{k \to \infty} \left[\frac{k \, N^2}{k +N} + \frac{(-k) \, N^2}{(-k) +N}\right] = \lim_{k \to \infty}\frac{2\,k^2 \,N^2}{k^2-N^2} = 2 N^2,
\eeq
which indeed contains no information about a partition.

The only similarity between $b_{6d}$ and $\bfree$ is that $b_{6d}$ contains terms scaling as $N^2$ at large $N$, though generically with different coefficients, as obvious in the examples of eqs.~\eqref{eq:b6dsymm} and~\eqref{eq:b6dantisymm}. In some cases the $N^2$ term in $b_{6d}$ dominates. For example, a totally symmetric representation in the large-$M$ and $N \gg M$ limits has $b_{6d} \approx - \frac{3}{5} N^2$, which however clearly disagrees with $\bfree$, even in sign.

Superficially, $b_{3d}$ is more similar to $\bfree$ than $b_{6d}$ is. Indeed, the examples in eqs.~\eqref{eq:b3d-cousins-symm} and~\eqref{eq:b3d-cousins-asymm} clearly scale as $N^2$ at large $N$. In fact, the coefficient even agrees in one case, namely $b_{3d}$ for the ``anti-symmetric'' partition in eq.~\eqref{eq:b3d-cousins-asymm}, with $\gamma = -1/4$.

However, we can identify at least two reasons why $b_{6d}$ or $b_{3d}$ and $\bfree$ need not agree. First, similar to the previous comparisons, for $b_{3d}$ the 3d BCFT has a different super-group from the maximally SUSY ABJM BCFT. Second, we calculated $b_{6d}$ and $b_{3d}$ via holography, and thus in the limit of large 't Hooft coupling $\lambda = N/k\gg 1$ and hence $N\gg 1$, whereas obviously we calculated $\bfree$ in the free limit. As a result, any agreement between $b_{6d}$ or $b_{3d}$ and $\bfree$ is likely just a coincidence, and should probably be treated with skepticism.

\section{Discussion and Outlook}
\label{sec:discussion}

We used the 11d SUGRA solutions of refs.~\cite{Bachas:2013vza,DHoker:2008rje,D'Hoker:2008wc,Estes:2012vm} to compute holographically the EE in two cases. First was a spherical region centered on a Wilson surface in the M5-brane theory, describing $N$ M2-branes ending on $M$ M5-branes at large $M$. Second was the EE of a semi-circular region centered on the 2d boundary in cousins of the maximally SUSY ABJM BCFT, also at large $N$. The Wilson surface or 2d boundary is characterized by a partition $\rho$ of $N$, which for the Wilson surface determines a representation ${\cal R}$ of the M5-branes' worldvolume $\mathfrak{su}(M)$ gauge algebra. From our result for EE we extracted a central charge, $b_{6d}$ or $b_{3d}$, as a contribution from the Wilson surface or 2d boundary to the coefficient of the term logarithmic in the cutoff. Presumably $b_{6d}$ or $b_{3d}$ provides one measure of the number of massless degrees of freedom on the Wilson surface or 2d boundary, respectively.

Our main result for $b_{6d}$ in eq.~\eqref{eq:b6dcompact} is written compactly in terms of ${\cal R}$'s highest weight vector, $\lambda$, and the $\mathfrak{su}(M)$ Weyl vector, $\vr$, and is manifestly invariant under the action of the Weyl group, including complex conjugation, ${\cal R} \to \overline{\cal R}$. We found that neither $b_{6d}$ nor $b_{3d}$ naturally scales with the number of degrees of freedom of the M5- or M2-brane theories at large $M$ or $N$, namely $M^3$ or $N^{3/2}$, respectively. Instead, for several examples of $\rho$ we found that $b_{6d}$ and $b_{3d}$ typically scale as $M^2$ or $N^2$. Our results also do not meaningfully agree with previous calculations of self-dual string central charges by Berman and Harvey~\cite{Berman:2004ew} and Niarchos and Siampos~\cite{Niarchos:2012cy} (nor do the results in those two references agree with each other), and indeed we provided a long list of reasons why. For example, a chiral anomaly was calculated in ref.~\cite{Berman:2004ew} while ref.~\cite{Niarchos:2012cy} computed thermodynamic entropy. The relation of either of these quantities to EE for a 2d defect or boundary is unclear. Moreover, neither of the previous putative central charges contain any information about the partition of $N$, and so it is unclear which $\rho$ to chose for our comparisons.

Indeed, we know of several more promising possible comparisons, which could provide more details about the origin and implications of our results. Examples include holographic EE in the asymptotically locally $AdS_4 \times S^7$ solutions dual to M2-branes ending on M5-branes~\cite{Bachas:2013vza} (which have potentially dangerous singularities), elliptic genera of M2-branes suspended between M5-branes~\cite{Haghighat:2013gba}, or the superconformal index of the M5-brane theory compactified on $S^1 \times S^5$---where upon dimensional reduction along the $S^1$ the Wilson surface reduces to a Wilson line of the effective 5d maximally SUSY Yang-Mills (SYM) theory on $S^5$~\cite{Bullimore:2014upa}. In some sense all of these should be counting the same massless degrees of freedom that contribute to our $b_{6d}$, though perhaps in different limits.

Our result for $b_{6d}$ in eq.~\eqref{eq:b6dcompact} also suggests a tantalizing potential connection to Toda CFTs. In particular, an $A_{M-1}$ Toda CFT has primaries labeled by representations ${\cal R}$ which have scaling dimensions $\frac{1}{2} \left [ 2Q \left(\lambda,\varrho\right)-\left(\lambda,\lambda\right)\right],$ with background charge $Q$. Our $b_{6d}$ is functionally similar to that with $Q=8$, but with an overall $3/5$ factor instead of $1/2$.

A natural question is thus whether we can reproduce our result for $b_{6d}$ via the AGT correspondence~\cite{Alday:2009aq}. Imagine compactifying M5-branes on a Riemann surface times a squashed $S^4$. The low-energy effective theory on the squashed $S^4$ will be an $\N=2$ SYM theory whose field content depends on the Riemann surface's genus and punctures. The AGT correspondence is the statement that the 4d SYM theory's partition function on the squashed $S^4$, which can be computed via SUSY localization~\cite{Pestun:2007rz,Hama:2012bg}, is equivalent to a certain correlator in a Toda CFT on the Riemann surface, with background charge $Q$ determined by the $S^4$'s squashing parameters. A 2d defect in the M5-brane theory that descends to a 2d defect in the 4d SYM theory appears in the Toda theory as a degenerate operator~\cite{Alday:2009fs,Kozcaz:2010af}. A key question is thus whether and how the dimension of that degenerate operator determines our $b_{6d}$. In other words, can we reproduce our result for $b_{6d}$, in whole or in part, from a calculation either in the Toda CFT with degenerate operator or in the 4d SYM theory with 2d defect~\cite{Bullimore:2014nla,Gomis:2014eya,Gomis:2016ljm}? In particular, does the dimension of the degenerate operator in the Toda CFT determine the EE of the 2d defect in the 4d $\N=2$ SYM? Or is the similarity we found merely a coincidence?

We are currently investigating all of the above possible comparisons. However, more generally we hope our results may help shed light on the mysterious degrees of freedom of self-dual strings, the 6d $\N=(2,0)$ CFT, and most of all, M-theory.

\acknowledgments

We would like to thank Costas Bachas, David Berman, Matthew Buican, Nadav Drukker, Christopher Herzog, Matti J\"arvinen, Kristan Jensen, S.~Prem Kumar, Bruno Le Floch, Vasilis Niarchos, and Kostas Skenderis for useful discussions and correspondence. D.~K. acknowledges support of PSC-CUNY Research Award. A.~O'B. is a Royal Society University Research Fellow. B.~R. and R.~R. acknowledge support from STFC through Consolidated Grant ST/L000296/1. A.~O'B., B.~R., and R.~R. thank the Institut Henri Poincar\'e for hospitality while this work was in progress.

\appendix

\section{Integrals for the Entanglement Entropy}

In this appendix we perform the integrals for ${\cal I}$ in the asymptotically locally $AdS_4 \times S^7$ case, eq.~\eqref{integralcomplex}, obatining the result in eq.~\eqref{eq:ads4int2}. The integrals in eq.~\eqref{integralcomplex} are
\beq
\label{integralcomplex2}
{\cal I}  = \frac{ (2 \pi^2)^2}{4 \gn} \frac{4}{c_1 c_2^4 c_3^4}  \int_0^{\pi} d\theta \sin\theta \int_{0}^{r_c(\varepsilon_v,\theta)} dr \,
r^2 \left(1 - \sum_{j,k=1}^{2n+1} (-1)^{j+k} \frac{r e^{i\theta} - \xi_j}{|r e^{i\theta} - \xi_j|} \frac{r e^{-i\theta} - \xi_k}{|r e^{i\theta} - \xi_k|} \right),
\eeq
with the cutoff $r_c(\varepsilon_v,\theta)$ in eq.~\eqref{eq:AdS4cutoff}. We re-write the sums in eq.~\eqref{integralcomplex2} as
\begin{align}
\label{eq:newsums}
\sum_{j,k=1}^{2n+1} (-1)^{j+k} \frac{r e^{i\theta} - \xi_j}{|r e^{i\theta} - \xi_j|} \frac{r e^{-i\theta} - \xi_k}{|r e^{i\theta} - \xi_k|} &=  2n  + 1
\nonumber \\ &\phantom{=}
+ 2\sum_{\underset{j<k}{j,k=1}}^{2n+1} \frac{(-1)^{j+k} \left(r^2 + \xi_j \xi_k-r\cos\theta(\xi_k+ \xi_j)\right)}{\sqrt{r^2 + \xi_j^2-2r\xi_j\cos\theta}\sqrt{r^2 + \xi_k^2-2r\xi_k\cos\theta}}
.
\end{align}
After substituting, the integral eq.~\eqref{integralcomplex} becomes
\beq
\label{integralcomplex2a}
{\cal I}  = - \frac{ (2 \pi^2)^2}{4 \gn} \frac{8}{c_1 c_2^4 c_3^4}  \int_0^{\pi} d\theta \sin\theta \int_{0}^{r_c(\varepsilon_v,\theta)} dr \,
r^2 \left(n + \sum_{\underset{j<k}{j,k=1}}^{2n+1} (-1)^{j+k} \frac{r e^{i\theta} - \xi_j}{|r e^{i\theta} - \xi_j|} \frac{r e^{-i\theta} - \xi_k}{|r e^{i\theta} - \xi_k|} \right).
\eeq
We decompose the integral into the following pieces
\beq
\label{integralcomplex3}
{\cal I}  = - \frac{ (2 \pi^2)^2}{4 \gn} \frac{8}{c_1 c_2^4 c_3^4}  \left( K_0+ K_1 + K_2 + K_3\right),
\eeq
where
\beq
K_0 \equiv n\int_0^{\pi} d\theta \sin\theta  \int_{0}^{r_c} dr \, r^2 =n\int_0^{\pi} d\theta \sin\theta \, \frac{r_c^3}{3},
\eeq
and we will not need to perform the $\theta$ integration because $K_0$ will cancel against a term in $K_1+K_2+K_3$ (specifically a term in $K_1$: see eq.~\eqref{eq:k0k1}), where we define $K_1$, $K_2$, and $K_3$ as follows.

If we plug eq.~\eqref{eq:newsums} into eq.~\eqref{integralcomplex2} then we obtain integrals nearly identical to those of ref.~\cite{Gentle:2015jma}. In fact, our results will differ from those of ref.~\cite{Gentle:2015jma} only because we have $2n+1$ branch points $\xi_j$ rather than $2n+2$, we will leave $m_1$ arbitrary rather than restricting to $m_1=2$, and our cutoff $r_c(\varepsilon_v,\theta)$ will be different. Explicitly, if we plug eq.~\eqref{eq:newsums} into eq.~\eqref{integralcomplex2} then from the terms involving the sums over $\xi_j$ and $\xi_k$ we obtain the integrals
\beq
\int_0^{\pi} d\theta \sin\theta  \int_{0}^{r_c} dr \, r^2 \, \sum_{\underset{j<k}{j,k=1}}^{2n+1} \frac{(-1)^{j+k} \left(r^2 + \xi_j \xi_k-r\cos\theta(\xi_k+ \xi_j)\right)}{\sqrt{r^2 + \xi_j^2-2r\xi_j\cos\theta}\sqrt{r^2 + \xi_k^2-2r\xi_k\cos\theta}} \equiv K_1 + K_2 + K_3,
\eeq
where we define $K_1$, $K_2$, and $K_3$ by expansions in Legendre polynomials, $P_k\left(\cos \theta\right)\equiv P_k$,
\begin{align}
& K_1 \equiv  \nn\\
& \sum_{\underset{j<k}{j,k=1}}^{2n+1} \int_0^{\pi} d\theta \sin\theta \int_{|\xi_k|}^{r_c} dr \frac{(-1)^{j+k} r^2(r^2 + \xi_j \xi_k-r\cos\theta(\xi_k+ \xi_j))}{r^2}\sum_{l,m=0}^{\infty}P_lP_m\left(\frac{\xi_j}{r}\right)^l\left(\frac{\xi_k}{r}\right)^m,\nn\\
& K_2 \equiv  \nn\\
& \sum_{\underset{j<k}{j,k=1}}^{2n+1}\int_0^{\pi} d\theta \sin\theta \int_{|\xi_j|}^{|\xi_k|} dr \frac{(-1)^{j+k}r^2(r^2 + \xi_j \xi_k-r\cos\theta(\xi_k+ \xi_j))}{r|\xi_j|}\sum_{l,m=0}^{\infty}P_lP_m\left(\frac{\xi_j}{r}\right)^l\left(\frac{r}{\xi_k}\right)^m,\nn\\
& K_3 \equiv  \nn\\
& \sum_{\underset{j<k}{j,k=1}}^{2n+1} \int_0^{\pi} d\theta \sin\theta \int^{|\xi_j|}_0 dr \frac{(-1)^{j+k}r^2(r^2 + \xi_j \xi_k-r\cos\theta(\xi_k+ \xi_j))}{|\xi_j||\xi_k|}\sum_{l,m=0}^{\infty}P_lP_m\left(\frac{r}{\xi_j}\right)^l\left(\frac{r}{\xi_k}\right)^m.\nn
\end{align}
The integrals $K_1$ and $K_2$ are identical in form to those in ref.~\cite{Gentle:2015jma}, though with sums up to $2n+1$ rather than $2n+2$, while in $K_3$ we correct a few typos in ref.~\cite{Gentle:2015jma}, although we still reproduce the subsequent equations of ref.~\cite{Gentle:2015jma}. Performing the integrals over $r$ and $\theta$ in $K_2$ and $K_3$, we find
\begin{align}
&K_2= \nn\\
&-\sum_{\underset{j<k}{j,k=1}}^{2n+1} (-1)^{j+k}
\sum_{l=0}^{\infty}\frac{2}{2l+1}\mathrm{sign}(\xi_k)\frac{\xi_j^l}{\xi_k^{l+1}}\left(\frac{\xi_k^4-\xi_j^4}{4}+ \frac{\xi_k^2-\xi_j^2}{2}\xi_j\xi_k\right)
\left[1-\frac{l+1}{2l+3}\left(1+\frac{\xi_j}{\xi_k}\right)\right],\\
&K_3= \nn\\
&\sum_{\underset{j<k}{j,k=1}}^{2n+1} (-1)^{j+k}
\sum_{l=0}^{\infty}\frac{2}{2l+1}\mathrm{sign}(\xi_k)\frac{\xi_j^l}{\xi_k^{l}}\left\{\frac{\xi_j^4}{\xi_k(5+2l)}\left[1-\frac{l+1}{2l+3}\left(2+\frac{\xi_k}{\xi_j}+\frac{\xi_j}{\xi_k}\right)\right]+ \frac{\xi_j^3}{3+2l}\right\}.
\end{align}
Performing the integral over $r$ in $K_1$, we find
\begin{multline}
\label{eq:k1def}
K_1=\sum_{\underset{j<k}{j,k=1}}^{2n+1} (-1)^{j+k} \int_0^{\pi} d\theta \sin\theta
\left( \sum_{\underset{l+m\neq 3}{l,m=0}}^{\infty} \frac{P_l P_m \xi_j^l \xi^m_k r^{3-l-m}}{3-l-m} \right. \\
-\cos\theta(\xi_j+\xi_k)\sum_{\underset{l+m\neq 2}{l,m=0}}^{\infty}\frac{P_l P_m \xi_j^l \xi^m_k r^{2-l-m}}{2-l-m}+\xi_j\xi_k\sum_{\underset{l+m\neq 1}{l,m=0}}^{\infty}\frac{P_l P_m \xi_j^l \xi^m_k r^{1-l-m}}{1-l-m} +  \\ \left.
\left(\sum_{\underset{l+m=3}{l,m=0}}^{3} P_l P_m \xi_j^l \xi^m_k -\cos\theta(\xi_j+\xi_k)
\sum_{\underset{l+m=2}{l,m=0}}^{2} P_l P_m \xi_j^l \xi^m_k+\xi_j\xi_k\sum_{\underset{l+m=1}{l,m=0}}^{1}P_l P_m \xi_j^l \xi^m_k
\right) \log r  \right)^{r_c}_{|\xi_k|}.
\end{multline}
Let $K_1 \equiv K_1^{\mathrm{lower}} + K_1^{\mathrm{upper}}$, with $K_1^{\mathrm{lower}}$ and $K_1^{\mathrm{upper}}$ representing the contributions from the lower and upper endpoints of the $r$ integration in eq.~\eqref{eq:k1def}, respectively. The lower limit is independent of the cutoff $r_c(\varepsilon_v,\theta)$, hence upon performing the integration over $\theta$ our result for $K_1^{\mathrm{lower}}$ is nearly identical to that of ref.~\cite{Gentle:2015jma},
\begin{align}
K_1^{\mathrm{lower}}=-\sum_{\underset{j<k}{j,k=1}}^{2n+1} (-1)^{j+k}
\sum_{l=0}^{\infty}\frac{2}{2l+1}\mathrm{sign}(\xi_k)\frac{\xi_j^l}{\xi_k^{l}}\left(\frac{\xi_k^3}{3-2l}+ \frac{\xi_k^2\xi_j}{1-2l}-\frac{\xi_k(\xi_k+\xi_j)^2(l+1)}{(2l+3)(1-2l)}\right).
\end{align}
As in ref.~\cite{Gentle:2015jma}, miraculous cancellations occur, leading to a very simple expression for
\begin{align}
\label{simplecombo}
K_1^{\mathrm{lower}}+K_2+K_3=\frac{1}{3}\sum_{\underset{j<k}{j,k=1}}^{2n+1} (-1)^{j+k}|\xi_j-\xi_k|^3.
\end{align}
For $K_1^{\mathrm{upper}}$ we find
\begin{multline}
K_1^{\mathrm{upper}}=\sum_{\underset{j<k}{j,k=1}}^{2n+1} (-1)^{j+k} \int_0^{\pi} d\theta \sin\theta  \\
\left(\frac{r_c^3}{3} -(\xi_j-\xi_k)^2\frac{\sin^2\theta}{2}r_c
-(\xi_j-\xi_k)^2(\xi_j+\xi_k)\cos\theta\sin^2\theta \log r_c   \right).
\end{multline}
The term involving $r_c^3$ is independent of $j$ and $k$. Using the fact that $\sum_{\underset{j<k}{j,k=1}}^{2n+1} (-1)^{j+k}=-n$, we find that the term in $K_1$ involving $r_c^3$ exactly cancels against $K_0$. In the remaining terms of $K_1^{\mathrm{upper}}$, we plug in $r_c(\varepsilon_v,\theta)$ from eq.~\eqref{eq:AdS4cutoff} and perform the $\theta$ integration, with the result
\begin{align}
\label{eq:k0k1}
K_0+K_1^{\mathrm{upper}}=-\frac{1}{3}\frac{1+\gamma}{\sqrt{-\gamma}}\sqrt{-m_1^2-m_2}\sum_{\underset{j<k}{j,k=1}}^{2n+1} (-1)^{j+k} (\xi_j-\xi_k)^2\frac{1}{\varepsilon_v}.
\end{align}
We perform the sums over $j$ and $k$ using the definition of the $m_k$ in eq.~\eqref{eq:ads4legendre},
\begin{align}
\sum_{\underset{j<k}{j,k=1}}^{2n+1} (-1)^{j+k} (\xi_j-\xi_k)^2=-m_1^2-m_2.
\end{align}
We thus find
\begin{align}\label{eq:K1upper}
K_0+K_1^{\mathrm{upper}}=-\frac{1}{3}\frac{1+\gamma}{\sqrt{-\gamma}}(-m_1^2-m_2)^{3/2}\frac{1}{\varepsilon_v}.
\end{align}
Using eq.~\eqref{eq:s7radius} to replace $-m_1^2-m_2$ with $L_{S^7}$ we find
\beq
K_0+K_1^{\mathrm{upper}}=-\frac{1}{3}c_1^9\,\frac{\gamma^4}{\left(1+\gamma\right)^8}\frac{L_{S^7}^9}{2^6}\frac{1}{\varepsilon_v}.
\eeq

Plugging $K_0+K_1+K_2+K_3$ into eq.~\eqref{integralcomplex3}, we thus find
\beq
{\cal I}  =\frac{(2 \pi^2)^2}{4 \gn} \frac{c_1^8}{c_2^4c_3^4} \frac{\gamma^4}{\left(1+\gamma\right)^8}\frac{L_{S^7}^9}{24} \frac{1}{\varepsilon_v} - \frac{(2 \pi^2)^2}{4 \gn} \frac{1}{ c_1 c_2^4 c_3^4}\frac{8}{3}\sum_{\underset{j<k}{j,k=1}}^{2n+1} (-1)^{j+k}(\xi_j-\xi_k)^3.
\eeq
Using eq.~\eqref{eq:celim} to eliminate $c_1$, $c_2$, and $c_3$ in the first term, we find
\beq
{\cal I}  =\frac{(2 \pi^2)^2}{4 \gn} \frac{L_{S^7}^9}{24} \frac{1}{\varepsilon_v} - \frac{(2 \pi^2)^2}{4 \gn} \frac{1}{ c_1 c_2^4 c_3^4}\frac{8}{3}\sum_{\underset{j<k}{j,k=1}}^{2n+1} (-1)^{j+k}(\xi_j-\xi_k)^3,
\eeq
which we quote in eq.~\eqref{eq:ads4int2}.

\bibliographystyle{JHEP}
\bibliography{wilsonccv3.bib}

\end{document}